\documentclass[%
 aip,
amsmath,amssymb,
 reprint,%
]{revtex4-1}

\usepackage{graphicx}
\usepackage{dcolumn}
\usepackage{bm}

\usepackage[utf8]{inputenc}
\usepackage[T1]{fontenc}
\usepackage{mathptmx}
\usepackage{etoolbox}
\usepackage{xcolor}

\makeatletter
\def\@email#1#2{%
 \endgroup
 \patchcmd{\titleblock@produce}
  {\frontmatter@RRAPformat}
  {\frontmatter@RRAPformat{\produce@RRAP{*#1\href{mailto:#2}{#2}}}\frontmatter@RRAPformat}
  {}{}
}%
\makeatother
\begin{document}

\title{Real-time dynamics with bead-Fourier path integrals I: Bead-Fourier CMD}

\author{Nathan London}
\author{Mohammad R. Momeni}
\email{mmomenitaheri@umkc.edu}
\date{}
\affiliation{Division of Energy, Matter and Systems, School of Science and Engineering, University of Missouri $-$ Kansas City, Kansas City 64110, MO, United States.}

\date{\today}

\begin{abstract}
Developing new methods for the accurate and efficient calculations of real-time quantum correlation functions is deemed one of the most challenging problems of modern condensed matter theory. Many popular methods, such as centroid molecular dynamics (CMD), make use of Feynman path integrals (PIs) to efficiently introduce nuclear quantum effects into classical dynamical simulations. Conventional CMD methods use the discretized form of the PI formalism to represent a quantum particle using a series of replicas, or ``beads'', connected with harmonic springs to create an imaginary time ring polymer. The alternative Fourier PI methodology, instead, represents the imaginary time path using a Fourier sine series. Presented as an intermediary between the two formalisms, bead-Fourier PIs (BF-PIs) have been shown to reduce the number of beads needed to converge equilibrium properties by including a few terms of the Fourier series. Here, a new CMD method is presented where the effective potential is calculated using BF-PIs as opposed to the typical discretized PIs. We demonstrate the accuracy and efficiency of this new BF-CMD method for a series of 1D model systems and show that at low temperatures, one can achieve a 4-fold reduction in the number of beads with the addition of a single Fourier component. The developed methodology is general and can be extended to other closely related methods, such as ring polymer molecular dynamics (RPMD), as well as non-adiabatic PI methods. 
\end{abstract}

\maketitle

\section{Introduction}\label{sec:intro}
Nuclear quantum effects (NQEs) such as zero-point energy and quantum tunneling play a critical role in a number of chemical phenomena.\cite{Markland2018} Some examples include chemical reaction rates, especially those involving proton
transfers,\cite{Meisner2016} and vibrational spectroscopy.\cite{Marsalek2017,Kapil2024} Thus, in order to accurately study these processes, it is necessary to capture NQEs within theoretical methods and simulations.
One set of methods for including NQEs within dynamical simulations make direct use of the Schr\"{o}dinger equation to quantize the nuclei. Some methods, including multiconfigurational time-dependent Hartree (MCTDH)\cite{janBeck2000a} and variational multiconfigurational Gaussian (vMCG),\cite{aprRichings2015} propagate nuclear wavepackets according to
the time-dependent Schr\"{o}dinger equation. Other methods, like those based on nuclear-electronic orbitals (NEO),\cite{sepWebb2002} treat nuclei and electrons on equal footing in the context of electronic structure calculations. While these methods have proven to be highly accurate, they are typically not conducive to simulating large-scale, condensed-phase systems.
These types of large systems are often restricted to the realm of classical molecular dynamics (MD) simulations, where the typical use of analytical force fields results in favorable computational scaling with system size. It is thus reasonable to desire a way to introduce NQEs into the classical MD framework in a way that is computationally feasible, with one such way being to utilize Feynman's path integral (PI) formulation of quantum statistical mechanics.\cite{julFeynman2010}

Currently, the most utilized formalism of PIs is in the form of ``discretized'' PIs where a quantum particle is represented as a series of identical classical particles, referred to as beads connected via harmonic springs to form a closed loop, referred to as a ring polymer (RP).\cite{julFeynman2010,aprCeperley1995} Using PI Monte Carlo (PIMC) or PI molecular
dynamics (PIMD), NQEs can be included in producing accurate equilibrium properties.\cite{janParrinello1984a} For systems including light atoms and/or at low temperatures where RPs with a large number of beads are needed, sufficient sampling of the large, extended phase space can pose a challenge.\cite{sepRyu2022} One method intended to directly combat this is coarse-graining of PIs (CG-PI), which treats
each particle as two pseudoparticles instead of as an $n$ bead RP.\cite{Ryu2019,sepRyu2022} While the method can match standard PIMD results, it does require additional parameterizations specific for each system.

These discretized PIs have also been utilized to perform approximate real-time dynamics that can incorporate NQEs, with the exception of quantum coherence. One method, centroid MD (CMD), introduced by Voth and coworkers, propagates a classical particle under an effective potential of an imaginary time PI, which has its centroid constrained to the position of the classical particle.\cite{cmd1,cmd2,cmd3,cmd4,cmd5,augJang1999,Jang:1999}
Variations of CMD, such as partially adiabatic (PA)-CMD\cite{aprHone2006a}, allow the method to be applied to large-scale systems through the calculation of the effective potential being done ``on-the-fly'' during dynamical simulations. The masses of the non-centroid normal modes of the RP are scaled down and attached to a thermostat so that they are moving on a much faster scale than the centroid and are properly sampling the effective potential.\cite{aprHone2006a} 

While CMD methods have been useful, they do have issues with the calculation of vibrational spectra, where a large delocalization of the RP inside of the potential curve of the stretching modes leads to the flattening of the effective potential and red-shifting of the stretch peaks, commonly referred to as the ``curvature problem''.\cite{janIvanov2010} This problem becomes worse at lower temperatures as the number of beads needed to converge the effective potential increases. Several methods have been developed to work to overcome the curvature problem, including quasi-centroid (QC)MD, which uses a set of curvilinear coordinates to create a more compact RP distribution and thus remove the flattening of the potential.~\cite{augTrenins2019,novTrenins2022} Methods to make QCMD more viable for large-scale systems, like fast-QCMD,~\cite{decFletcher2021,octLawrence2023} and for complex interfacial systems, like hybrid-CMD,~\cite{janLimbu2025} have also been developed. Alternatively, Te-PIGS overcomes the curvature problem by training a neural network potential on the effective CMD potential at a temperature high enough that the curvature problem is negligible.~\cite{tepigs} While these methods have proven successful, they have their own limitations and computational complexities.

An alternative PI approach is to not use the discretized PIs of PIMC/PIMD. Instead, the quantum canonical partition function is calculated as a sum over all closed paths in imaginary time weighted by their imaginary time action.~\cite{marDoll1984,julDoll1985,sepLobaugh1992} The paths to be considered can be reduced by expressing them as small fluctuations about free particle trajectories through the use of a Fourier sine series, resulting in these methods being referred to as Fourier (F)-PI methods.~\cite{marDoll1984,julDoll1985} F-PIs should yield accurate equilibrium properties in the limit of an infinite Fourier series, but only a finite number can be included in a calculation. To combat this, the effects of the non-included terms can be incorporated through partial averaging~\cite{julDoll1985}. Additionally, the use of the full Fourier series instead of just the sine series has been proposed.~\cite{sepLobaugh1992}

The number of Fourier components needed even for systems with just heavy atoms can reach into the hundreds,~\cite{marTopper1993} making F-PI methods very computationally demanding. As previously discussed, discretized PI methods can also require a large number of beads at low temperatures.
As an intermediate between these two extremes of representing imaginary time PIs, Gorbunov and coworkers introduced bead-Fourier (BF-)PIMC.~\cite{febVorontsov1997} In BF-PIMC, the imaginary time paths of F-PIs are divided into $n$ segments in imaginary time.~\cite{febVorontsov1997} This results in an $n$ bead imaginary time RP with the paths between the beads include the fluctuations of the Fourier series. The BF-PI representation, both BF-PIMC~\cite{febVorontsov1997} and its extension to MD (BF-PIMD),~\cite{junIvanov2003,julIvanov2005} are able to achieve accurate results with a small number of beads through the addition of a few Fourier components.~\cite{febVorontsov1997,junIvanov2003,julIvanov2005}

Here, we capitalize on the fairly quick convergence of BF-PIs for approximate real-time dynamics. We introduce a new method, BF-CMD, where we use BF-PIs to calculate the effective CMD potential used in dynamical simulations. The rest of the paper is outlined as follows: in Section~\ref{sec:theory}, we outline the basics of BF-PIs and discretized CMD before discussing how we combine the two methodologies into BF-CMD. Section~\ref{sec:sim-dets} discusses the model systems on which we will evaluate the performance of BF-CMD as well as our simulation details. We present and discuss our results in Section~\ref{sec:results} and end with conclusions and future work.

\section{Theory}\label{sec:theory}

  \subsection{Bead-Fourier Path Integrals}
  The bead-Fourier method combines the discretized and Fourier approaches to imaginary time PIs into one method. The path of a particle in imaginary time is represented using a number of discrete slices, which are equivalent to beads in discretized PIs, connected using a Fourier sine series,\cite{febVorontsov1997,junIvanov2003}

  \begin{equation}
    q_{j}(\xi) = q_{j} + (q_{j+1}-q_{j})\xi + \sum^{k_{\mathrm{max}}}_{k=1}a_{jk}\sin(k\pi\xi), 
    \label{eq:path}
  \end{equation}
  where $k_{\mathrm{kmax}}$ is the maximum number of terms included in the Fourier series, $ q_{j} $ is the position of the $j^{th}$ bead, $ a_{jk} $ is the Fourier amplitude for the $k^{th}$ term in the series for the $j^{th}$ bead, and $ \xi $, which runs between 0 and 1, is the position along the imaginary time slice between beads $j$ and $j+1$.
   
  In this framework, the canonical partition function can be written as \cite{febVorontsov1997,junIvanov2003}
  \begin{equation}
    Z  \propto \int\mathrm{d}\mathbf{q}\int\mathrm{d}\mathbf{a}\ \mathrm{e}^{-\beta H_{\mathrm{BF}}(\mathbf{q},\mathbf{a})},
  \end{equation}
  where $\mathbf{q}$ is the set of bead positions and $ \mathbf{a} $ is the set of Fourier amplitudes.

  The corresponding BF Hamiltonian is written as,\cite{febVorontsov1997,junIvanov2003}
  \begin{multline}
    H_{\mathrm{BF}}(\mathbf{q},\mathbf{a})= \sum_{j=1}^{n} \left[ \frac{1}{2}\omega_{n}^2\left((q_{j+1}-q_j)^2 + \sum^{k_\mathrm{max}}_{k=1}
      \frac{(k\pi)^2}{2}a_{jk}^2\right) \right. \\ \left.   + \frac{1}{n}\int_0^1\mathrm{d}\xi\ V(q_j(\xi))\right],
      \label{eq:ham-mc}
  \end{multline}
  with $ \omega_{n} = \sqrt{\frac{mn}{\beta^2\hbar^2}} $, $m$ the particle mass, $n$ the number of beads, $\beta =
  (k_{B}T)^{-1}$ the inverse temperature, and $V(\mathbf{q})$ the external potential of the system.

 

  \subsection{Real-time dynamics with bead-Fourier path integrals}
  In this section, we present our extension to the BF methodology by using it as a method for approximate real-time dynamics. 
  We first provide a brief overview of CMD in its discretized form before discussing our adaptation of CMD to make use of BF-PIs.

  \subsubsection{Centroid molecular dynamics}

Originally presented by Voth and coworkers,~\cite{cmd1,cmd2,cmd3,cmd4,cmd5} CMD performs dynamics under the effective potential of an imaginary time PI whose centroid is constrained to the position of the classical particle. 

  In CMD, the centroid,
  \begin{equation}
    Q = \frac{1}{n} \sum^{n}_{j=1} q_{j},
    \label{eq:centroid}
  \end{equation}
  and its conjugate momentum, $ P $, evolve under the equations of motion,

  \begin{equation}
    \dot{Q} = \frac{P}{m},
    \label{eq:EOM-cent-Q}
  \end{equation}
  and~\cite{augTrenins2019}
  \begin{equation}
    \dot{P} = - \frac{\partial F(Q)}{\partial Q}.
    \label{eq:EOM-cent-P}
  \end{equation}
  
  The forces on the centroid momentum come from the free energy,~\cite{augTrenins2019}
  \begin{equation}
    F(Q) = - \frac{1}{\beta} \ln Z(Q),
    \label{eq:cent-freeEng}
  \end{equation}

  \noindent with $Z(Q)$ being the constrained canonical partition function for the discretized PI,~\cite{augTrenins2019}
  \begin{equation}
    Z(Q) = \int \mathrm{d}\mathbf{q}'\ \mathrm{e}^{-\beta H_{\mathrm{bd}}(\mathbf{q}')} \delta(Q'-Q),
    \label{eq:label}
  \end{equation}
  
  \noindent and
  \begin{equation}
    H_{\mathrm{bd}} = U(\mathbf{q}) + S(\mathbf{q}), 
    \label{eq:discHam}
  \end{equation}
  where
  \begin{equation}
    U(\mathbf{q}) = \frac{1}{n} \sum^{n}_{j=1} V(q_{j}),
    \label{eq:beadPot}
  \end{equation}
    and
    \begin{equation}
      S(\mathbf{q}) = \sum^{n}_{j=1} \frac{1}{2} \omega_{n}^{2} (q_{j+1} - q_{j})^2. 
      \label{eq:spring}
    \end{equation}
    
  The forces on the centroid momenta can be found through just the external potential, as the spring potential does not include the centroid position,~\cite{augTrenins2019}
  \begin{equation}
    - \frac{\partial F(Q)}{\partial Q} = - \left\langle \frac{\partial U(\mathbf{q})}{\partial Q} \right\rangle_{Q},
    \label{eq:freeCalc}
  \end{equation}
  where $ \langle \dots \rangle_{Q} $ is the ensemble average with the centroid of the ring polymer constrained to be the
  value $Q$. The right-hand side of Eq.~\ref{eq:freeCalc} can be pre-calculated on a grid before the dynamics simulation or ``on-the-fly'' in methods like adiabatic/partially adiabatic CMD.
 
  The calculation of the correlation function using CMD is done as\cite{augTrenins2019}
  \begin{equation}
    C_{AB}(t) = \frac{1}{(2\pi\hbar)^{2}Z} \int \mathrm{d}P\ \int \mathrm{d}Q\ \mathrm{e}^{-\beta H_{\mathrm{CMD}}(P,Q)}
      A(Q)B(Q(t)),     
    \label{eq:CMD-corr}
  \end{equation}
    where
    \begin{equation}
      H_{\mathrm{CMD}} = \frac{P^2}{2m} + F(Q),
      \label{eq:Ham-CMD}
    \end{equation}
    and $Q(t)$ is the evolved position of the centroid under the CMD Hamiltonian in Eq.~\ref{eq:Ham-CMD}.

  \subsection{Bead-Fourier CMD}
  To extend the CMD method described above to BF-PIs, we alter the form of the free energy in
  Eq.~\ref{eq:cent-freeEng} and how it is calculated (as in Eq.~\ref{eq:freeCalc}). The calculation of estimators for ensemble averages in BF-PIMD simulations has taken two different forms: ``continuous'' and ``pure bead''.\cite{junIvanov2003} The pure bead form only takes into account the positions of the beads when calculating the estimators, while the continuous form includes the path information between the beads. Previous studies have shown that for the calculation of average quantum energy, while both forms should converge to the same result, the pure bead form converges faster with respect to the number of beads.~\cite{junIvanov2003}
  We include both forms of the estimators in this work to observe their effects on the effective BF-CMD potential.
  
  The pure bead estimator for the BF-CMD forces very closely resembles that of the bead only CMD approach since the spring and Fourier terms (the first two terms of Eq.~\ref{eq:ham-mc}) do not include the centroid position, the forces are derived as
  \begin{equation}
    - \frac{\partial F_{\mathrm{BF-bd}}(Q)}{\partial Q} = - \left\langle \frac{\partial U(\mathbf{q})}{\partial Q} \right\rangle_{Q,\mathrm{BF}},
    \label{eq:freeCalcBFbead}
  \end{equation}
  with the constrained BF ensemble average,
  \begin{equation}
    \left\langle\dots\right\rangle_{Q,\mathrm{BF}} = \frac{1}{Z_{\mathrm{BF}}(Q)} \int\mathrm{d}\mathbf{q}' \int\mathrm{d}\mathbf{a}\
    \mathrm{e}^{-\beta H_{\mathrm{BF}}(\mathbf{q}',\mathbf{a})}(\dots) \delta(Q' - Q),
    \label{eq:BF-ensemble}
  \end{equation}
  and
  \begin{equation}
    Z_{\mathrm{BF}}(Q)= \int\mathrm{d}\mathbf{q}' \int\mathrm{d}\mathbf{a}\
    \mathrm{e}^{-\beta H_{\mathrm{BF}}(\mathbf{q}',\mathbf{a})}\delta(Q' - Q).
    \label{eq:BF-constZ}
  \end{equation}
  
  While the potential on the right-hand side of Eq.~\ref{eq:freeCalcBFbead} is the same as that from
  Eq.~\ref{eq:beadPot}, the phase-space average performed still includes the full path information with the Fourier amplitudes.

  This results in a BF-CMD Hamiltonian of the form,
  \begin{equation}
    H_{\mathrm{BF-bd}} = \frac{P^2}{2m} + F_{\mathrm{BF-bd}}(Q).
    \label{eq:BF-bd-ham}
  \end{equation}
 
  We define the continuous form of the forces as,
  \begin{equation} 
    - \frac{\partial F_{\mathrm{BF-cont}}(Q)}{\partial Q} = - \left\langle \frac{\partial 
    \sum_{j=1}^n\int_{0}^1\mathrm{d}\xi\ V(q_{j}(\xi))}{\partial Q} \right\rangle_{Q,\mathrm{BF}},
    \label{eq:freeCalcBFcont}
  \end{equation}
  and the continuous BF-CMD Hamiltonian, 
  \begin{equation}
    H_{\mathrm{BF-cont}} = \frac{P^2}{2m} + F_{\mathrm{BF-cont}}(Q).
    \label{eq:BF-cont-ham}
  \end{equation}

  The BF-CMD correlation functions have the same form as Eq.~\ref{eq:CMD-corr}, but with the Hamiltonian in the
  Boltzmann weighting replaced with Eq.~\ref{eq:BF-bd-ham} and Eq.~\ref{eq:BF-cont-ham} for the pure bead and continuous estimators, respectively. Similarly, the centroid is evolved under the respective Hamiltonian.

\section{Simulation Details}\label{sec:sim-dets}
\subsection{Model Systems}
    To study the newly introduced BF-CMD method, we apply it to several one-dimensional model systems that have been previously used to benchmark approximate real-time PI methods like RPMD\cite{augCraig2004} and CMD\cite{Jang:1999}. 
    These models include the harmonic oscillator,
    \begin{equation}
        V(x) = \frac{1}{2}x^2,
    \end{equation}
    the mildly anharmonic oscillator,
    \begin{equation}
        V(x) = \frac{1}{2}x^2 + \frac{1}{10}x^3 + \frac{1}{100}x^4,
    \end{equation}
    and the quartic potential,
    \begin{equation}
        V(x) = \frac{1}{4}x^4.
    \end{equation}
    We take $m=\hbar=1$a.u. and consider two temperatures: $\beta=1$ and $\beta=8$.

    These systems are simple enough to allow comparisons with exact quantum mechanical results, but the two anharmonic models prove difficult enough that classical mechanics cannot accurately reproduce the position autocorrelation functions, especially at low temperatures.

\subsection{BF-CMD Effective Potential}

For all model systems, the free energies were pre-calculated on a grid of $Q$ values (see Supplementary Material Table S1). The included range varies for each model in order to ensure covering a sufficient position space. The forces of Eqns.~\ref{eq:freeCalcBFbead} and
~\ref{eq:freeCalcBFcont} for each value of $Q$ are calculated through the constrained ensemble average using a Metropolis MC scheme.

We base our MC scheme on similar approaches from Fourier-PIMC studies.\cite{janMielke2001a} In particular, we have two types of moves: a bead move and a Fourier amplitude move. For a bead move, we randomly pick a bead and move it.
This is repeated $n$ times before shifting the bead positions to constrain the new centroid to the required value and then calculating the new potential energy of Eq.~\ref{eq:ham-mc} and determining whether to accept or reject the move. We are thus on average moving each bead with every bead move, without strictly moving all of them in each step. In these moves, the Fourier amplitudes are left unchanged, but they are included in the energy calculations. For a Fourier amplitude move, we also randomly pick a bead, but we then move all the Fourier amplitudes associated with that bead. Again, this is repeated $n$ times before determining to accept or reject the move. When choosing to make a bead or Fourier amplitude move, bead moves are weighted more favorably, with them being chosen $\approx 90\% $ of the time. This choice reduces computational cost, and previous studies have shown that this has no effect on the accuracy of the sampling.\cite{janMielke2001a,marTopper1993} Having separate moves also allows for simplicity in
determining the appropriate step size for each type of MC move, as we can change them separately. 
We alter the step sizes so that the acceptance ratio for both types of moves is simultaneously $\approx50\%$ for efficient sampling of the full configuration space.

The ensemble average for each grid point is calculated using 500,000 MC points with a decorrelation length of 50. To study convergence, we include up to $n=4$ beads for
$\beta=1$ and $n=32$ for $\beta=8$, and from $k_{\mathrm{max}}=0$ to $k_{\mathrm{max}}=5$. For the systems with $k_{\mathrm{max}}=0$, we still include the linear path between the beads, which results in a different value for continuously estimated, non-linear operators. The case of $n=1$ is of special note here, as when using the bead estimator, the resulting Hamiltonian is purely that of the classical system due to constraining the position of the single bead. The continuous estimator yields the free energy associated with a pure Fourier-CMD method. One area of convergence we did not include in this study is integration over paths. We calculate the integrals over $\xi$ using the trapezoid rule with 20 segments. This is much larger than what is typically used in
BF(and Fourier)-PIMC calculations where the number used is based upon the number of Fourier components and the curvature of the potential.\cite{junIvanov2003}
After calculating the forces on the grid, we calculate the free energy as the potential of mean force by integrating the forces along the centroid coordinate.

\begin{figure}[t]
\begin{center}
  \includegraphics[width=0.99\linewidth]{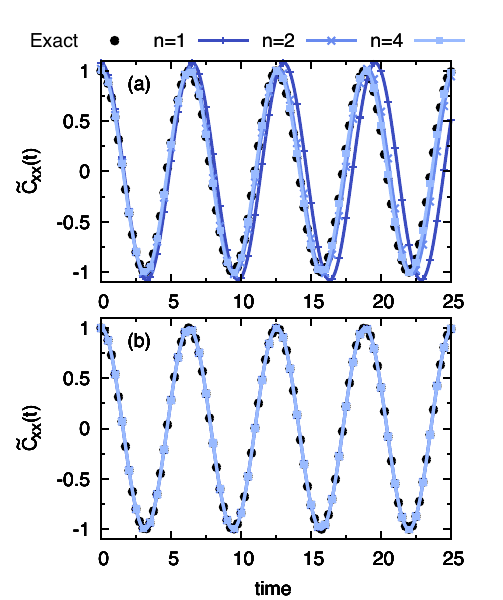}
\end{center}
\caption{Kubo-transformed position autocorrelation function for the harmonic oscillator with $\beta=1$ and free energy calculated using the (a) continuous estimator with $k_{\mathrm{max}}=1$ and (b) bead estimator with $k_{\mathrm{max}}=0$ for all numbers of beads. Exact results are given in black dots.}
\label{fig:harm-corr1}
\end{figure}
\begin{figure*}[t]
\begin{center}
  \includegraphics[scale=1]{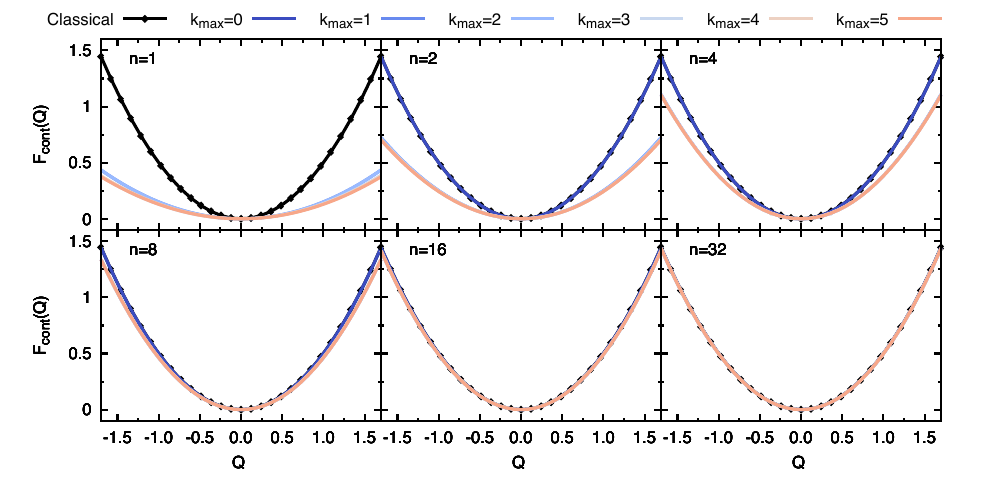}
\end{center}
\caption{Free energy for the harmonic oscillator at $\beta=8$ using the continuous estimator with varying number of beads and Fourier components. Standard classical potential shown in black dots is given as reference.}
\label{fig:harm8-conv}
\end{figure*}
\subsection{Position Autocorrelation Functions}
To check the performance of our BF-CMD method, we calculate the position autocorrelation function for each of the model systems. We obtain the initial conditions for our dynamics simulations through a similar MC scheme as described above.
For this case, we only have ``bead'' moves, but reduced down to $n=1$, and we have no constraints on its position. The energy used to determine accepting or rejecting a move comes from the free energy previously calculated, where we use 3-point interpolation to find the value for points not on the grid. The initial momenta are obtained separately from the Boltzmann distribution.

The dynamics are performed using $1\times 10^6$ trajectories taken from the MC sampling with a decorrelation length of 500 between initial conditions. Trajectories are evolved using a time step of 0.001 a.u. according to the Hamiltonians of Eqns.~\ref{eq:BF-bd-ham} and~\ref{eq:BF-cont-ham} for the case using the pure bead and continuous effective potentials, respectively. The same interpolation method is used when calculating the forces on the centroids.
We compare these correlation functions with the exact quantum mechanical results. It must be noted that, as CMD correlation functions have the same form as a classical correlation function, they cannot be directly compared to the quantum mechanical ones, but instead are more related to the
Kubo-transformed correlation function.\cite{augCraig2004}

\section{Results and Discussion}\label{sec:results}

Here we showcase the performance of our BF-CMD method in calculating the position autocorrelation function for the model systems outlined above. Our focus is on the behavior of the two ways of estimating the free energy and the convergence with respect to the number of beads and Fourier components.

\subsection{Harmonic Oscillator}
The first system we analyze is the harmonic oscillator. It is important to note that a purely classical simulation of the system will give the exact result for the Kubo-transformed autocorrelation function,~\cite{augCraig2004} putting a clear standard on the accuracy the method should strive to meet.

Fig.~\ref{fig:harm-corr1} shows the high temperature, $\beta=1$, results using different numbers of beads compared to the exact results. For the continuous estimated potential, one Fourier component is enough to reach convergence, while for the bead estimated potential, just having the linear path between beads ($k_{\mathrm{max}}=0$) is enough to converge the free energy (full convergence plots are given in Supplementary Material Figs. S1-S6). For the case where the free energy is found using the bead estimator, the exact results are achieved for any number of beads. This can be easily rationalized through the nature of the potential. For the harmonic oscillator, the force is simply proportional to the position, and averaging over the bead estimated force when they are constrained to a given centroid value gives the same force as that of just the centroid. The resulting free energy is then the same as the classical potential, which, as mentioned previously, gives the exact Kubo-transformed correlation function.

The case where the free energy is obtained using the continuous estimator yields more interesting results. First, the case of $n=1$, which is the pure Fourier-PI limit, has oscillations with larger amplitude and frequency compared to the exact results (see Fig. \ref{fig:harm-corr1}). The same phenomenon can be seen for $n=2$ beads, but to a far lesser extent. 
This can be attributed to the widening of the potential compared to the standard potential (see Supplementary Material Figs. S1 and S2), which is similar in effect to the position distributions seen in BF-PIMD simulations of the harmonic oscillator.\cite{junIvanov2003} Finally, we do see convergence to the exact results for $n=4$ beads. These results indicate that the impact of the Fourier components, and particularly the path that is included in the continuous estimator, on the effective potential diminishes with increasing number of beads.

To demonstrate the convergence of the free energy, we present the calculated free energies for the harmonic oscillator at $\beta=8$ using the continuous estimator in Fig.~\ref{fig:harm8-conv}. We see clearly that when using the continuous estimator, the effective potential can become much flatter than the classical potential, which would result in a wider position distribution and lower frequency oscillations. The effect is more pronounced for fewer numbers of beads, with only a small difference seen for 16 and 32 beads. This is likely due to the fact that a larger number of beads results in more slices in the imaginary time interval between 0 and $\beta\hbar$, thus the path between the beads becomes much shorter and closer to the linear path. The number of Fourier components needed to converge the free energy also decreases with increasing beads, further indicating that the impact of the Fourier components diminishes with increasing beads.

\begin{figure}
\begin{center}
  \includegraphics[width=0.99\linewidth]{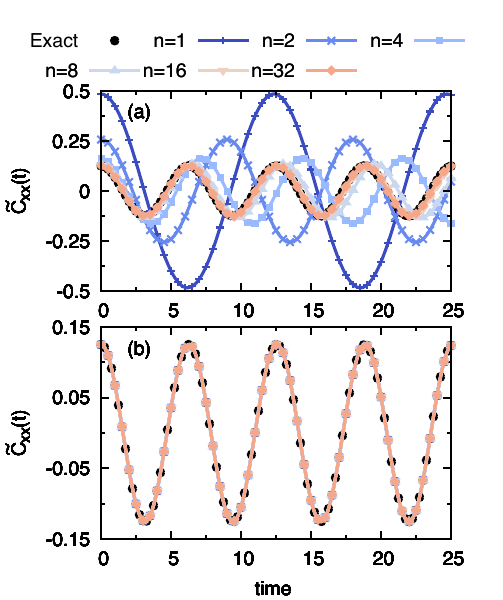}
\end{center}
\caption{Kubo-transformed position correlation function for the harmonic oscillator with $\beta=8$. Free energy
estimated with (a) continuous estimator. Number of Fourier components used for $n=1:5$; $n=2$, $n=4$, and $n=8:3$; and 
$n=16$ and $n=32:1$. (b) Free energy estimated with bead estimator and all beads using $k_{\mathrm{max}}=0$. Exact
results given as black dots.}
\label{fig:harm-corr8}
\end{figure}

When looking at the correlation functions, we see similar behavior as for $\beta=1$. Fig.~\ref{fig:harm-corr8}b shows that the bead estimator recovers the exact results for any number of beads, as expected. The errors from the continuous estimator are enhanced with lower temperatures, as is the number of Fourier components needed to converge the free energy.
Fig.~\ref{fig:harm-corr8}a includes the results with $k_{\mathrm{max}}=5$ for $n=1$, $k_{\mathrm{max}}=3$ for $n=2$, $n=4$, and $n=8$, and $k_{\mathrm{max}}=1$ for $n=16$ and $n=32$. At this lower temperature, even the largest number of beads considered here does not fully converge to the exact results.
After considering these results, we recommend using the free energy from the bead estimator and not the continuous estimator. The results for the other two systems will only include the bead estimator, but for completeness, the continuous estimator results are provided in the Supplementary Material.

\subsection{Mildly Anharmonic Oscillator}
\begin{figure}
\begin{center}
  \includegraphics[width=0.99\linewidth]{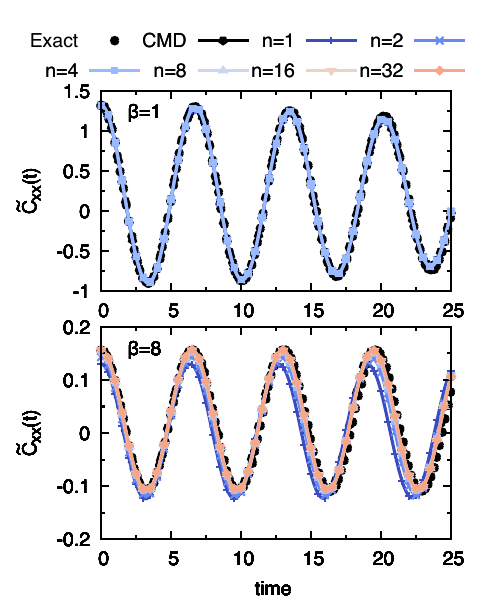}
\end{center}
\caption{Kubo-transformed position autocorrelation function for the mildly anharmonic oscillator with the free energy found using the bead estimator at (top) $\beta=1$ and (bottom) $\beta=8$. All beads have $k_{\mathrm{max}}=0$ for $\beta=1$. For $\beta=8$, $n=2$, $n=4$, and $n=8$ use one Fourier component while $n=16$ and $n=32$ use only the linear path. $n=1$ corresponds to classical results. Exact results are given in black dots, and CMD results are given in black pentagons.}
\label{fig:mild-corr}
\end{figure}

For the remaining systems, we include the CMD results, which are found using the same methodology as the BF-CMD, with the pure bead estimated free energy with no Fourier components and no linear path. We use $n=4$ beads and $n=32$ beads for $\beta=1$ and $\beta=8$, respectively, when calculating the forces and free energy according to Eq.~\ref{eq:freeCalc}. The same number of trajectories and time steps is used for dynamics simulations.

The results for the mildly anharmonic oscillator are shown in Fig.~\ref{fig:mild-corr}. The high temperature results, which include up to 4 beads, only include the linear path between the beads to reach convergence. Again, the $n=1$ results correspond to classical simulations due to only considering the bead estimator. When comparing to the CMD results, we see that BF-CMD gives the same results using only 2 beads. At this high temperature, the number of beads to converge PI results in general is fairly low, so the improvement of BF-CMD is minimal.

The benefits of using BF-PIs to generate the effective potential are more pronounced at lower temperatures. The bottom panel of Fig.~\ref{fig:mild-corr} displays that BF-CMD results converge to the CMD with as few as 8 beads. The results for $n=2$, $n=4$, and $n=8$ include one Fourier component, and only the linear paths are considered for $n=16$ and $n=32$. The inclusion of just 1 Fourier component being able to reduce the number of beads by 4 times compared to other PI methods~\cite{augCraig2004,aprHone2006a} is a remarkable benefit of the BF-CMD method.

\subsection{Quartic Oscillator}
\begin{figure}
\begin{center}
  \includegraphics[width=0.99\linewidth]{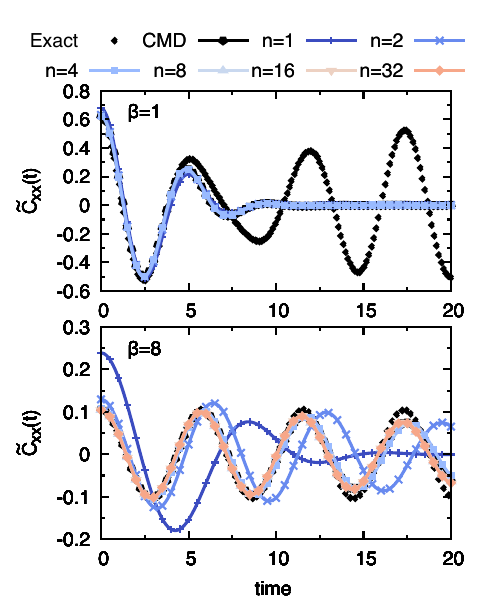}
\end{center}
\caption{Kubo-transformed position autocorrelation function for the quartic oscillator with the free energy found using the bead estimator at (top) $\beta=1$ and (bottom) $\beta=8$. All beads have $k_{\mathrm{max}}=0$ for $\beta=1$. For $\beta=8$, $n=2$ and $n=4$ use 5 Fourier components, $n=8$ and $n=16$ use 1 Fourier component, and $n=32$ uses only the linear path. $n=1$ corresponds to classical results. Exact results are given in black dots and CMD results in black pentagons.}
\label{fig:quart-corr}
\end{figure}

The quartic oscillator presents the largest challenge of these models for any PI method, where quantum coherence is very pronounced. Our motivations for BF-CMD for these model systems are to be able to achieve results on par with CMD with fewer beads through the inclusion of the Fourier components. As such, we do not expect to more accurately reproduce the exact correlation functions for this system. The resulting correlation functions are shown in Fig.~\ref{fig:quart-corr}. The high temperature results include only the linear path between beads. We see the same behavior as CMD, with good agreement with the exact results for short times before quickly losing all correlated motion. Additionally, we see that we need to go to $n=4$ before converging with the CMD results. While we do see the desired improvement in this case in terms of bead convergence, these high temperature systems are unlikely to suffer from the problems of PI methods discussed previously. The cases of more importance are those where current methods require a large number of beads to converge.

The $\beta=8$ results are given in the bottom panel of Fig.~\ref{fig:quart-corr}. For these correlation functions, we include the results with $k_{\mathrm{max}}=5$ for $n=2$ and $n=4$, $k_{\mathrm{max}}=1$ for $n=8$ and $n=16$, and only the linear paths for $n=32$. From these convergence results, we see that for a given number of beads, lower temperatures and more anharmonic potentials require a larger number of Fourier components. 
As with the mildly anharmonic oscillator, the BF-CMD results are converged with $n=8$ beads. We do see a slightly higher frequency of oscillations compared to CMD (which is also seen in PA-CMD\cite{aprHone2006a} and RPMD\cite{augCraig2004}) that is closer to the exact frequency, but the amplitude does remain the same. This small difference derives from a very slightly different shape of the BF-CMD effective potential (See Supplementary Material Fig. S49)

\section{Conclusions}
We have presented here a new CMD method where the effective potential is calculated using BF-PIs as opposed to the typical discretized PIs. We have explored how the inclusion of Fourier components reduces the number of beads needed to reach convergence. Our results for the mildly anharmonic and quartic oscillators at low temperature show that the inclusion of just one Fourier component reduces the number of beads by 4-fold. Since we did not explore the full convergence space in terms of efficiency by using a large number of segments when integrating over paths, we cannot comment on the computational savings of the method, as a far larger number of force calculations are required for this purpose. Additionally, due to the few number of Fourier components needed for these systems, we do not see it necessary to explore methods used to improve the efficiency and convergence of Fourier-PIs like partial averaging\cite{julDoll1985} or using the full Fourier series\cite{sepLobaugh1992} as the additional computational complexity would likely yield negligible improvements.
The current form of the BF-CMD method is not applicable to anything beyond simple model systems due to our implementation of pre-calculating the forces and free energy before performing sampling and dynamics simulations. Thus, for future works, we plan to explore how the method can be expanded into an adiabatic (or partially adiabatic) form to calculate the effective potential during the dynamics. We are also looking to explore how the reduction in the number of beads to reach a converged effective potential would affect the curvature problem in the CMD calculated vibrational spectra.

\section*{Supplementary Material}
Convergence analyses for each model system with respect to the estimator used, the number of beads, and the number of Fourier components are included in the Supplementary Material.

\begin{acknowledgments}
M.R.M. would like to thank Robert Q. Topper for helpful discussions. Simulations presented in this work used resources from Bridges-2 at Pittsburgh Supercomputing Center through allocations PHY230099 and CHE240103 from the Advanced Cyberinfrastructure Coordination Ecosystem: Services \& Support (ACCESS) program,\cite{access} which is supported by National Science Foundation grants \#2138259, \#2138286, \#2138307, \#2137603, and \#2138296. The authors also acknowledge the HPC center at UMKC for providing computing resources and support.
\end{acknowledgments}

\subsection*{Conflict of Interest}

\noindent The authors have no conflicts to disclose.

\subsection*{Author Contributions}
\textbf{Nathan London}: Method (equal); Writing - original draft (lead); Writing - review \& editing (equal). 
\textbf{Mohammad R. Momeni}: Conceptualization (lead); Funding acquisition (lead); Supervision (lead); Method (equal); Writing - original draft (supporting); Writing - review \& editing (equal). 

\section*{\label{sec_contrib} DATA AVAILABILITY}
\noindent  The data that support the findings of this study are provided in the main text and the accompanying Supplementary Material. Additional data are available from the corresponding author
upon reasonable request.

\section*{REFERENCES}
\bibliography{bib}

\begin{thebibliography}{37}%
\makeatletter
\providecommand \@ifxundefined [1]{%
 \@ifx{#1\undefined}
}%
\providecommand \@ifnum [1]{%
 \ifnum #1\expandafter \@firstoftwo
 \else \expandafter \@secondoftwo
 \fi
}%
\providecommand \@ifx [1]{%
 \ifx #1\expandafter \@firstoftwo
 \else \expandafter \@secondoftwo
 \fi
}%
\providecommand \natexlab [1]{#1}%
\providecommand \enquote  [1]{``#1''}%
\providecommand \bibnamefont  [1]{#1}%
\providecommand \bibfnamefont [1]{#1}%
\providecommand \citenamefont [1]{#1}%
\providecommand \href@noop [0]{\@secondoftwo}%
\providecommand \href [0]{\begingroup \@sanitize@url \@href}%
\providecommand \@href[1]{\@@startlink{#1}\@@href}%
\providecommand \@@href[1]{\endgroup#1\@@endlink}%
\providecommand \@sanitize@url [0]{\catcode `\\12\catcode `\$12\catcode
  `\&12\catcode `\#12\catcode `\^12\catcode `\_12\catcode `\%12\relax}%
\providecommand \@@startlink[1]{}%
\providecommand \@@endlink[0]{}%
\providecommand \url  [0]{\begingroup\@sanitize@url \@url }%
\providecommand \@url [1]{\endgroup\@href {#1}{\urlprefix }}%
\providecommand \urlprefix  [0]{URL }%
\providecommand \Eprint [0]{\href }%
\providecommand \doibase [0]{http://dx.doi.org/}%
\providecommand \selectlanguage [0]{\@gobble}%
\providecommand \bibinfo  [0]{\@secondoftwo}%
\providecommand \bibfield  [0]{\@secondoftwo}%
\providecommand \translation [1]{[#1]}%
\providecommand \BibitemOpen [0]{}%
\providecommand \bibitemStop [0]{}%
\providecommand \bibitemNoStop [0]{.\EOS\space}%
\providecommand \EOS [0]{\spacefactor3000\relax}%
\providecommand \BibitemShut  [1]{\csname bibitem#1\endcsname}%
\let\auto@bib@innerbib\@empty
\bibitem [{\citenamefont {Markland}\ and\ \citenamefont
  {Ceriotti}(2018)}]{Markland2018}%
  \BibitemOpen
  \bibfield  {author} {\bibinfo {author} {\bibfnamefont {T.}~\bibnamefont
  {Markland}}\ and\ \bibinfo {author} {\bibfnamefont {M.}~\bibnamefont
  {Ceriotti}},\ }\bibfield  {title} {\enquote {\bibinfo {title} {{Nuclear
  quantum effects enter the mainstream}},}\ }\href
  {https://doi.org/10.1038/s41570-017-0109} {\bibfield  {journal} {\bibinfo
  {journal} {Nat. Rev. Chem.}\ }\textbf {\bibinfo {volume} {2}},\ \bibinfo
  {pages} {0109} (\bibinfo {year} {2018})}\BibitemShut {NoStop}%
\bibitem [{\citenamefont {Meisner}\ and\ \citenamefont
  {K\"astner}()}]{Meisner2016}%
  \BibitemOpen
  \bibfield  {author} {\bibinfo {author} {\bibfnamefont {J.}~\bibnamefont
  {Meisner}}\ and\ \bibinfo {author} {\bibfnamefont {J.}~\bibnamefont
  {K\"astner}},\ }\bibfield  {title} {\enquote {\bibinfo {title} {Atom
  {{Tunneling}} in {{Chemistry}}},}\ }\href {\doibase 10.1002/anie.201511028}
  {\ \textbf {\bibinfo {volume} {55}},\ \bibinfo {pages}
  {5400--5413}}\BibitemShut {NoStop}%
\bibitem [{\citenamefont {Marsalek}\ and\ \citenamefont
  {Markland}(2017)}]{Marsalek2017}%
  \BibitemOpen
  \bibfield  {author} {\bibinfo {author} {\bibfnamefont {O.}~\bibnamefont
  {Marsalek}}\ and\ \bibinfo {author} {\bibfnamefont {T.~E.}\ \bibnamefont
  {Markland}},\ }\bibfield  {title} {\enquote {\bibinfo {title} {Quantum
  {{Dynamics}} and {{Spectroscopy}} of {{Ab Initio Liquid Water}}: {{The
  Interplay}} of {{Nuclear}} and {{Electronic Quantum Effects}}},}\ }\href
  {\doibase 10.1021/acs.jpclett.7b00391} {\bibfield  {journal} {\bibinfo
  {journal} {J. Phys. Chem. Lett.}\ }\textbf {\bibinfo {volume} {8}},\ \bibinfo
  {pages} {1545--1551} (\bibinfo {year} {2017})}\BibitemShut {NoStop}%
\bibitem [{\citenamefont {Kapil}\ \emph {et~al.}(2024)\citenamefont {Kapil},
  \citenamefont {Kov{\'a}cs}, \citenamefont {Cs{\'a}nyi},\ and\ \citenamefont
  {Michaelides}}]{Kapil2024}%
  \BibitemOpen
  \bibfield  {author} {\bibinfo {author} {\bibfnamefont {V.}~\bibnamefont
  {Kapil}}, \bibinfo {author} {\bibfnamefont {D.~P.}\ \bibnamefont
  {Kov{\'a}cs}}, \bibinfo {author} {\bibfnamefont {G.}~\bibnamefont
  {Cs{\'a}nyi}}, \ and\ \bibinfo {author} {\bibfnamefont {A.}~\bibnamefont
  {Michaelides}},\ }\bibfield  {title} {\enquote {\bibinfo {title}
  {First-principles spectroscopy of aqueous interfaces using machine-learned
  electronic and quantum nuclear effects},}\ }\href {\doibase
  10.1039/D3FD00113J} {\bibfield  {journal} {\bibinfo  {journal} {Faraday
  Discuss.}\ }\textbf {\bibinfo {volume} {249}},\ \bibinfo {pages} {50--68}
  (\bibinfo {year} {2024})}\BibitemShut {NoStop}%
\bibitem [{\citenamefont {Beck}\ \emph {et~al.}()\citenamefont {Beck},
  \citenamefont {J\"ackle}, \citenamefont {Worth},\ and\ \citenamefont
  {Meyer}}]{janBeck2000a}%
  \BibitemOpen
  \bibfield  {author} {\bibinfo {author} {\bibfnamefont {M.~H.}\ \bibnamefont
  {Beck}}, \bibinfo {author} {\bibfnamefont {A.}~\bibnamefont {J\"ackle}},
  \bibinfo {author} {\bibfnamefont {G.~A.}\ \bibnamefont {Worth}}, \ and\
  \bibinfo {author} {\bibfnamefont {H.~D.}\ \bibnamefont {Meyer}},\ }\bibfield
  {title} {\enquote {\bibinfo {title} {The multiconfiguration time-dependent
  hartree (mctdh) method: A highly efficient algorithm for propagating
  wavepackets},}\ }\href {\doibase 10.1016/S0370-1573(99)00047-2} {\ \textbf
  {\bibinfo {volume} {324}},\ \bibinfo {pages} {1--105}}\BibitemShut {NoStop}%
\bibitem [{\citenamefont {Richings}\ \emph {et~al.}()\citenamefont {Richings},
  \citenamefont {~}, \citenamefont {~}, \citenamefont {~}, \citenamefont {~},\
  and\ \citenamefont {family=Lasorne}}]{aprRichings2015}%
  \BibitemOpen
  \bibfield  {author} {\bibinfo {author} {\bibfnamefont {G.}~\bibnamefont
  {Richings}}, \bibinfo {author} {\bibfnamefont {P.}~\bibnamefont {~},
  \bibfnamefont {I.}}, \bibinfo {author} {\bibfnamefont {S.}~\bibnamefont {~},
  \bibfnamefont {K.E.}}, \bibinfo {author} {\bibfnamefont {W.}~\bibnamefont
  {~}, \bibfnamefont {G.A.}}, \bibinfo {author} {\bibfnamefont
  {B.}~\bibnamefont {~}, \bibfnamefont {I.}}, \ and\ \bibinfo {author}
  {\bibfnamefont {p.~u.}\ \bibnamefont {family=Lasorne}, \bibfnamefont
  {given=B.}},\ }\bibfield  {title} {\enquote {\bibinfo {title} {Quantum
  dynamics simulations using {{Gaussian}} wavepackets: The {{vMCG}} method},}\
  }\href {\doibase 10.1080/0144235X.2015.1051354} {\ \textbf {\bibinfo {volume}
  {34}},\ \bibinfo {pages} {269--308}}\BibitemShut {NoStop}%
\bibitem [{\citenamefont {Webb}, \citenamefont {Iordanov},\ and\ \citenamefont
  {Hammes-Schiffer}()}]{sepWebb2002}%
  \BibitemOpen
  \bibfield  {author} {\bibinfo {author} {\bibfnamefont {S.~P.}\ \bibnamefont
  {Webb}}, \bibinfo {author} {\bibfnamefont {T.}~\bibnamefont {Iordanov}}, \
  and\ \bibinfo {author} {\bibfnamefont {S.}~\bibnamefont {Hammes-Schiffer}},\
  }\bibfield  {title} {\enquote {\bibinfo {title} {Multiconfigurational
  nuclear-electronic orbital approach: {{Incorporation}} of nuclear quantum
  effects in electronic structure calculations},}\ }\href {\doibase
  10.1063/1.1494980} {\ \textbf {\bibinfo {volume} {117}},\ \bibinfo {pages}
  {4106--4118}}\BibitemShut {NoStop}%
\bibitem [{\citenamefont {Feynman}, \citenamefont {Hibbs},\ and\ \citenamefont
  {Styer}(2010)}]{julFeynman2010}%
  \BibitemOpen
  \bibfield  {author} {\bibinfo {author} {\bibfnamefont {R.~P.}\ \bibnamefont
  {Feynman}}, \bibinfo {author} {\bibfnamefont {A.~R.}\ \bibnamefont {Hibbs}},
  \ and\ \bibinfo {author} {\bibfnamefont {D.~F.}\ \bibnamefont {Styer}},\
  }\href@noop {} {\emph {\bibinfo {title} {Quantum {{Mechanics}} and {{Path
  Integrals}}}}}\ (\bibinfo  {publisher} {Courier Corporation},\ \bibinfo
  {year} {2010})\BibitemShut {NoStop}%
\bibitem [{\citenamefont {Ceperley}(1995)}]{aprCeperley1995}%
  \BibitemOpen
  \bibfield  {author} {\bibinfo {author} {\bibfnamefont {D.~M.}\ \bibnamefont
  {Ceperley}},\ }\bibfield  {title} {\enquote {\bibinfo {title} {Path integrals
  in the theory of condensed helium},}\ }\href {\doibase
  10.1103/RevModPhys.67.279} {\bibfield  {journal} {\bibinfo  {journal}
  {Reviews of Modern Physics}\ }\textbf {\bibinfo {volume} {67}},\ \bibinfo
  {pages} {279--355} (\bibinfo {year} {1995})}\BibitemShut {NoStop}%
\bibitem [{\citenamefont {Parrinello}\ and\ \citenamefont
  {Rahman}(1984)}]{janParrinello1984a}%
  \BibitemOpen
  \bibfield  {author} {\bibinfo {author} {\bibfnamefont {M.}~\bibnamefont
  {Parrinello}}\ and\ \bibinfo {author} {\bibfnamefont {A.}~\bibnamefont
  {Rahman}},\ }\bibfield  {title} {\enquote {\bibinfo {title} {Study of an
  \emph{F} center in molten kcl},}\ }\href {\doibase 10.1063/1.446740}
  {\bibfield  {journal} {\bibinfo  {journal} {The Journal of Chemical Physics}\
  }\textbf {\bibinfo {volume} {80}},\ \bibinfo {pages} {860--867} (\bibinfo
  {year} {1984})}\BibitemShut {NoStop}%
\bibitem [{\citenamefont {Ryu}\ and\ \citenamefont {Voth}()}]{sepRyu2022}%
  \BibitemOpen
  \bibfield  {author} {\bibinfo {author} {\bibfnamefont {W.~H.}\ \bibnamefont
  {Ryu}}\ and\ \bibinfo {author} {\bibfnamefont {G.~A.}\ \bibnamefont {Voth}},\
  }\bibfield  {title} {\enquote {\bibinfo {title} {Coarse-{{Graining}} of
  {{Imaginary Time Feynman Path Integrals}}: {{Inclusion}} of {{Intramolecular
  Interactions}} and {{Bottom-up Force-Matching}}},}\ }\href {\doibase
  10.1021/acs.jpca.2c04349} {\ \textbf {\bibinfo {volume} {126}},\ \bibinfo
  {pages} {6004--6019}}\BibitemShut {NoStop}%
\bibitem [{\citenamefont {Ryu}, \citenamefont {Han},\ and\ \citenamefont
  {Voth}()}]{Ryu2019}%
  \BibitemOpen
  \bibfield  {author} {\bibinfo {author} {\bibfnamefont {H.}~\bibnamefont
  {Ryu}}, \bibinfo {author} {\bibfnamefont {Y.}~\bibnamefont {Han}}, \ and\
  \bibinfo {author} {\bibfnamefont {G.~A.}\ \bibnamefont {Voth}},\ }\bibfield
  {title} {\enquote {\bibinfo {title} {Coarse-graining of many-body path
  integrals: {{Theory}} and numerical approximations},}\ }\href {\doibase
  10.1063/1.5097141} {\ \textbf {\bibinfo {volume} {150}},\ \bibinfo {pages}
  {244103}}\BibitemShut {NoStop}%
\bibitem [{\citenamefont {Cao}\ and\ \citenamefont
  {Voth}(1994{\natexlab{a}})}]{cmd1}%
  \BibitemOpen
  \bibfield  {author} {\bibinfo {author} {\bibfnamefont {J.}~\bibnamefont
  {Cao}}\ and\ \bibinfo {author} {\bibfnamefont {G.~A.}\ \bibnamefont {Voth}},\
  }\bibfield  {title} {\enquote {\bibinfo {title} {{The formulation of quantum
  statistical mechanics based on the Feynman path centroid density. I.
  Equilibrium properties}},}\ }\href {\doibase 10.1063/1.467175} {\bibfield
  {journal} {\bibinfo  {journal} {J. Chem. Phys.}\ }\textbf {\bibinfo {volume}
  {100}},\ \bibinfo {pages} {5093--5105} (\bibinfo {year}
  {1994}{\natexlab{a}})}\BibitemShut {NoStop}%
\bibitem [{\citenamefont {Cao}\ and\ \citenamefont
  {Voth}(1994{\natexlab{b}})}]{cmd2}%
  \BibitemOpen
  \bibfield  {author} {\bibinfo {author} {\bibfnamefont {J.}~\bibnamefont
  {Cao}}\ and\ \bibinfo {author} {\bibfnamefont {G.~A.}\ \bibnamefont {Voth}},\
  }\bibfield  {title} {\enquote {\bibinfo {title} {{The formulation of quantum
  statistical mechanics based on the Feynman path centroid density. II.
  Dynamical properties}},}\ }\href {\doibase 10.1063/1.467176} {\bibfield
  {journal} {\bibinfo  {journal} {J. Chem. Phys.}\ }\textbf {\bibinfo {volume}
  {100}},\ \bibinfo {pages} {5106--5117} (\bibinfo {year}
  {1994}{\natexlab{b}})}\BibitemShut {NoStop}%
\bibitem [{\citenamefont {Cao}\ and\ \citenamefont
  {Voth}(1994{\natexlab{c}})}]{cmd3}%
  \BibitemOpen
  \bibfield  {author} {\bibinfo {author} {\bibfnamefont {J.}~\bibnamefont
  {Cao}}\ and\ \bibinfo {author} {\bibfnamefont {G.~A.}\ \bibnamefont {Voth}},\
  }\bibfield  {title} {\enquote {\bibinfo {title} {{The formulation of quantum
  statistical mechanics based on the Feynman path centroid density. III. Phase
  space formalism and analysis of centroid molecular dynamics}},}\ }\href
  {\doibase 10.1063/1.468503} {\bibfield  {journal} {\bibinfo  {journal} {J.
  Chem. Phys.}\ }\textbf {\bibinfo {volume} {101}},\ \bibinfo {pages}
  {6157--6167} (\bibinfo {year} {1994}{\natexlab{c}})}\BibitemShut {NoStop}%
\bibitem [{\citenamefont {Cao}\ and\ \citenamefont
  {Voth}(1994{\natexlab{d}})}]{cmd4}%
  \BibitemOpen
  \bibfield  {author} {\bibinfo {author} {\bibfnamefont {J.}~\bibnamefont
  {Cao}}\ and\ \bibinfo {author} {\bibfnamefont {G.~A.}\ \bibnamefont {Voth}},\
  }\bibfield  {title} {\enquote {\bibinfo {title} {The formulation of quantum
  statistical mechanics based on the {{Feynman}} path centroid density. {{IV}}.
  {{Algorithms}} for centroid molecular dynamics},}\ }\href@noop {} {\bibfield
  {journal} {\bibinfo  {journal} {Journal of Chemical Physics}\ }\textbf
  {\bibinfo {volume} {101}},\ \bibinfo {pages} {6168--6183} (\bibinfo {year}
  {1994}{\natexlab{d}})}\BibitemShut {NoStop}%
\bibitem [{\citenamefont {Cao}\ and\ \citenamefont {Voth}(1999)}]{cmd5}%
  \BibitemOpen
  \bibfield  {author} {\bibinfo {author} {\bibfnamefont {J.}~\bibnamefont
  {Cao}}\ and\ \bibinfo {author} {\bibfnamefont {G.~A.}\ \bibnamefont {Voth}},\
  }\bibfield  {title} {\enquote {\bibinfo {title} {A derivation of centroid
  molecular dynamics and other approximate time evolution methods for path
  integral centroid variables},}\ }\href@noop {} {\bibfield  {journal}
  {\bibinfo  {journal} {Journal of Chemical Physics}\ }\textbf {\bibinfo
  {volume} {111}},\ \bibinfo {pages} {2371--2384} (\bibinfo {year}
  {1999})}\BibitemShut {NoStop}%
\bibitem [{\citenamefont {Jang}\ and\ \citenamefont
  {Voth}(1999{\natexlab{a}})}]{augJang1999}%
  \BibitemOpen
  \bibfield  {author} {\bibinfo {author} {\bibfnamefont {S.}~\bibnamefont
  {Jang}}\ and\ \bibinfo {author} {\bibfnamefont {G.~A.}\ \bibnamefont
  {Voth}},\ }\bibfield  {title} {\enquote {\bibinfo {title} {Path integral
  centroid variables and the formulation of their exact real time dynamics},}\
  }\href {\doibase 10.1063/1.479514} {\bibfield  {journal} {\bibinfo  {journal}
  {Journal of Chemical Physics}\ }\textbf {\bibinfo {volume} {111}},\ \bibinfo
  {pages} {2357--2370} (\bibinfo {year} {1999}{\natexlab{a}})}\BibitemShut
  {NoStop}%
\bibitem [{\citenamefont {Jang}\ and\ \citenamefont
  {Voth}(1999{\natexlab{b}})}]{Jang:1999}%
  \BibitemOpen
  \bibfield  {author} {\bibinfo {author} {\bibfnamefont {S.}~\bibnamefont
  {Jang}}\ and\ \bibinfo {author} {\bibfnamefont {G.~A.}\ \bibnamefont
  {Voth}},\ }\bibfield  {title} {\enquote {\bibinfo {title} {A derivation of
  centroid molecular dynamics and other approximate time evolution methods for
  path integral centroid variables},}\ }\href {\doibase 10.1063/1.479515}
  {\bibfield  {journal} {\bibinfo  {journal} {Journal of Chemical Physics}\
  }\textbf {\bibinfo {volume} {111}},\ \bibinfo {pages} {2371--2384} (\bibinfo
  {year} {1999}{\natexlab{b}})}\BibitemShut {NoStop}%
\bibitem [{\citenamefont {Hone}, \citenamefont {Rossky},\ and\ \citenamefont
  {Voth}(2006)}]{aprHone2006a}%
  \BibitemOpen
  \bibfield  {author} {\bibinfo {author} {\bibfnamefont {T.~D.}\ \bibnamefont
  {Hone}}, \bibinfo {author} {\bibfnamefont {P.~J.}\ \bibnamefont {Rossky}}, \
  and\ \bibinfo {author} {\bibfnamefont {G.~A.}\ \bibnamefont {Voth}},\
  }\bibfield  {title} {\enquote {\bibinfo {title} {A comparative study of
  imaginary time path integral based methods for quantum dynamics},}\ }\href
  {\doibase 10.1063/1.2186636} {\bibfield  {journal} {\bibinfo  {journal} {The
  Journal of Chemical Physics}\ }\textbf {\bibinfo {volume} {124}},\ \bibinfo
  {pages} {154103} (\bibinfo {year} {2006})}\BibitemShut {NoStop}%
\bibitem [{\citenamefont {Ivanov}\ \emph {et~al.}(2010)\citenamefont {Ivanov},
  \citenamefont {Witt}, \citenamefont {Shiga},\ and\ \citenamefont
  {Marx}}]{janIvanov2010}%
  \BibitemOpen
  \bibfield  {author} {\bibinfo {author} {\bibfnamefont {S.~D.}\ \bibnamefont
  {Ivanov}}, \bibinfo {author} {\bibfnamefont {A.}~\bibnamefont {Witt}},
  \bibinfo {author} {\bibfnamefont {M.}~\bibnamefont {Shiga}}, \ and\ \bibinfo
  {author} {\bibfnamefont {D.}~\bibnamefont {Marx}},\ }\bibfield  {title}
  {\enquote {\bibinfo {title} {Communications: {{On}} artificial frequency
  shifts in infrared spectra obtained from centroid molecular dynamics:
  {{Quantum}} liquid water},}\ }\href {\doibase 10.1063/1.3290958} {\bibfield
  {journal} {\bibinfo  {journal} {J. Chem. Phys.}\ }\textbf {\bibinfo {volume}
  {132}},\ \bibinfo {pages} {031101} (\bibinfo {year} {2010})}\BibitemShut
  {NoStop}%
\bibitem [{\citenamefont {Trenins}, \citenamefont {Willatt},\ and\
  \citenamefont {Althorpe}(2019)}]{augTrenins2019}%
  \BibitemOpen
  \bibfield  {author} {\bibinfo {author} {\bibfnamefont {G.}~\bibnamefont
  {Trenins}}, \bibinfo {author} {\bibfnamefont {M.~J.}\ \bibnamefont
  {Willatt}}, \ and\ \bibinfo {author} {\bibfnamefont {S.~C.}\ \bibnamefont
  {Althorpe}},\ }\bibfield  {title} {\enquote {\bibinfo {title} {Path-integral
  dynamics of water using curvilinear centroids},}\ }\href {\doibase
  10.1063/1.5100587} {\bibfield  {journal} {\bibinfo  {journal} {J. Chem.
  Phys}\ }\textbf {\bibinfo {volume} {151}},\ \bibinfo {pages} {054109}
  (\bibinfo {year} {2019})}\BibitemShut {NoStop}%
\bibitem [{\citenamefont {Trenins}, \citenamefont {Haggard},\ and\
  \citenamefont {Althorpe}(2022)}]{novTrenins2022}%
  \BibitemOpen
  \bibfield  {author} {\bibinfo {author} {\bibfnamefont {G.}~\bibnamefont
  {Trenins}}, \bibinfo {author} {\bibfnamefont {C.}~\bibnamefont {Haggard}}, \
  and\ \bibinfo {author} {\bibfnamefont {S.~C.}\ \bibnamefont {Althorpe}},\
  }\bibfield  {title} {\enquote {\bibinfo {title} {Improved torque estimator
  for condensed-phase quasicentroid molecular dynamics},}\ }\href {\doibase
  10.1063/5.0129482} {\bibfield  {journal} {\bibinfo  {journal} {The Journal of
  Chemical Physics}\ }\textbf {\bibinfo {volume} {157}},\ \bibinfo {pages}
  {174108} (\bibinfo {year} {2022})}\BibitemShut {NoStop}%
\bibitem [{\citenamefont {Fletcher}\ \emph {et~al.}(2021)\citenamefont
  {Fletcher}, \citenamefont {Zhu}, \citenamefont {Lawrence},\ and\
  \citenamefont {Manolopoulos}}]{decFletcher2021}%
  \BibitemOpen
  \bibfield  {author} {\bibinfo {author} {\bibfnamefont {T.}~\bibnamefont
  {Fletcher}}, \bibinfo {author} {\bibfnamefont {A.}~\bibnamefont {Zhu}},
  \bibinfo {author} {\bibfnamefont {J.~E.}\ \bibnamefont {Lawrence}}, \ and\
  \bibinfo {author} {\bibfnamefont {D.~E.}\ \bibnamefont {Manolopoulos}},\
  }\bibfield  {title} {\enquote {\bibinfo {title} {Fast quasi-centroid
  molecular dynamics},}\ }\href {\doibase 10.1063/5.0076704} {\bibfield
  {journal} {\bibinfo  {journal} {J. Chem. Phys.}\ }\textbf {\bibinfo {volume}
  {155}},\ \bibinfo {pages} {231101} (\bibinfo {year} {2021})}\BibitemShut
  {NoStop}%
\bibitem [{\citenamefont {Lawrence}\ \emph {et~al.}(2023)\citenamefont
  {Lawrence}, \citenamefont {Lieberherr}, \citenamefont {Fletcher},\ and\
  \citenamefont {Manolopoulos}}]{octLawrence2023}%
  \BibitemOpen
  \bibfield  {author} {\bibinfo {author} {\bibfnamefont {J.~E.}\ \bibnamefont
  {Lawrence}}, \bibinfo {author} {\bibfnamefont {A.~Z.}\ \bibnamefont
  {Lieberherr}}, \bibinfo {author} {\bibfnamefont {T.}~\bibnamefont
  {Fletcher}}, \ and\ \bibinfo {author} {\bibfnamefont {D.~E.}\ \bibnamefont
  {Manolopoulos}},\ }\bibfield  {title} {\enquote {\bibinfo {title} {Fast
  {{Quasi-Centroid Molecular Dynamics}} for {{Water}} and {{Ice}}},}\ }\href
  {\doibase 10.1021/acs.jpcb.3c05028} {\bibfield  {journal} {\bibinfo
  {journal} {J. Phys. Chem. B}\ }\textbf {\bibinfo {volume} {127}},\ \bibinfo
  {pages} {9172--9180} (\bibinfo {year} {2023})}\BibitemShut {NoStop}%
\bibitem [{\citenamefont {Limbu}\ \emph {et~al.}(2025)\citenamefont {Limbu},
  \citenamefont {London}, \citenamefont {Faruque},\ and\ \citenamefont
  {Momeni}}]{janLimbu2025}%
  \BibitemOpen
  \bibfield  {author} {\bibinfo {author} {\bibfnamefont {D.~K.}\ \bibnamefont
  {Limbu}}, \bibinfo {author} {\bibfnamefont {N.}~\bibnamefont {London}},
  \bibinfo {author} {\bibfnamefont {M.~O.}\ \bibnamefont {Faruque}}, \ and\
  \bibinfo {author} {\bibfnamefont {M.~R.}\ \bibnamefont {Momeni}},\ }\bibfield
   {title} {\enquote {\bibinfo {title} {H-{{CMD}}: {{An}} efficient hybrid fast
  centroid and quasi-centroid molecular dynamics method for the simulation of
  vibrational spectra},}\ }\href {\doibase 10.1063/5.0248115} {\bibfield
  {journal} {\bibinfo  {journal} {J. Chem. Phys.}\ }\textbf {\bibinfo {volume}
  {162}},\ \bibinfo {pages} {014111} (\bibinfo {year} {2025})}\BibitemShut
  {NoStop}%
\bibitem [{\citenamefont {Musil}\ \emph {et~al.}(2022)\citenamefont {Musil},
  \citenamefont {Zaporozhets}, \citenamefont {Noé}, \citenamefont {Clementi},\
  and\ \citenamefont {Kapil}}]{tepigs}%
  \BibitemOpen
  \bibfield  {author} {\bibinfo {author} {\bibfnamefont {F.}~\bibnamefont
  {Musil}}, \bibinfo {author} {\bibfnamefont {I.}~\bibnamefont {Zaporozhets}},
  \bibinfo {author} {\bibfnamefont {F.}~\bibnamefont {Noé}}, \bibinfo {author}
  {\bibfnamefont {C.}~\bibnamefont {Clementi}}, \ and\ \bibinfo {author}
  {\bibfnamefont {V.}~\bibnamefont {Kapil}},\ }\bibfield  {title} {\enquote
  {\bibinfo {title} {{Quantum dynamics using path integral coarse-graining}},}\
  }\href {\doibase 10.1063/5.0120386} {\bibfield  {journal} {\bibinfo
  {journal} {J. Chem. Phys.}\ }\textbf {\bibinfo {volume} {157}},\ \bibinfo
  {pages} {181102} (\bibinfo {year} {2022})}\BibitemShut {NoStop}%
\bibitem [{\citenamefont {Doll}\ and\ \citenamefont
  {Freeman}(1984)}]{marDoll1984}%
  \BibitemOpen
  \bibfield  {author} {\bibinfo {author} {\bibfnamefont {J.~D.}\ \bibnamefont
  {Doll}}\ and\ \bibinfo {author} {\bibfnamefont {D.~L.}\ \bibnamefont
  {Freeman}},\ }\bibfield  {title} {\enquote {\bibinfo {title} {A {{Monte
  Carlo}} method for quantum {{Boltzmann}} statistical mechanics},}\ }\href
  {\doibase 10.1063/1.446919} {\bibfield  {journal} {\bibinfo  {journal} {J.
  Chem. Phys.}\ }\textbf {\bibinfo {volume} {80}},\ \bibinfo {pages}
  {2239--2240} (\bibinfo {year} {1984})}\BibitemShut {NoStop}%
\bibitem [{\citenamefont {Doll}, \citenamefont {Coalson},\ and\ \citenamefont
  {Freeman}(1985)}]{julDoll1985}%
  \BibitemOpen
  \bibfield  {author} {\bibinfo {author} {\bibfnamefont {J.~D.}\ \bibnamefont
  {Doll}}, \bibinfo {author} {\bibfnamefont {R.~D.}\ \bibnamefont {Coalson}}, \
  and\ \bibinfo {author} {\bibfnamefont {D.~L.}\ \bibnamefont {Freeman}},\
  }\bibfield  {title} {\enquote {\bibinfo {title} {Fourier path-integral monte
  carlo methods: Partial averaging},}\ }\href {\doibase
  10.1103/PhysRevLett.55.1} {\bibfield  {journal} {\bibinfo  {journal} {Phys.
  Rev. Lett.}\ }\textbf {\bibinfo {volume} {55}},\ \bibinfo {pages} {1--4}
  (\bibinfo {year} {1985})}\BibitemShut {NoStop}%
\bibitem [{\citenamefont {Lobaugh}\ and\ \citenamefont
  {Voth}(1992)}]{sepLobaugh1992}%
  \BibitemOpen
  \bibfield  {author} {\bibinfo {author} {\bibfnamefont {J.}~\bibnamefont
  {Lobaugh}}\ and\ \bibinfo {author} {\bibfnamefont {G.~A.}\ \bibnamefont
  {Voth}},\ }\bibfield  {title} {\enquote {\bibinfo {title} {A partial
  averaging strategy for low temperature fourier path integral monte carlo
  calculations},}\ }\href {\doibase 10.1063/1.463923} {\bibfield  {journal}
  {\bibinfo  {journal} {J. Chem. Phys.}\ }\textbf {\bibinfo {volume} {97}},\
  \bibinfo {pages} {4205--4214} (\bibinfo {year} {1992})}\BibitemShut {NoStop}%
\bibitem [{\citenamefont {Topper}\ \emph {et~al.}(1993)\citenamefont {Topper},
  \citenamefont {Zhang}, \citenamefont {Liu},\ and\ \citenamefont
  {Truhlar}}]{marTopper1993}%
  \BibitemOpen
  \bibfield  {author} {\bibinfo {author} {\bibfnamefont {R.~Q.}\ \bibnamefont
  {Topper}}, \bibinfo {author} {\bibfnamefont {Q.}~\bibnamefont {Zhang}},
  \bibinfo {author} {\bibfnamefont {Y.-P.}\ \bibnamefont {Liu}}, \ and\
  \bibinfo {author} {\bibfnamefont {D.~G.}\ \bibnamefont {Truhlar}},\
  }\bibfield  {title} {\enquote {\bibinfo {title} {Quantum steam tables. free
  energy calculations for h2o, d2o, h2s, and h2se by adaptively optimized monte
  carlo fourier path integrals},}\ }\href {\doibase 10.1063/1.464953}
  {\bibfield  {journal} {\bibinfo  {journal} {J. Chem. Phys.}\ }\textbf
  {\bibinfo {volume} {98}},\ \bibinfo {pages} {4991--5005} (\bibinfo {year}
  {1993})}\BibitemShut {NoStop}%
\bibitem [{\citenamefont {{Vorontsov-Velyaminov}}, \citenamefont {Nesvit},\
  and\ \citenamefont {Gorbunov}(1997)}]{febVorontsov1997}%
  \BibitemOpen
  \bibfield  {author} {\bibinfo {author} {\bibfnamefont {P.~N.}\ \bibnamefont
  {{Vorontsov-Velyaminov}}}, \bibinfo {author} {\bibfnamefont {M.~O.}\
  \bibnamefont {Nesvit}}, \ and\ \bibinfo {author} {\bibfnamefont {R.~I.}\
  \bibnamefont {Gorbunov}},\ }\bibfield  {title} {\enquote {\bibinfo {title}
  {Bead-{{Fourier}} path-integral {{Monte Carlo}} method applied to systems of
  identical particles},}\ }\href {\doibase 10.1103/PhysRevE.55.1979} {\bibfield
   {journal} {\bibinfo  {journal} {Phys. Rev. E}\ }\textbf {\bibinfo {volume}
  {55}},\ \bibinfo {pages} {1979--1997} (\bibinfo {year} {1997})}\BibitemShut
  {NoStop}%
\bibitem [{\citenamefont {Ivanov}, \citenamefont {Lyubartsev},\ and\
  \citenamefont {Laaksonen}(2003)}]{junIvanov2003}%
  \BibitemOpen
  \bibfield  {author} {\bibinfo {author} {\bibfnamefont {S.~D.}\ \bibnamefont
  {Ivanov}}, \bibinfo {author} {\bibfnamefont {A.~P.}\ \bibnamefont
  {Lyubartsev}}, \ and\ \bibinfo {author} {\bibfnamefont {A.}~\bibnamefont
  {Laaksonen}},\ }\bibfield  {title} {\enquote {\bibinfo {title}
  {Bead-{{Fourier}} path integral molecular dynamics},}\ }\href {\doibase
  10.1103/PhysRevE.67.066710} {\bibfield  {journal} {\bibinfo  {journal} {Phys.
  Rev. E}\ }\textbf {\bibinfo {volume} {67}},\ \bibinfo {pages} {066710}
  (\bibinfo {year} {2003})}\BibitemShut {NoStop}%
\bibitem [{\citenamefont {Ivanov}\ and\ \citenamefont
  {Lyubartsev}(2005)}]{julIvanov2005}%
  \BibitemOpen
  \bibfield  {author} {\bibinfo {author} {\bibfnamefont {S.~D.}\ \bibnamefont
  {Ivanov}}\ and\ \bibinfo {author} {\bibfnamefont {A.~P.}\ \bibnamefont
  {Lyubartsev}},\ }\bibfield  {title} {\enquote {\bibinfo {title} {Simulations
  of one- and two-electron systems by {{Bead-Fourier}} path integral molecular
  dynamics},}\ }\href {\doibase 10.1063/1.1961312} {\bibfield  {journal}
  {\bibinfo  {journal} {J. Chem. Phys.}\ }\textbf {\bibinfo {volume} {123}},\
  \bibinfo {pages} {034105} (\bibinfo {year} {2005})}\BibitemShut {NoStop}%
\bibitem [{\citenamefont {Craig}\ and\ \citenamefont
  {Manolopoulos}(2004)}]{augCraig2004}%
  \BibitemOpen
  \bibfield  {author} {\bibinfo {author} {\bibfnamefont {I.~R.}\ \bibnamefont
  {Craig}}\ and\ \bibinfo {author} {\bibfnamefont {D.~E.}\ \bibnamefont
  {Manolopoulos}},\ }\bibfield  {title} {\enquote {\bibinfo {title} {Quantum
  statistics and classical mechanics: {{Real}} time correlation functions from
  ring polymer molecular dynamics},}\ }\href {\doibase 10.1063/1.1777575}
  {\bibfield  {journal} {\bibinfo  {journal} {The Journal of Chemical Physics}\
  }\textbf {\bibinfo {volume} {121}},\ \bibinfo {pages} {3368--3373} (\bibinfo
  {year} {2004})},\ \Eprint
  {http://arxiv.org/abs/https://doi.org/10.1063/1.1777575}
  {https://doi.org/10.1063/1.1777575} \BibitemShut {NoStop}%
\bibitem [{\citenamefont {Mielke}\ and\ \citenamefont
  {Truhlar}(2001)}]{janMielke2001a}%
  \BibitemOpen
  \bibfield  {author} {\bibinfo {author} {\bibfnamefont {S.~L.}\ \bibnamefont
  {Mielke}}\ and\ \bibinfo {author} {\bibfnamefont {D.~G.}\ \bibnamefont
  {Truhlar}},\ }\bibfield  {title} {\enquote {\bibinfo {title} {A new
  {{Fourier}} path integral method, a more general scheme for extrapolation,
  and comparison of eight path integral methods for the quantum mechanical
  calculation of free energies},}\ }\href {\doibase 10.1063/1.1290476}
  {\bibfield  {journal} {\bibinfo  {journal} {J. Chem. Phys.}\ }\textbf
  {\bibinfo {volume} {114}},\ \bibinfo {pages} {621--630} (\bibinfo {year}
  {2001})}\BibitemShut {NoStop}%
\bibitem [{\citenamefont {Boerner}\ \emph {et~al.}(2023)\citenamefont
  {Boerner}, \citenamefont {Deems}, \citenamefont {Furlani}, \citenamefont
  {Knuth},\ and\ \citenamefont {Towns}}]{access}%
  \BibitemOpen
  \bibfield  {author} {\bibinfo {author} {\bibfnamefont {T.~J.}\ \bibnamefont
  {Boerner}}, \bibinfo {author} {\bibfnamefont {S.}~\bibnamefont {Deems}},
  \bibinfo {author} {\bibfnamefont {T.~R.}\ \bibnamefont {Furlani}}, \bibinfo
  {author} {\bibfnamefont {S.~L.}\ \bibnamefont {Knuth}}, \ and\ \bibinfo
  {author} {\bibfnamefont {J.}~\bibnamefont {Towns}},\ }\bibfield  {title}
  {\enquote {\bibinfo {title} {{{ACCESS}}: {{Advancing}} innovation: {{NSF}}'s
  advanced cyberinfrastructure coordination ecosystem: {{Services}} \&
  support},}\ }in\ \href {\doibase 10.1145/3569951.3597559} {\emph {\bibinfo
  {booktitle} {Practice and Experience in Advanced Research Computing}}},\
  \bibinfo {series and number} {Pearc '23}\ (\bibinfo  {publisher} {Association
  for Computing Machinery},\ \bibinfo {address} {New York, NY, USA},\ \bibinfo
  {year} {2023})\ pp.\ \bibinfo {pages} {173--176}\BibitemShut {NoStop}%
\end{thebibliography}%


\providecommand{\latin}[1]{#1}
\makeatletter
\providecommand{\doi}
  {\begingroup\let\do\@makeother\dospecials
  \catcode`\{=1 \catcode`\}=2 \doi@aux}
\providecommand{\doi@aux}[1]{\endgroup\texttt{#1}}
\makeatother
\providecommand*\mcitethebibliography{\thebibliography}
\csname @ifundefined\endcsname{endmcitethebibliography}
  {\let\endmcitethebibliography\endthebibliography}{}
\begin{mcitethebibliography}{0}
\providecommand*\natexlab[1]{#1}
\providecommand*\mciteSetBstSublistMode[1]{}
\providecommand*\mciteSetBstMaxWidthForm[2]{}
\providecommand*\mciteBstWouldAddEndPuncttrue
  {\def\EndOfBibitem{\unskip.}}
\providecommand*\mciteBstWouldAddEndPunctfalse
  {\let\EndOfBibitem\relax}
\providecommand*\mciteSetBstMidEndSepPunct[3]{}
\providecommand*\mciteSetBstSublistLabelBeginEnd[3]{}
\providecommand*\EndOfBibitem{}
\mciteSetBstSublistMode{f}
\mciteSetBstMaxWidthForm{subitem}{(\alph{mcitesubitemcount})}
\mciteSetBstSublistLabelBeginEnd
  {\mcitemaxwidthsubitemform\space}
  {\relax}
  {\relax}

\end{mcitethebibliography}
  \end{document}


\clearpage

\section{Section S1. Effective Potential Sampling Details}
\begin{table}[]
		\centering
		\caption{Effective potential grid details for all model systems considered in this work.}
		\label{tab:grid}
		\begin{tabular}{|c|c|c|c|}
			\hline
      Model & $Q_{\mathrm{min}}$ & $Q_{\mathrm{max}}$ & $n_{\mathrm{points}}$   \\
			\hline
			 Harmonic & -4.5 & 4.5 & 101  \\
			 Mildly anharmonic & -6.5 & 3.5 & 101  \\
			 Quartic & -3 & 3 & 241  \\
			\hline
		\end{tabular}
	\end{table} 
\clearpage
\section{Section S2. Covergence data for harmonic oscillator}
\begin{figure}[h!]
\begin{center}
  \includegraphics[scale=1]{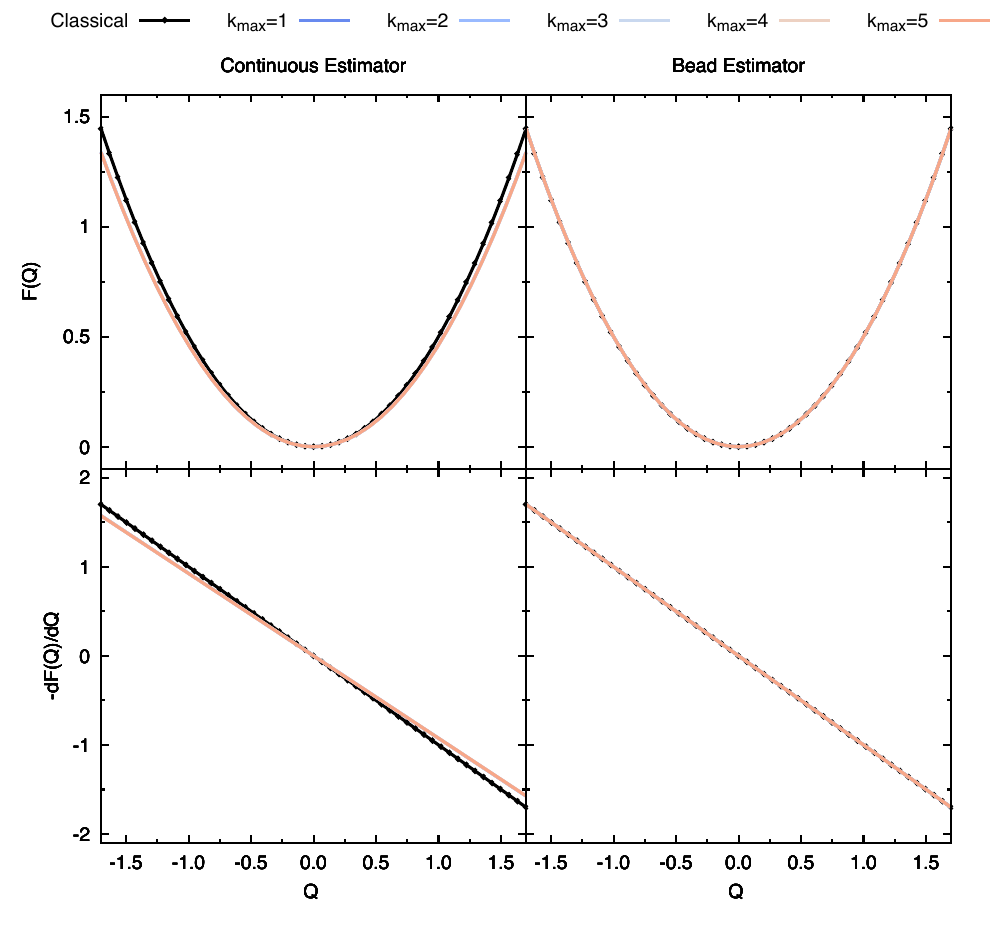}
\end{center}
\caption{Harmonic oscillator free energy (top) and force (bottom) for 1 bead at $\beta=1$.}
\label{fig:harm-pot-beta1-1bd}
\end{figure}

\clearpage
\begin{figure}[h!]
\begin{center}
  \includegraphics[scale=1]{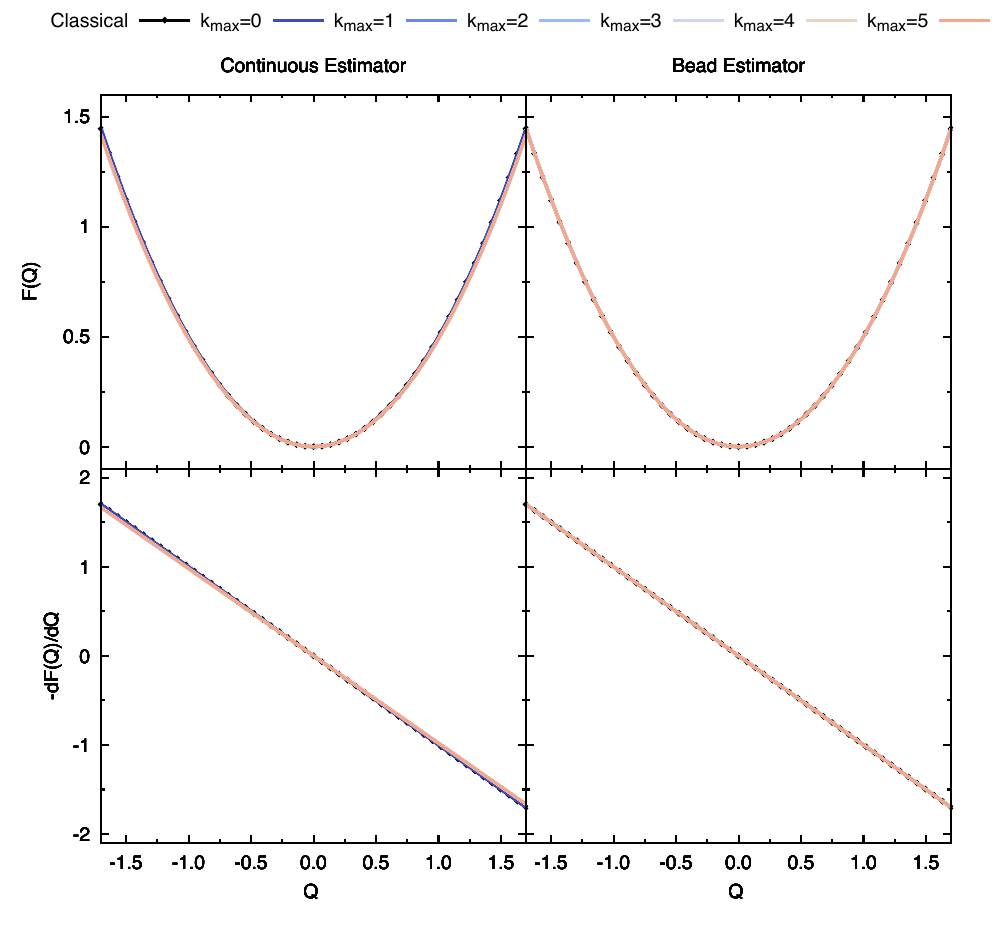}
\end{center}
\caption{Harmonic oscillator free energy (top) and force (bottom) for 2 beads at $\beta=1$.}
\label{fig:harm-pot-beta1-2bd}
\end{figure}

\newpage
\begin{figure}[h!]
\begin{center}
  \includegraphics[scale=1]{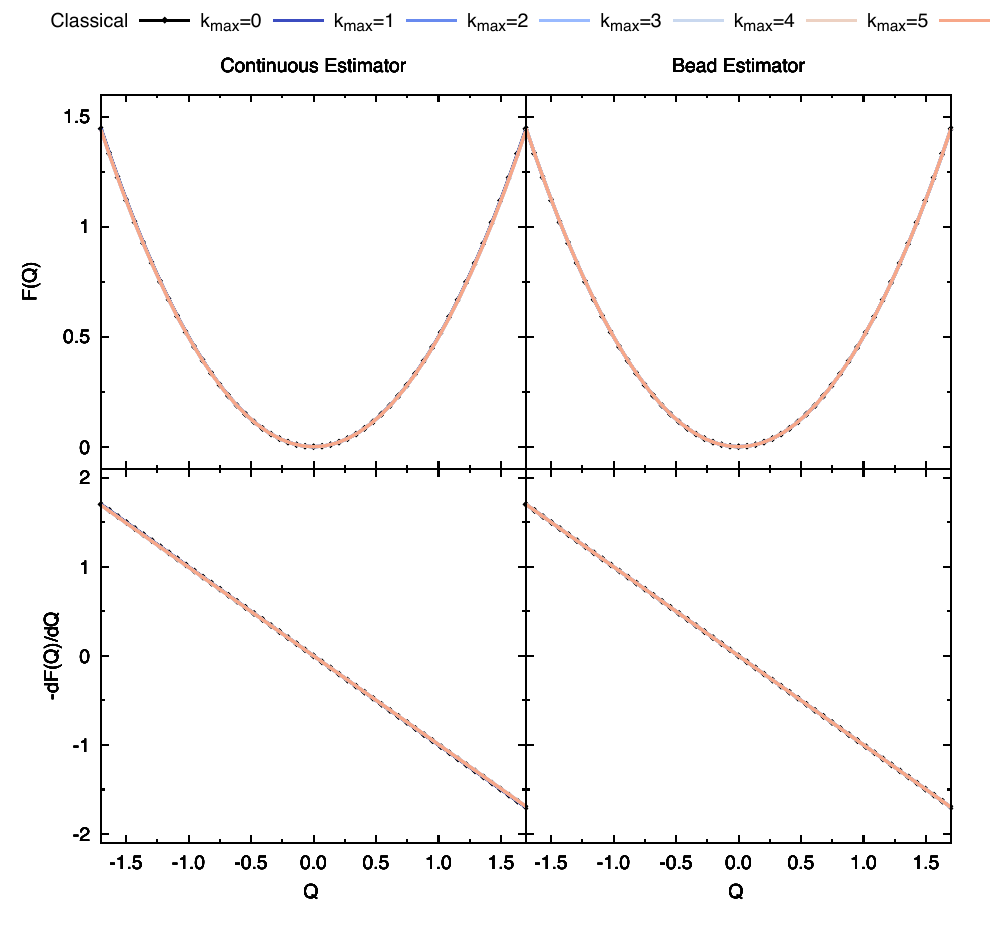}
\end{center}
\caption{Harmonic oscillator free energy (top) and force (bottom) for 4 beads at $\beta=1$.}
\label{fig:harm-pot-beta1-4bd}
\end{figure}

\newpage
\begin{figure}[h!]
\begin{center}
  \includegraphics[scale=1]{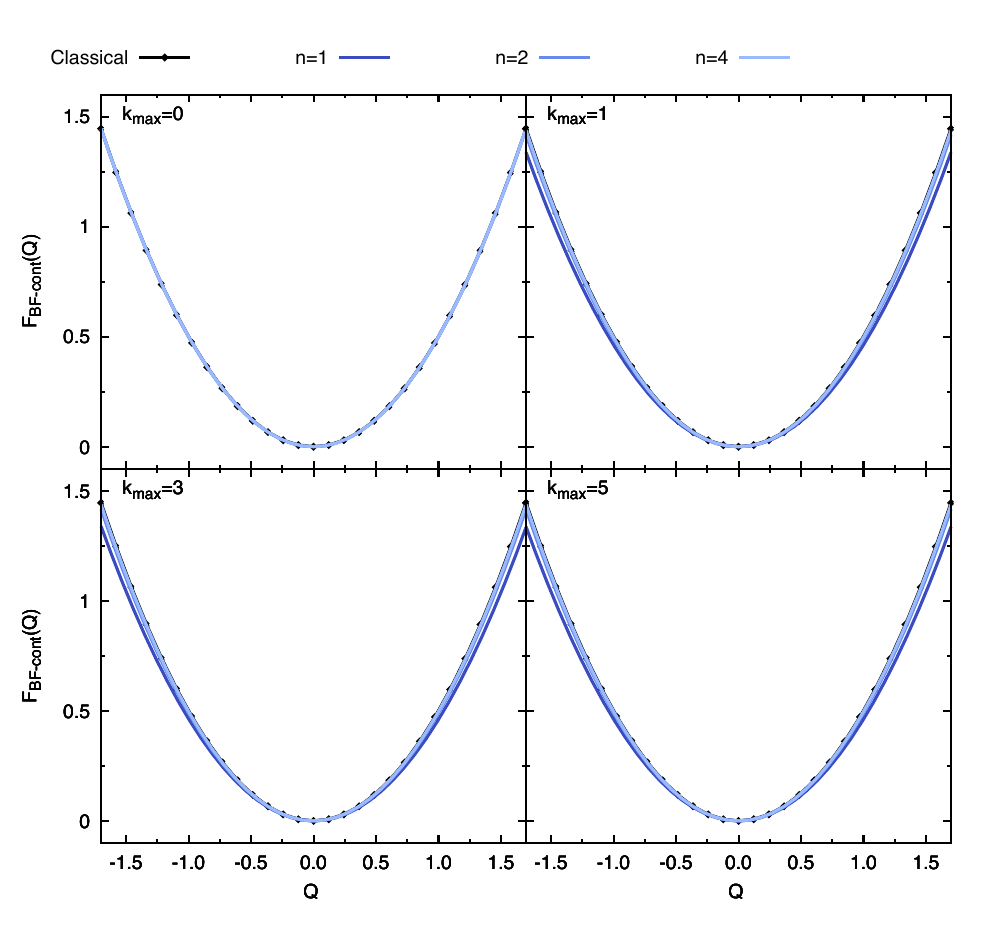}
\end{center}
\caption{Continuous estimated free energy for harmonic oscillator $ \beta=1 $.}
\label{fig:harm-pot-beta1-cont}
\end{figure}

\newpage
\begin{figure}[h!]
\begin{center}
  \includegraphics[scale=1]{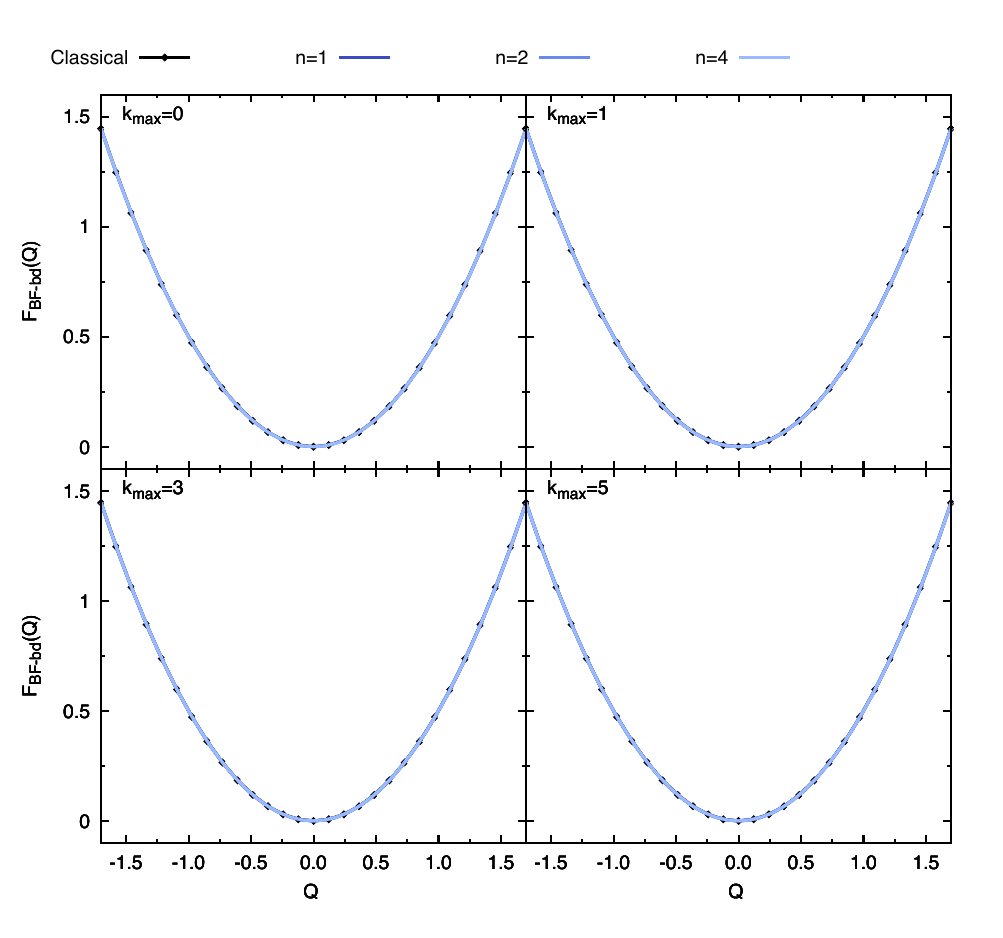}
\end{center}
\caption{Bead estimated free energy for harmonic oscillator $ \beta=1 $.}
\label{fig:harm-pot-beta1-bead}
\end{figure}


\newpage

\begin{figure}[h!]
\begin{center}
  \includegraphics[scale=1]{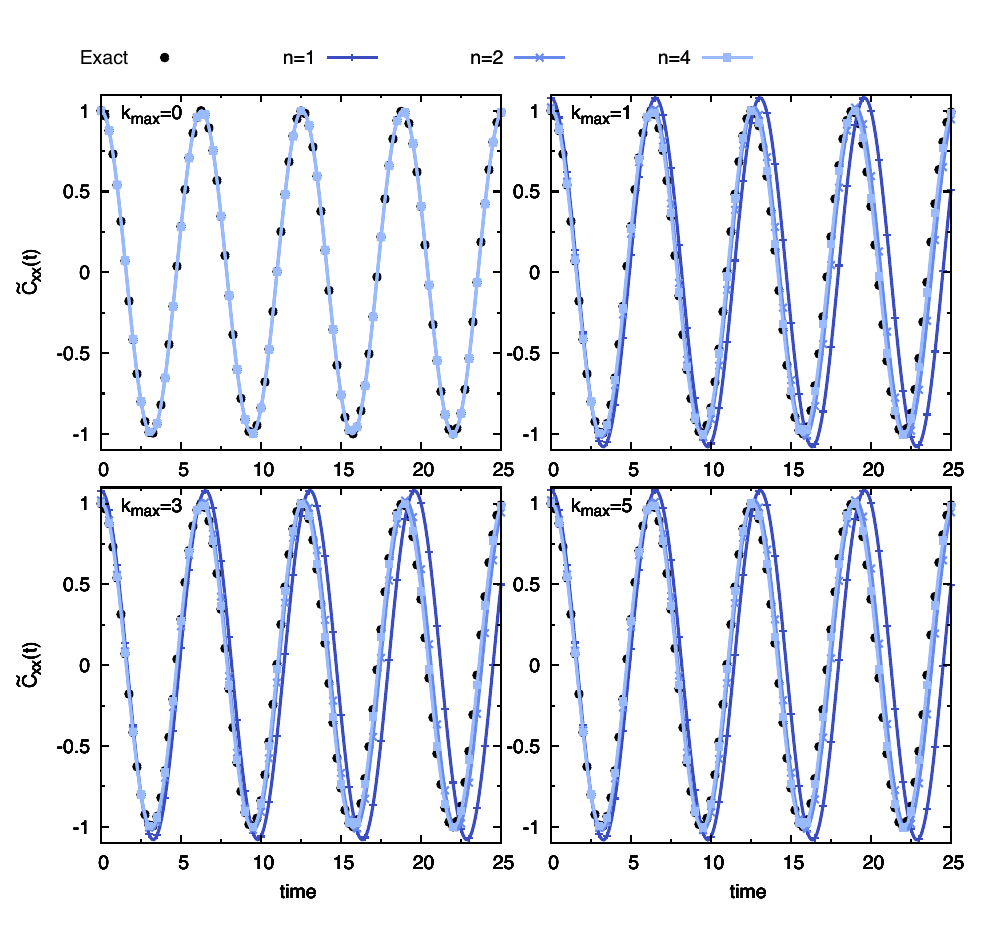}
\end{center}
\caption{Position autocorrelation function from continuous estimated free energy for harmonic oscillator $ \beta=1 $.}
\label{fig:harm-corr-beta1-cont}
\end{figure}

\newpage
\begin{figure}[h!]
\begin{center}
  \includegraphics[scale=1]{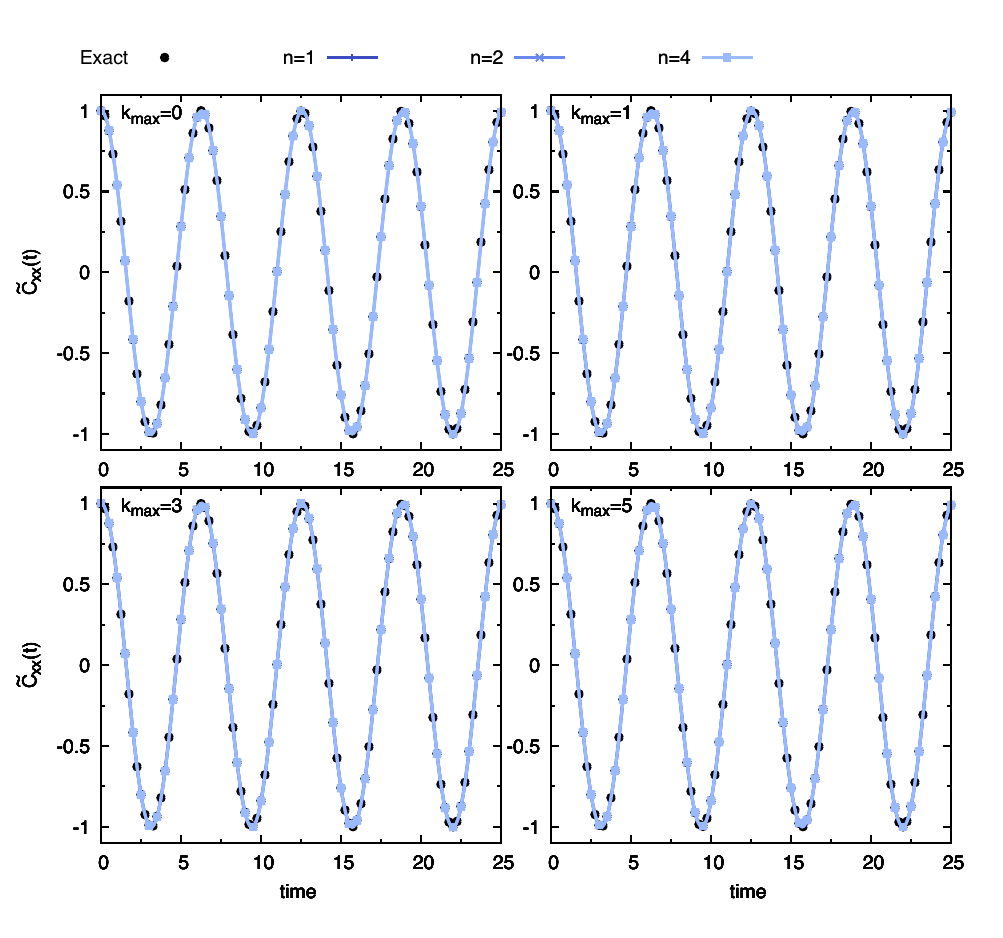}
\end{center}
\caption{Position autocorrelation function from bead estimated free energy for harmonic oscillator $ \beta=1 $.}
\label{fig:harm-corr-beta1-bead}
\end{figure}

\newpage
\begin{figure}[h!]
\begin{center}
  \includegraphics[scale=1]{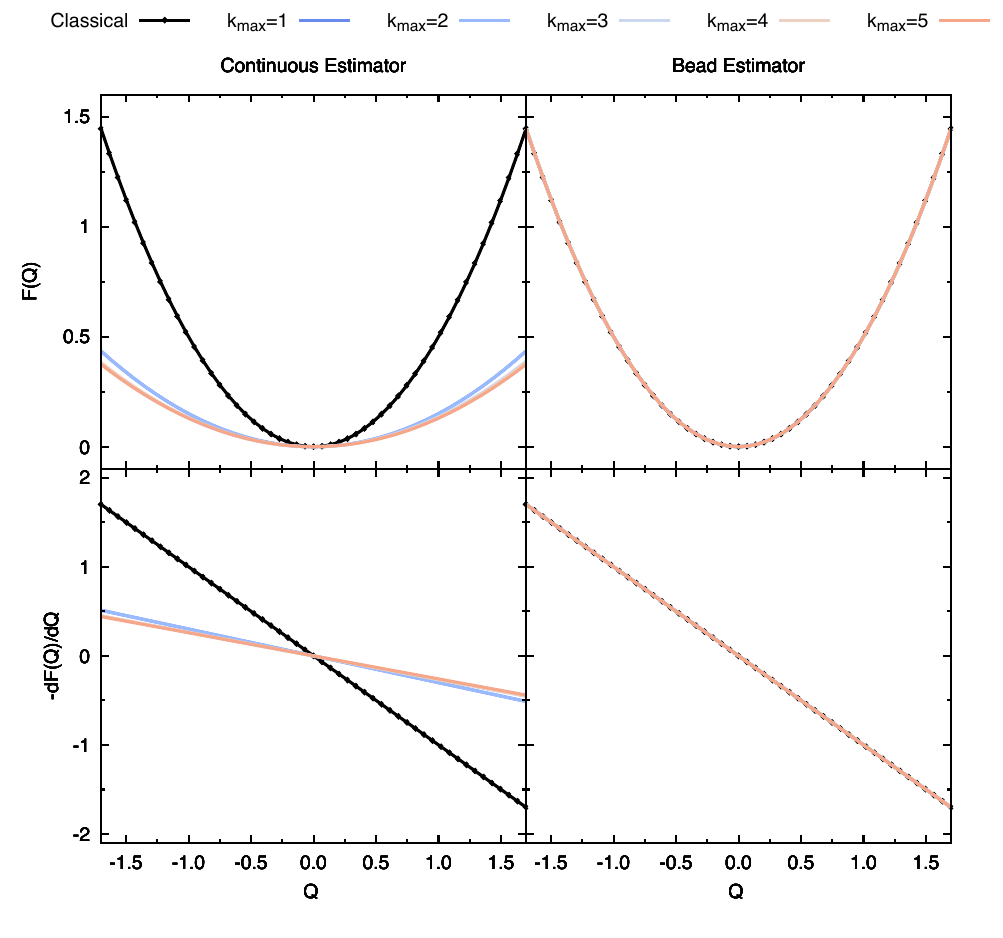}
\end{center}
\caption{Harmonic oscillator free energy (top) and force (bottom) for 1 bead at $\beta=8$.}
\label{fig:harm-pot-beta8-1bd}
\end{figure}

\newpage
\begin{figure}[h!]
\begin{center}
  \includegraphics[scale=1]{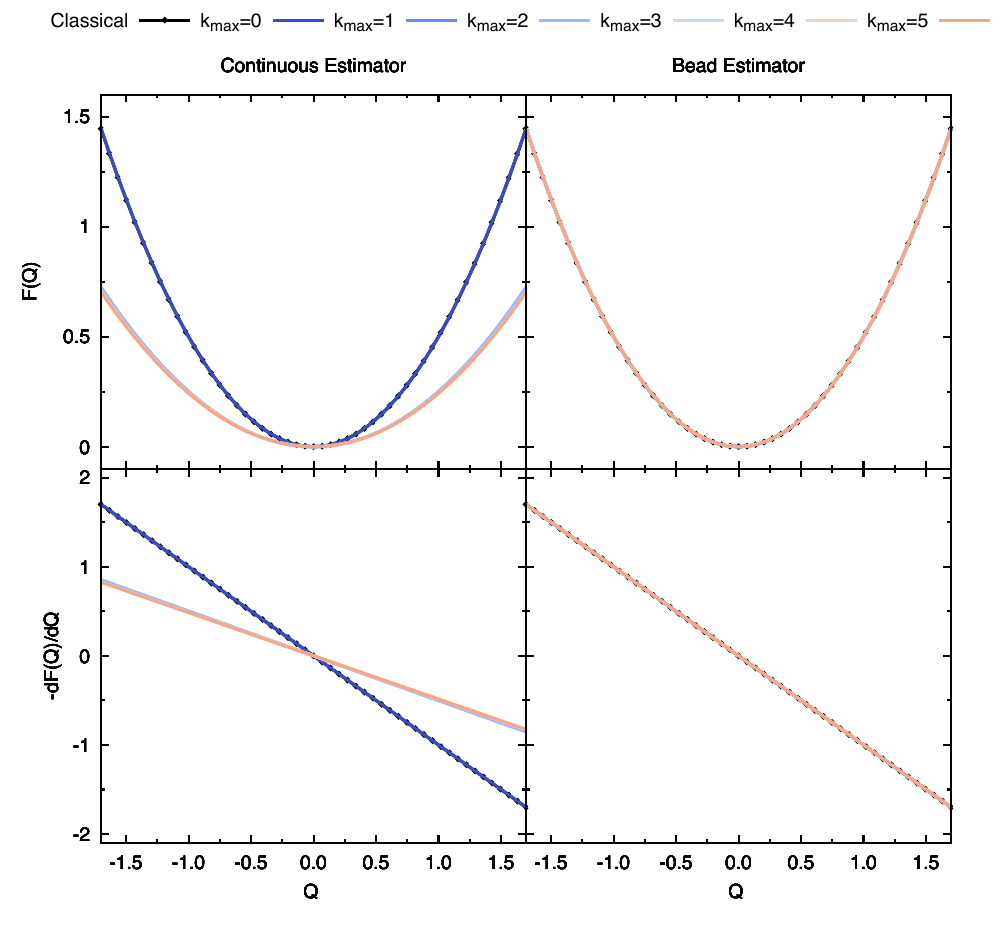}
\end{center}
\caption{Harmonic oscillator free energy (top) and force (bottom) for 2 beads at $\beta=8$.}
\label{fig:harm-pot-beta8-2bd}
\end{figure}

\newpage
\begin{figure}[h!]
\begin{center}
  \includegraphics[scale=1]{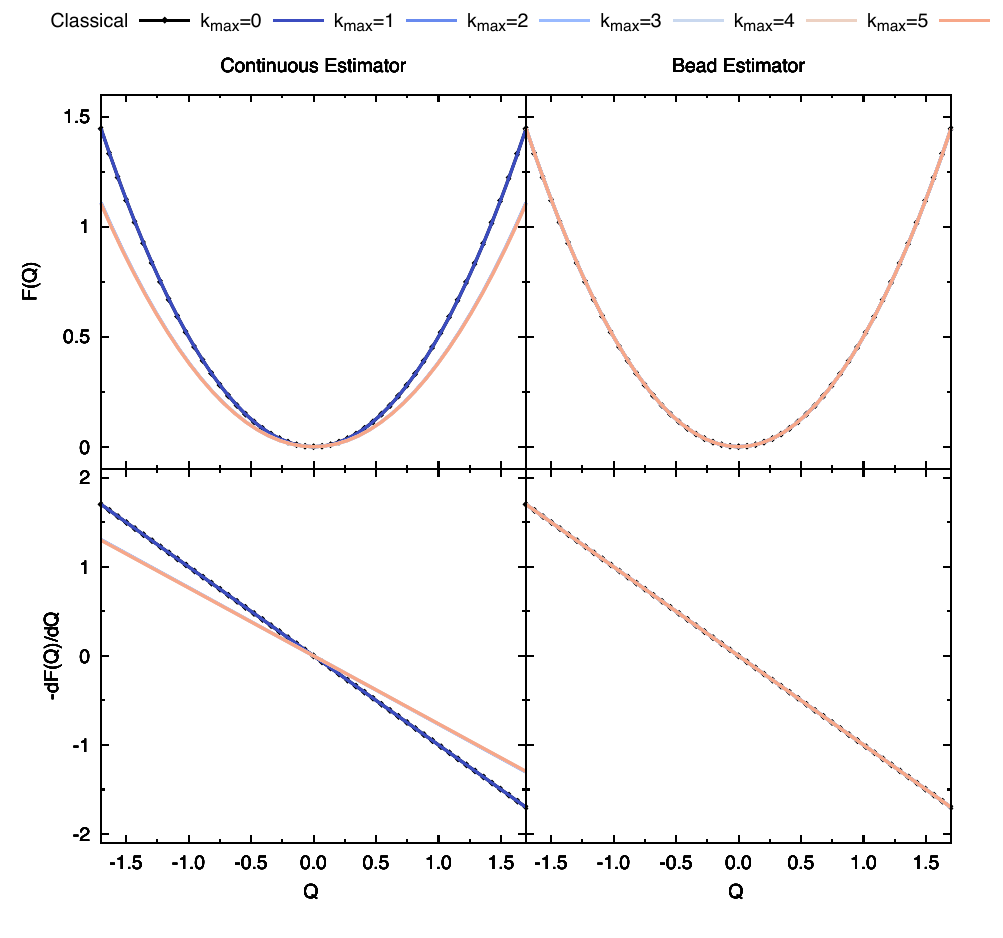}
\end{center}
\caption{Harmonic oscillator free energy (top) and force (bottom) for 4 beads at $\beta=8$.}
\label{fig:harm-pot-beta8-4bd}
\end{figure}

\newpage
\begin{figure}[h!]
\begin{center}
  \includegraphics[scale=1]{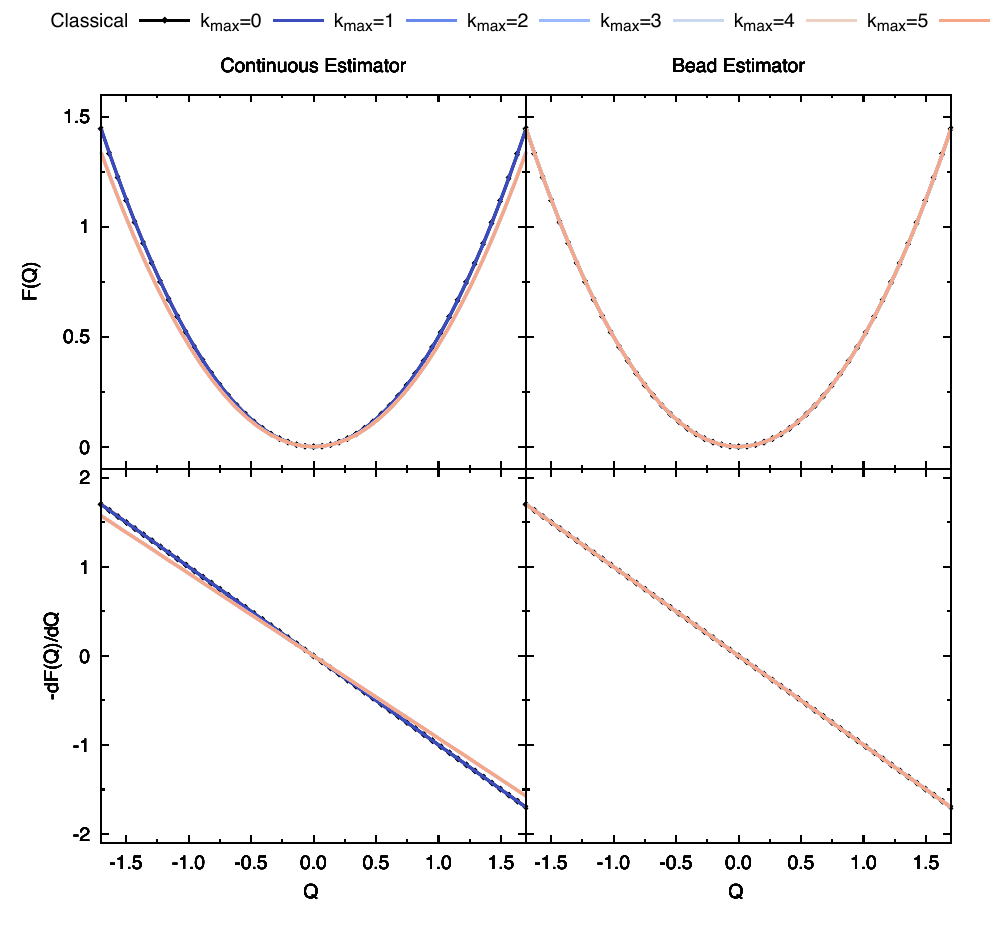}
\end{center}
\caption{Harmonic oscillator free energy (top) and force (bottom) for 8 beads at $\beta=8$.}
\label{fig:harm-pot-beta8-8bd}
\end{figure}

\newpage
\begin{figure}[h!]
\begin{center}
  \includegraphics[scale=1]{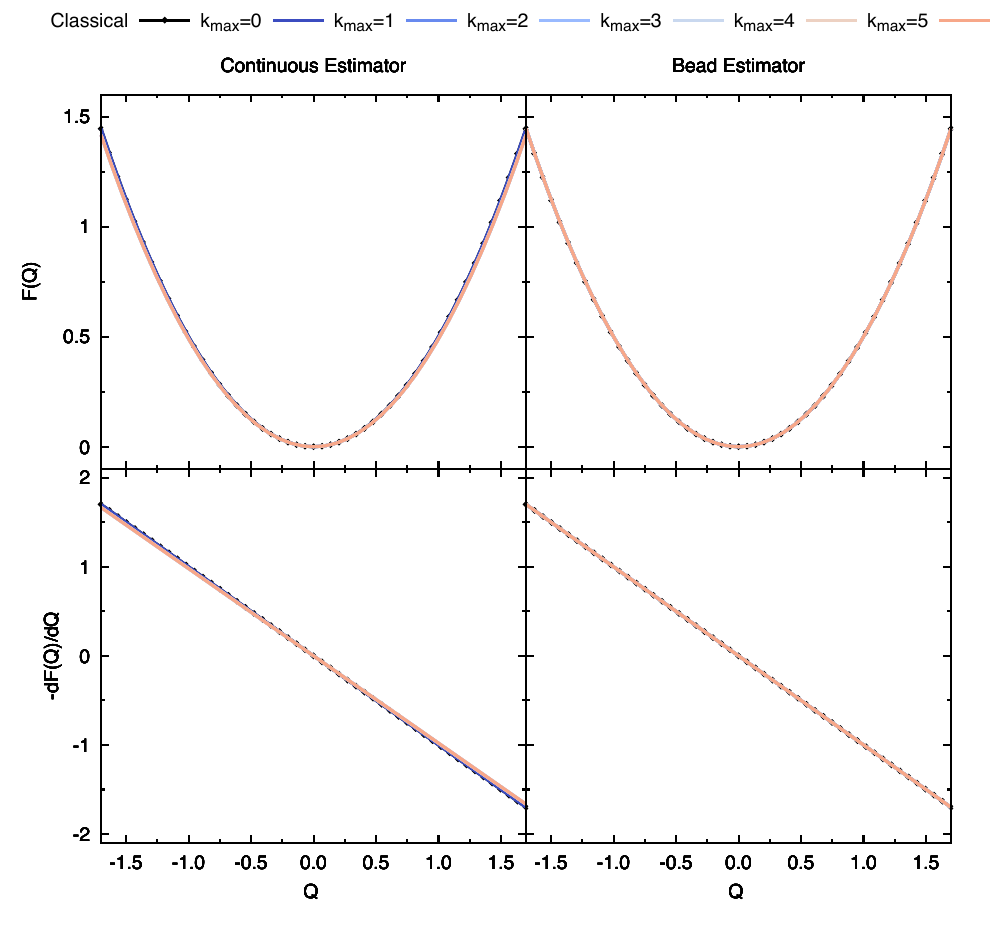}
\end{center}
\caption{Harmonic oscillator free energy (top) and force (bottom) for 16 beads at $\beta=8$.}
\label{fig:harm-pot-beta8-16bd}
\end{figure}

\newpage
\begin{figure}[h!]
\begin{center}
  \includegraphics[scale=1]{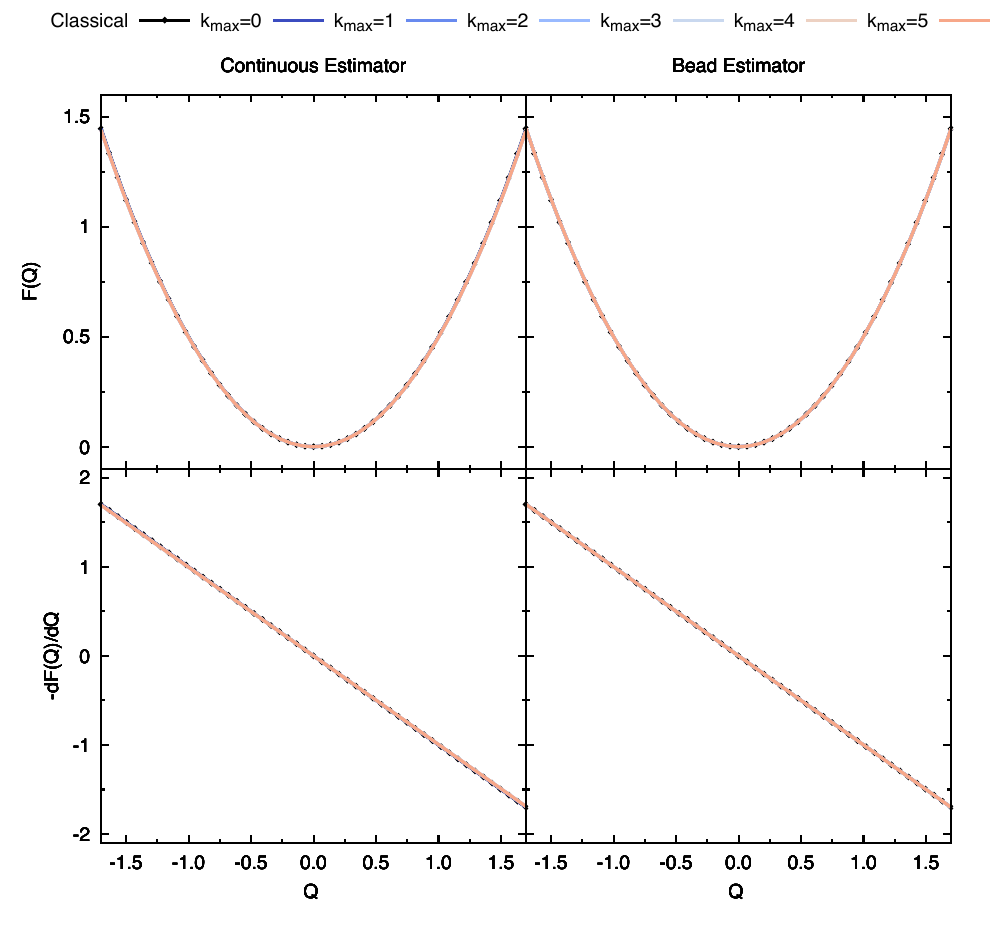}
\end{center}
\caption{Harmonic oscillator free energy (top) and force (bottom) for 32 beads at $\beta=8$.}
\label{fig:harm-pot-beta8-32bd}
\end{figure}

\newpage
\begin{figure}[h!]
\begin{center}
  \includegraphics[scale=1]{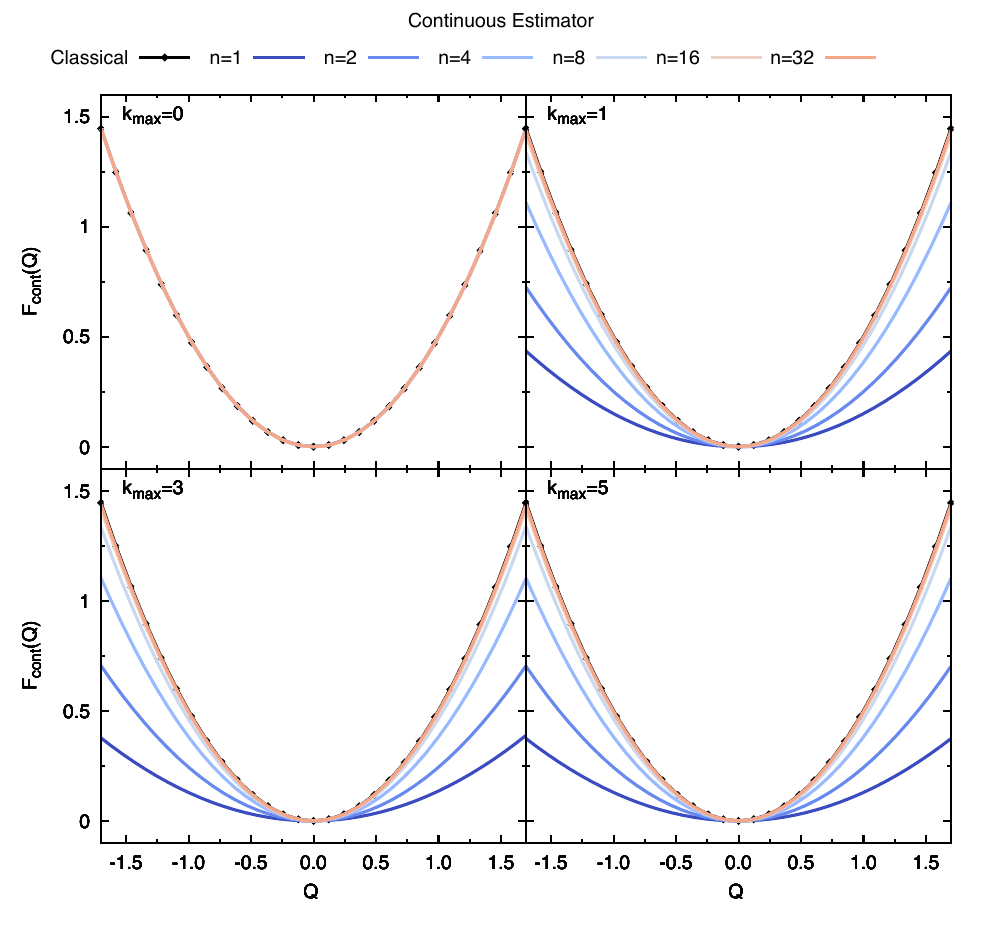}
\end{center}
\caption{Continuous estimated free energy for harmonic oscillator $ \beta=8 $.}
\label{fig:harm-pot-beta8-cont}
\end{figure}

\newpage
\begin{figure}[h!]
\begin{center}
  \includegraphics[scale=1]{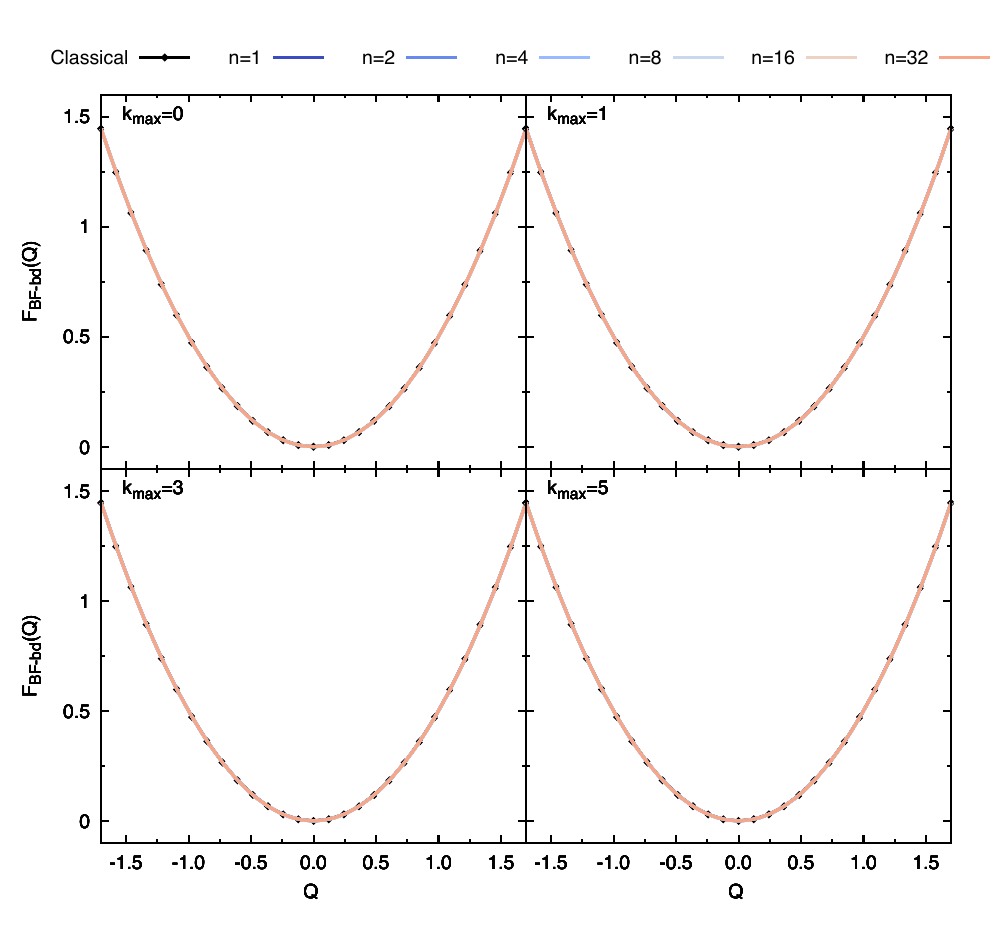}
\end{center}
\caption{Bead estimated free energy for harmonic oscillator $ \beta=8 $.}
\label{fig:harm-pot-beta8-bead}
\end{figure}

\newpage

\newpage
\begin{figure}[h!]
\begin{center}
  \includegraphics[scale=1]{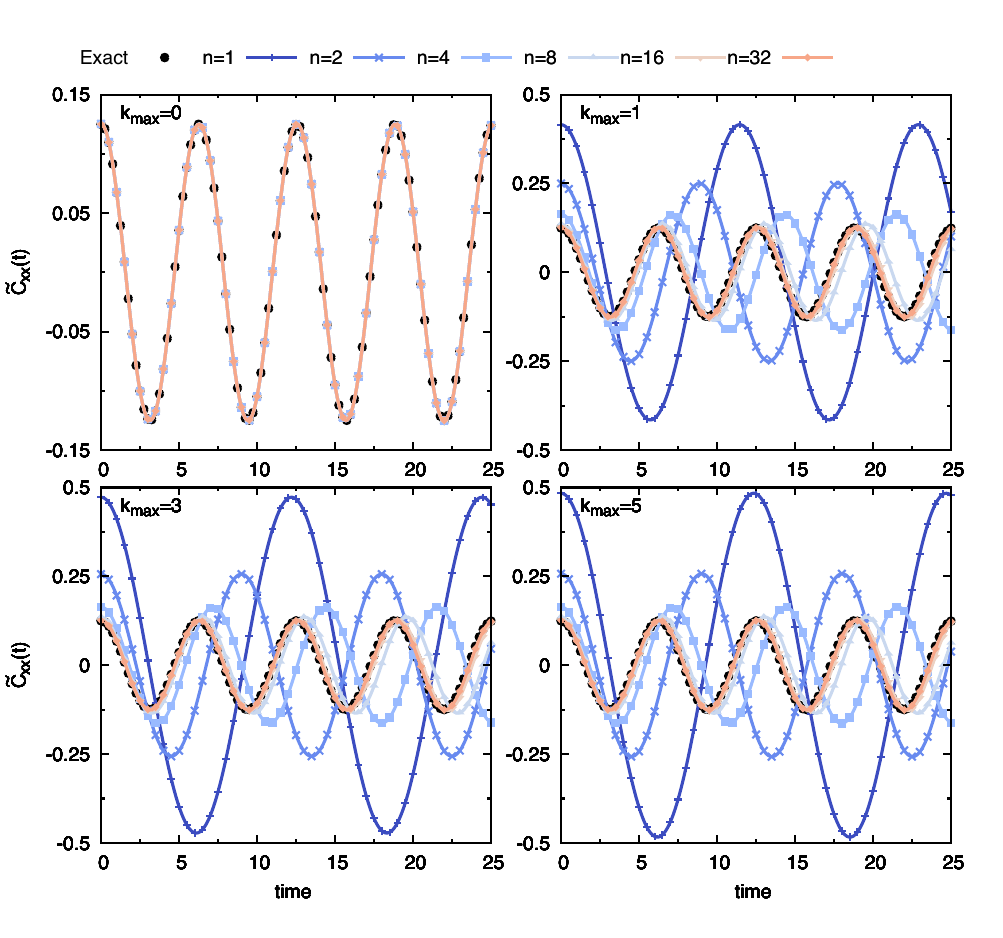}
\end{center}
\caption{Position autocorrelation function from continuous estimated free energy for harmonic oscillator $ \beta=8 $.}
\label{fig:harm-corr-beta8-cont}
\end{figure}

\newpage
\begin{figure}[h!]
\begin{center}
  \includegraphics[scale=1]{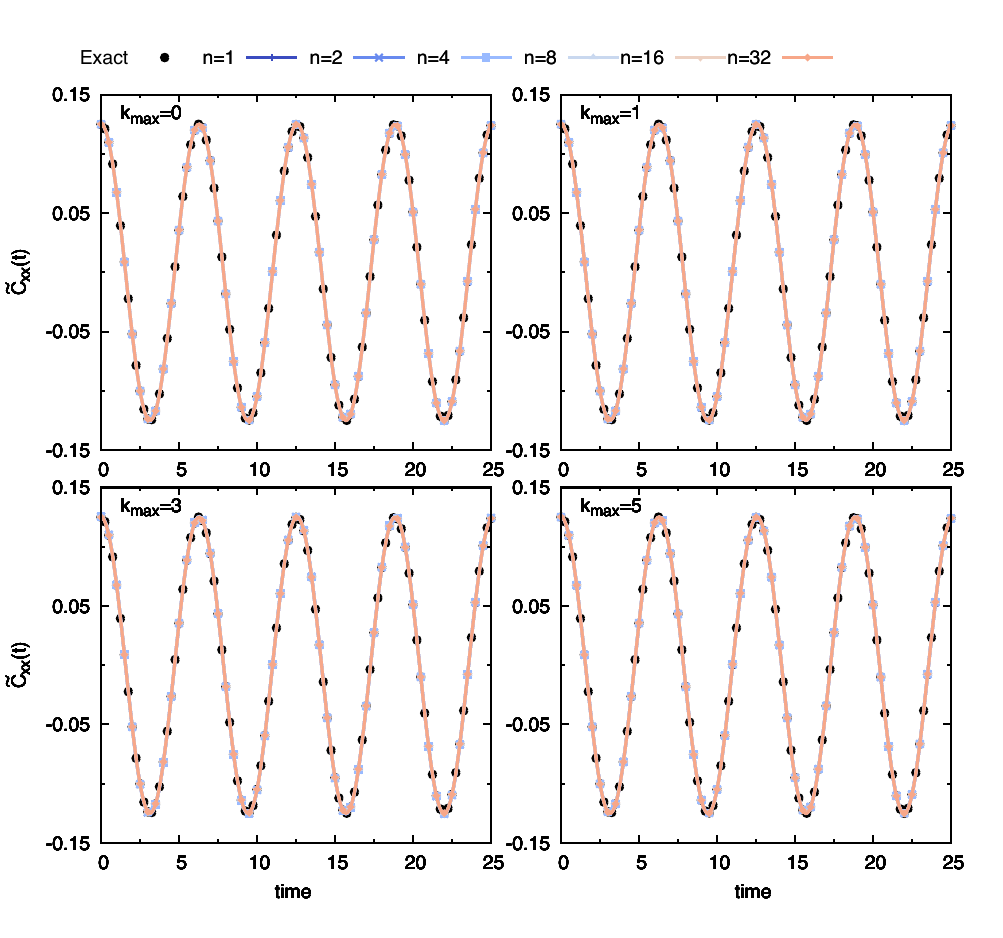}
\end{center}
\caption{Position autocorrelation function from bead estimated free energy for harmonic oscillator $ \beta=8 $.}
\label{fig:harm-corr-beta8-bead}
\end{figure}

\clearpage
\section{Section S3. Convergence data for the mildly anharmonic oscillator}
\begin{figure}[h!]
\begin{center}
  \includegraphics[scale=1]{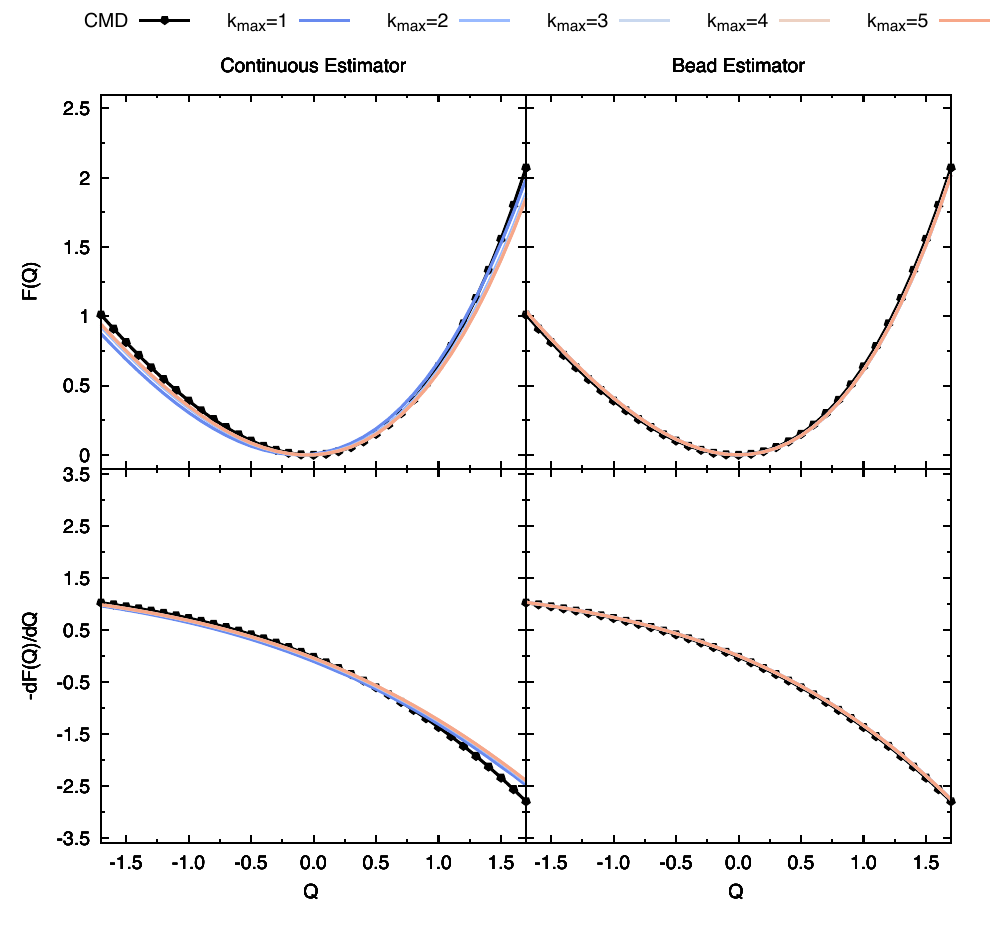}
\end{center}
\caption{Mildly anharmonic oscillator free energy (top) and force (bottom) for 1 bead at $\beta=1$.}
\label{fig:mild-pot-beta1-1bd}
\end{figure}

\newpage
\begin{figure}[h!]
\begin{center}
  \includegraphics[scale=1]{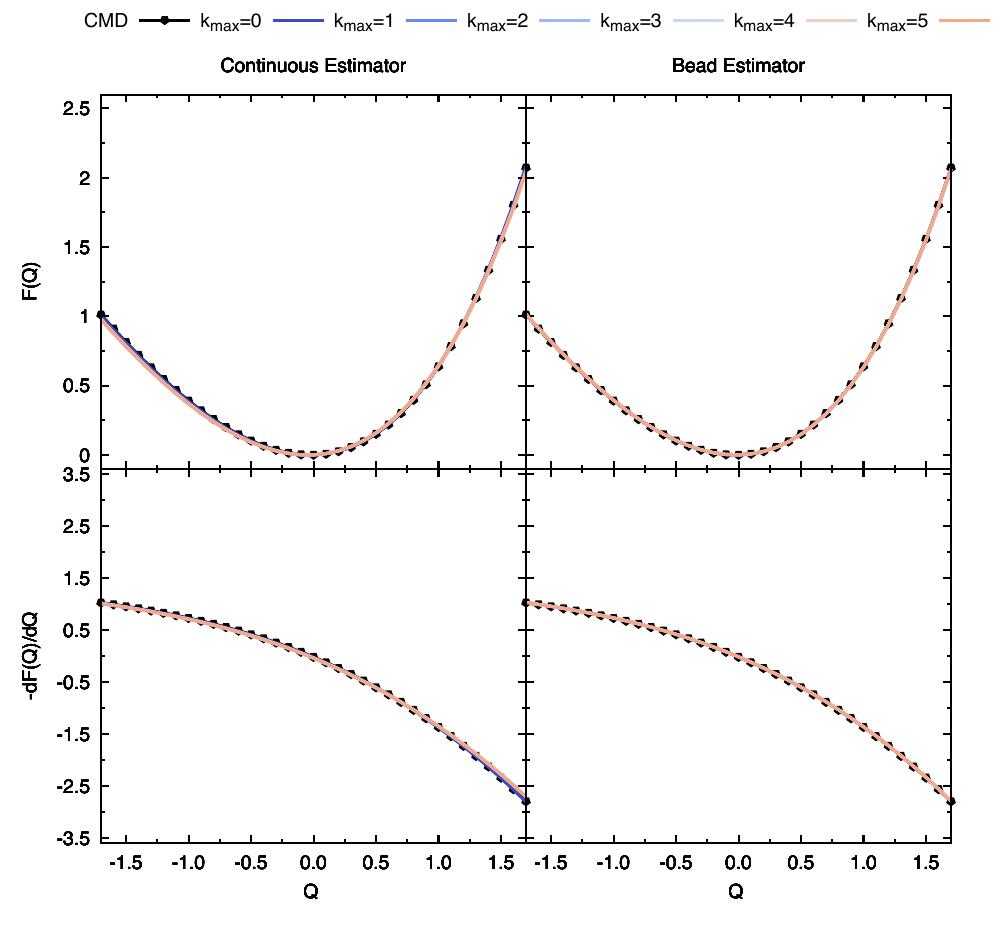}
\end{center}
\caption{Mildly anharmonic oscillator free energy (top) and force (bottom) for 2 beads at $\beta=1$.}
\label{fig:mild-pot-beta1-2bd}
\end{figure}

\newpage
\begin{figure}[h!]
\begin{center}
  \includegraphics[scale=1]{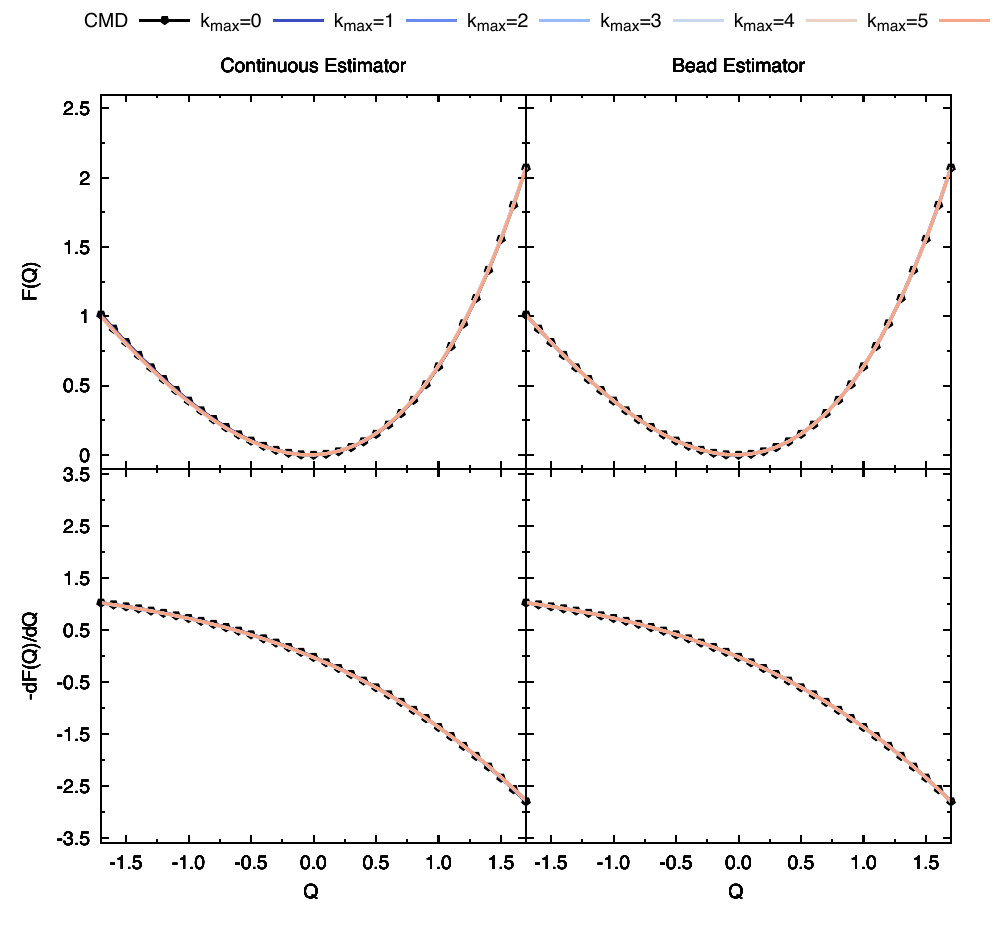}
\end{center}
\caption{Mildly anharmonic oscillator free energy (top) and force (bottom) for 4 beads at $\beta=1$.}
\label{fig:mild-pot-beta1-4bd}
\end{figure}

\newpage
\begin{figure}[h!]
\begin{center}
  \includegraphics[scale=1]{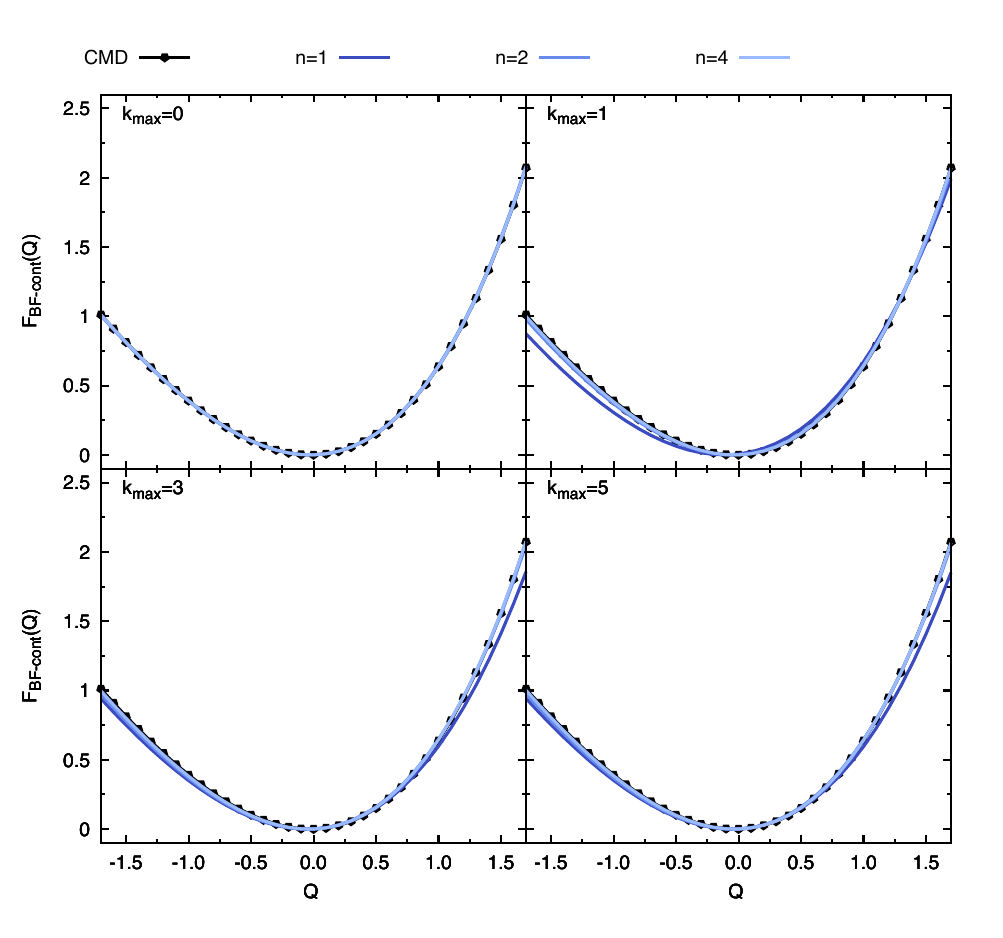}
\end{center}
\caption{Continuous estimated free energy for mildly anharmonic oscillator $ \beta=1 $.}
\label{fig:mild-pot-beta1-cont}
\end{figure}

\newpage
\begin{figure}[h!]
\begin{center}
  \includegraphics[scale=1]{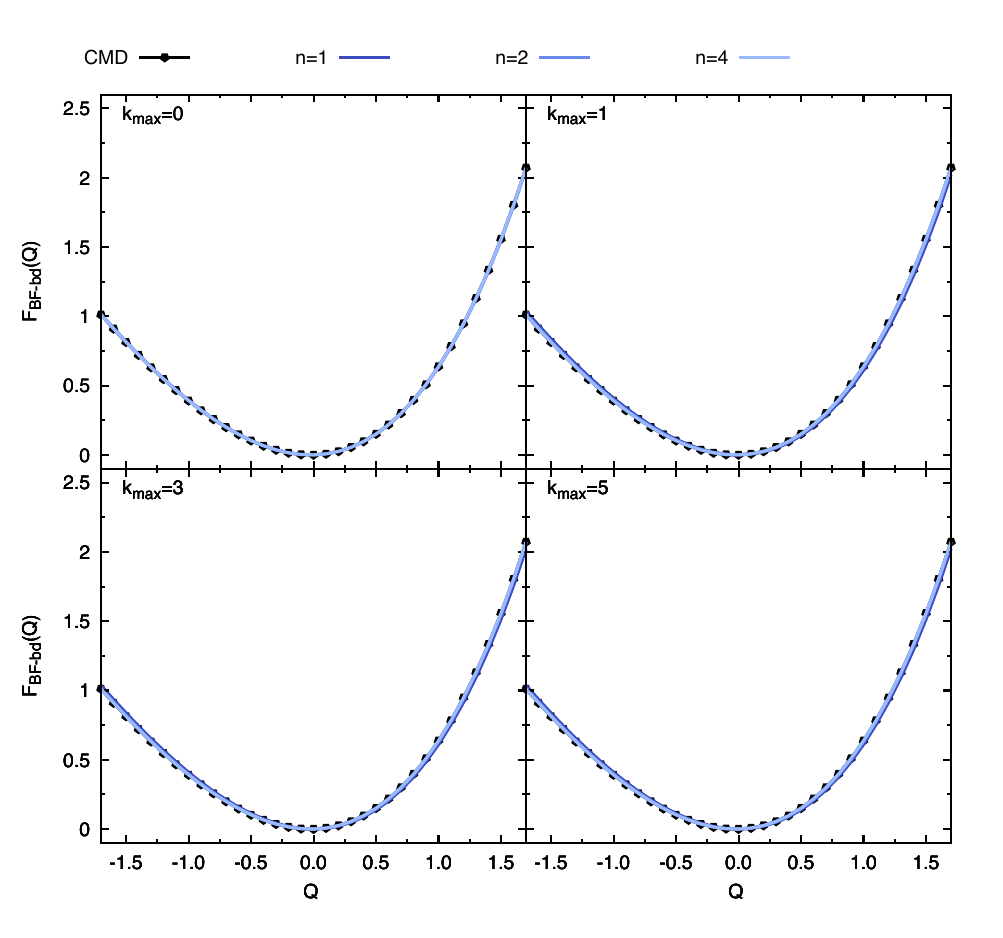}
\end{center}
\caption{Bead estimated free energy for mildly anharmonic oscillator $ \beta=1 $.}
\label{fig:mild-pot-beta1-bead}
\end{figure}


\newpage

\newpage
\begin{figure}[h!]
\begin{center}
  \includegraphics[scale=1]{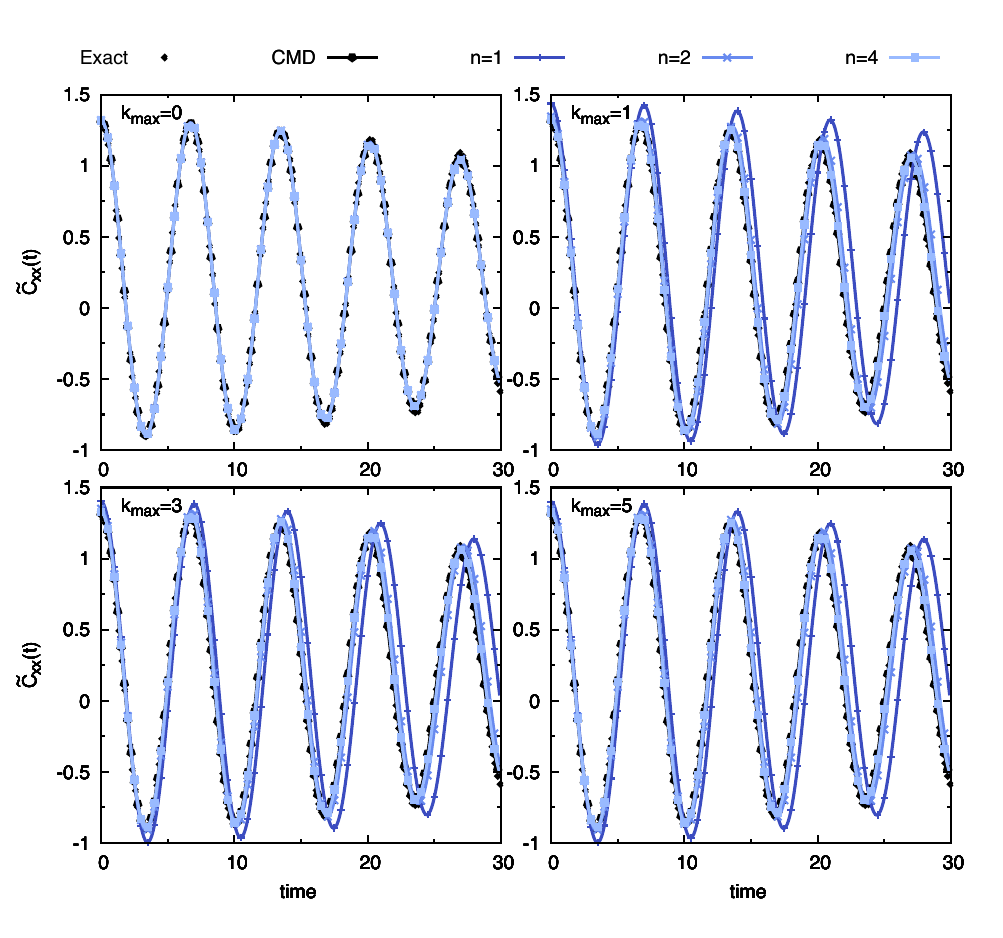}
\end{center}
\caption{Position autocorrelation function from continuous estimated free energy for mildly anharmonic oscillator $ \beta=1 $.}
\label{fig:mild-corr-beta1-cont}
\end{figure}

\newpage
\begin{figure}[h!]
\begin{center}
  \includegraphics[scale=1]{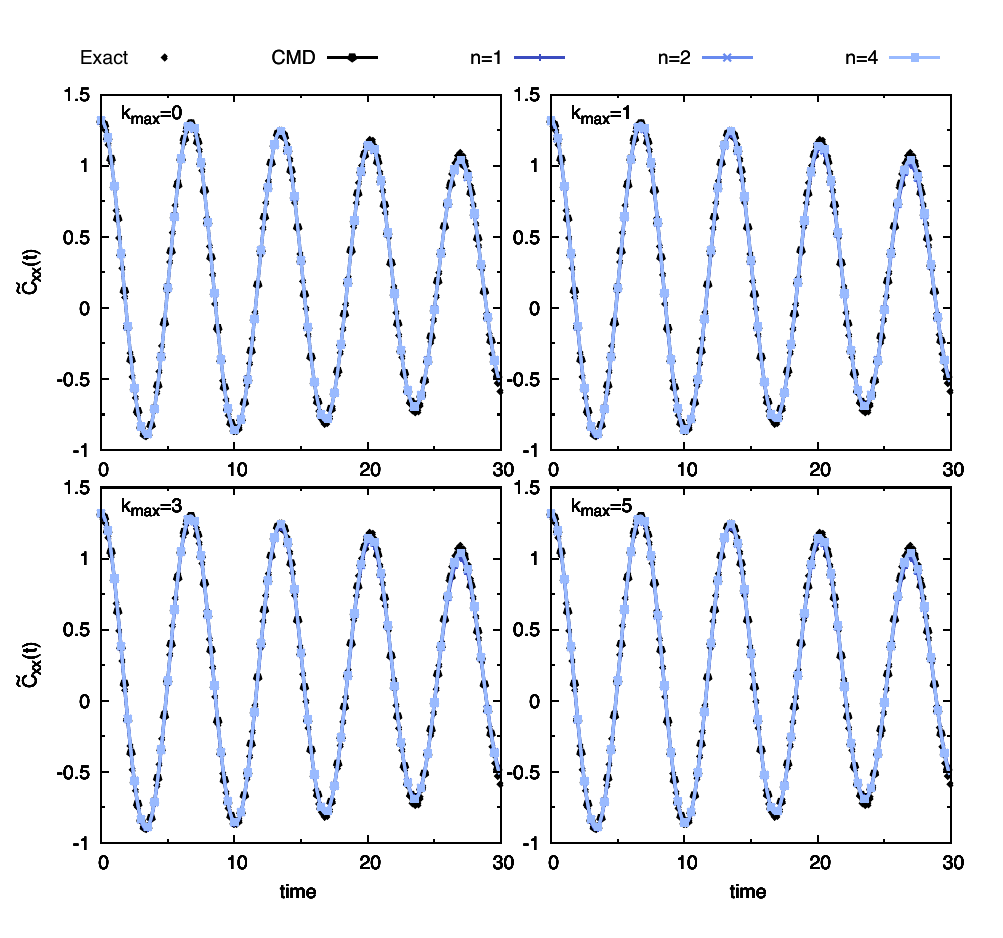}
\end{center}
\caption{Position autocorrelation function from bead estimated free energy for mildly anharmonic oscillator $ \beta=1 $.}
\label{fig:mild-corr-beta1-bead}
\end{figure}

\newpage
\begin{figure}[h!]
\begin{center}
  \includegraphics[scale=1]{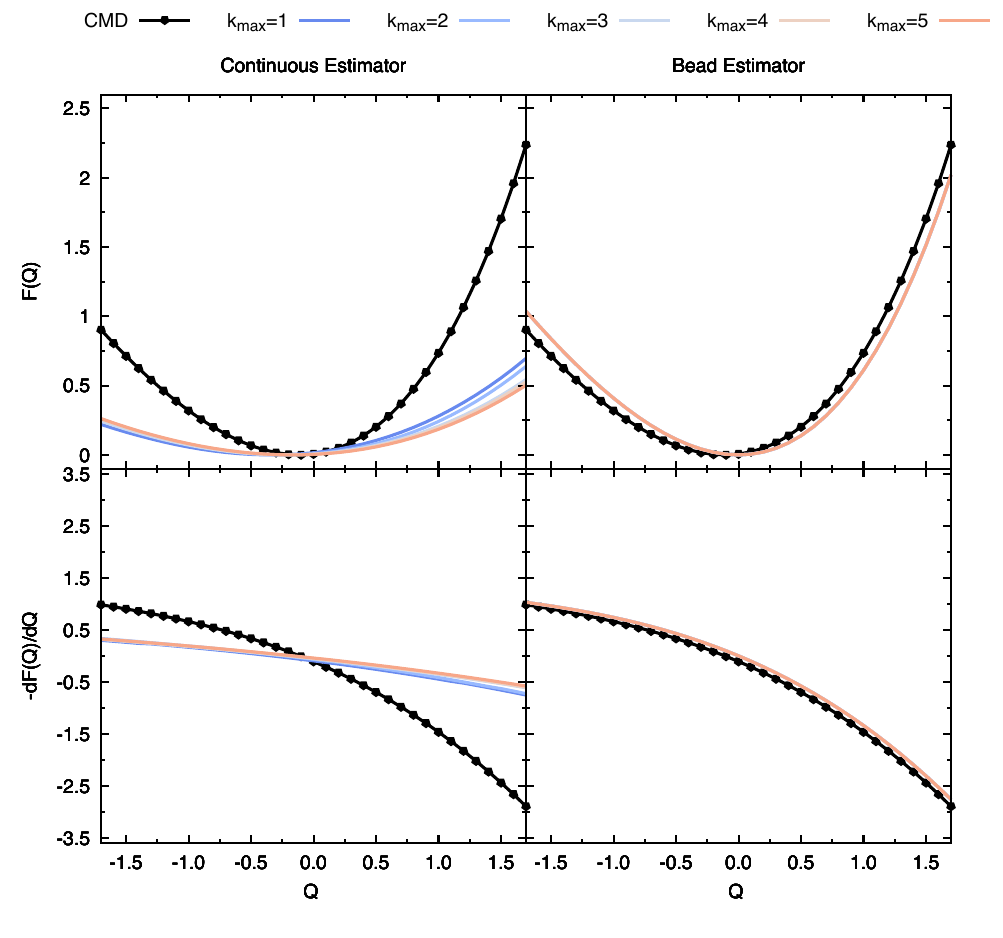}
\end{center}
\caption{Mildly anharmonic oscillator free energy (top) and force (bottom) for 1 bead at $\beta=8$.}
\label{fig:mild-pot-beta8-1bd}
\end{figure}

\newpage
\begin{figure}[h!]
\begin{center}
  \includegraphics[scale=1]{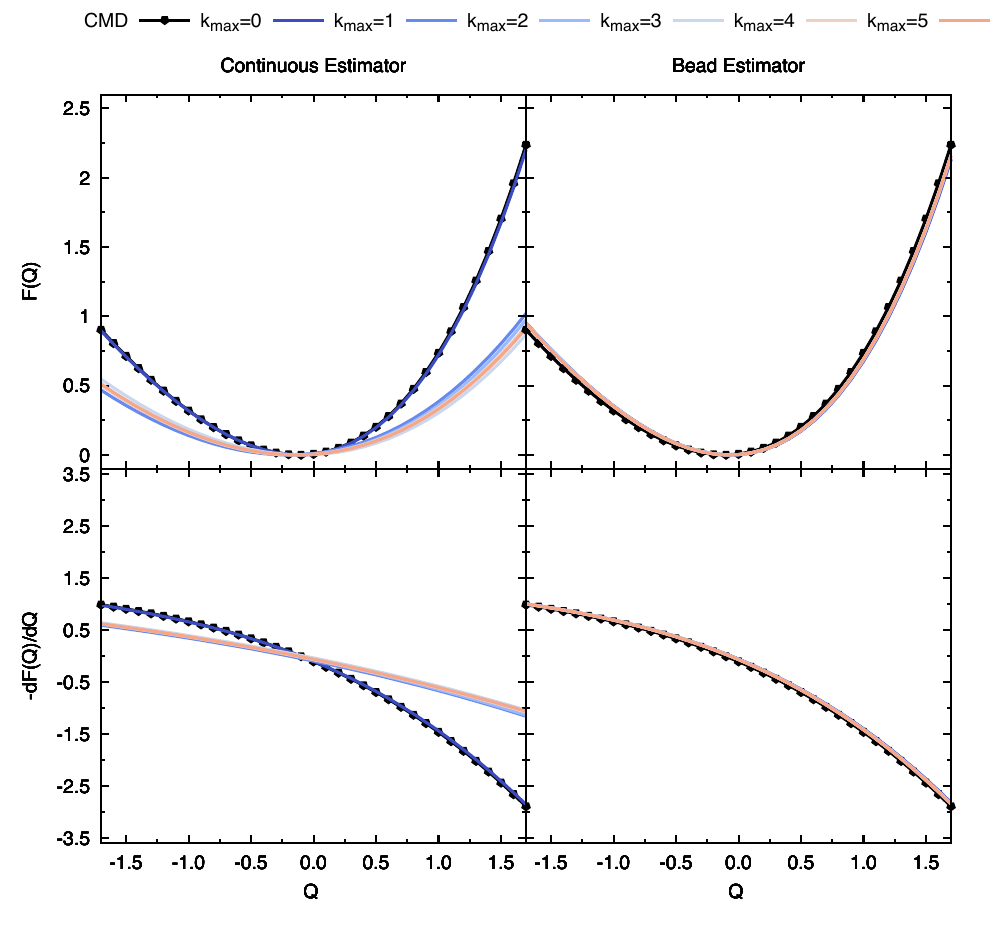}
\end{center}
\caption{Mildly anharmonic oscillator free energy (top) and force (bottom) for 2 beads at $\beta=8$.}
\label{fig:mild-pot-beta8-2bd}
\end{figure}

\newpage
\begin{figure}[h!]
\begin{center}
  \includegraphics[scale=1]{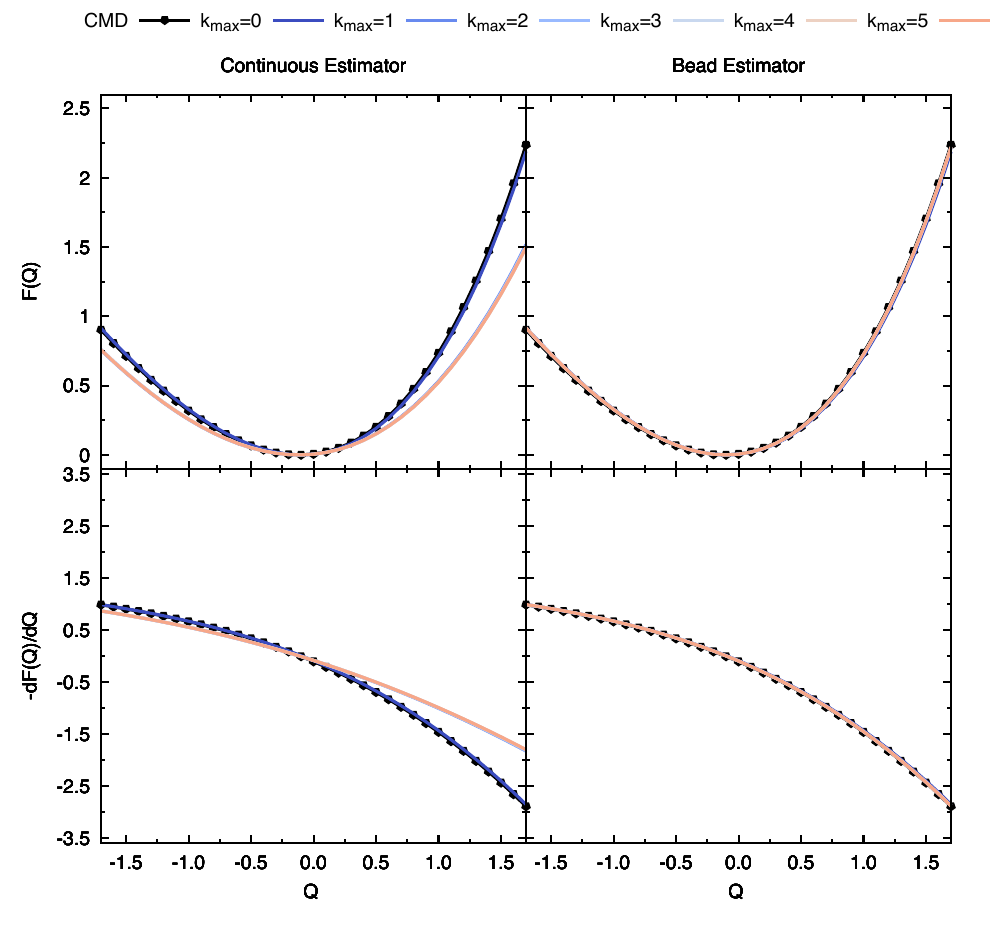}
\end{center}
\caption{Mildly anharmonic oscillator free energy (top) and force (bottom) for 4 beads at $\beta=8$.}
\label{fig:mild-pot-beta8-4bd}
\end{figure}

\newpage
\begin{figure}[h!]
\begin{center}
  \includegraphics[scale=1]{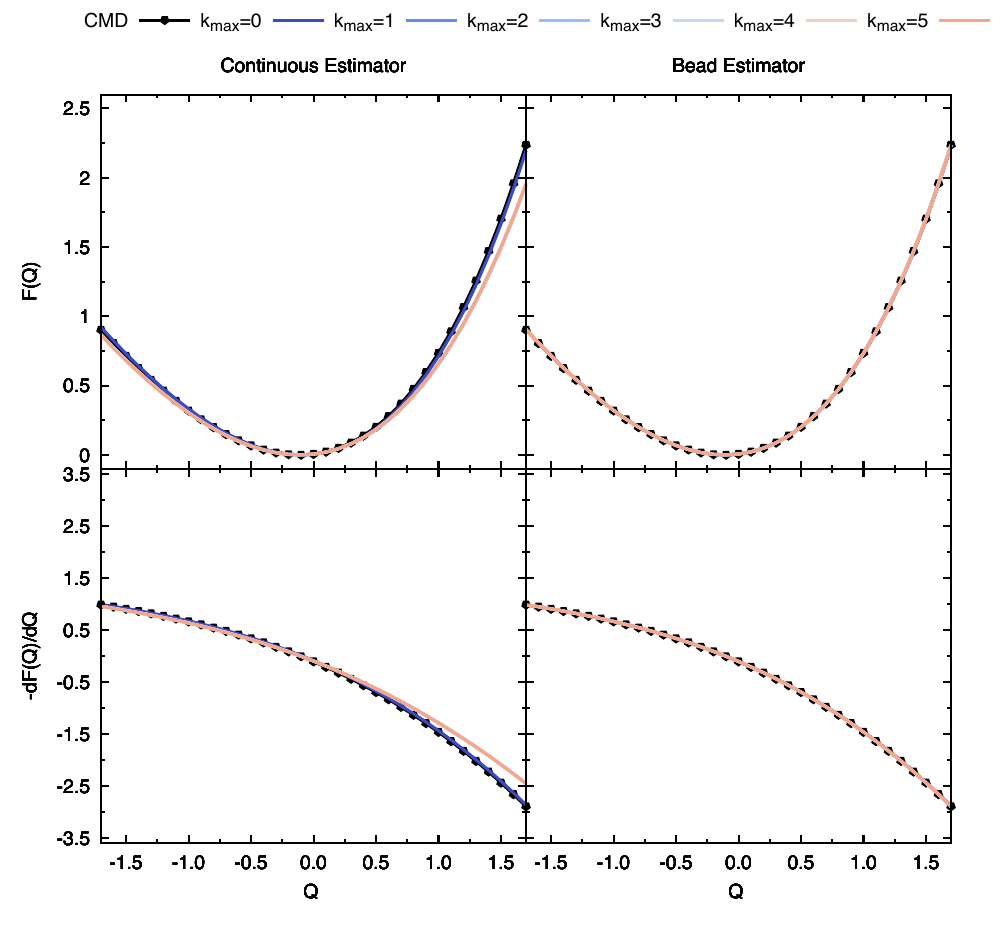}
\end{center}
\caption{Mildly anharmonic oscillator free energy (top) and force (bottom) for 8 beads at $\beta=8$.}
\label{fig:mild-pot-beta8-8bd}
\end{figure}

\newpage
\begin{figure}[h!]
\begin{center}
  \includegraphics[scale=1]{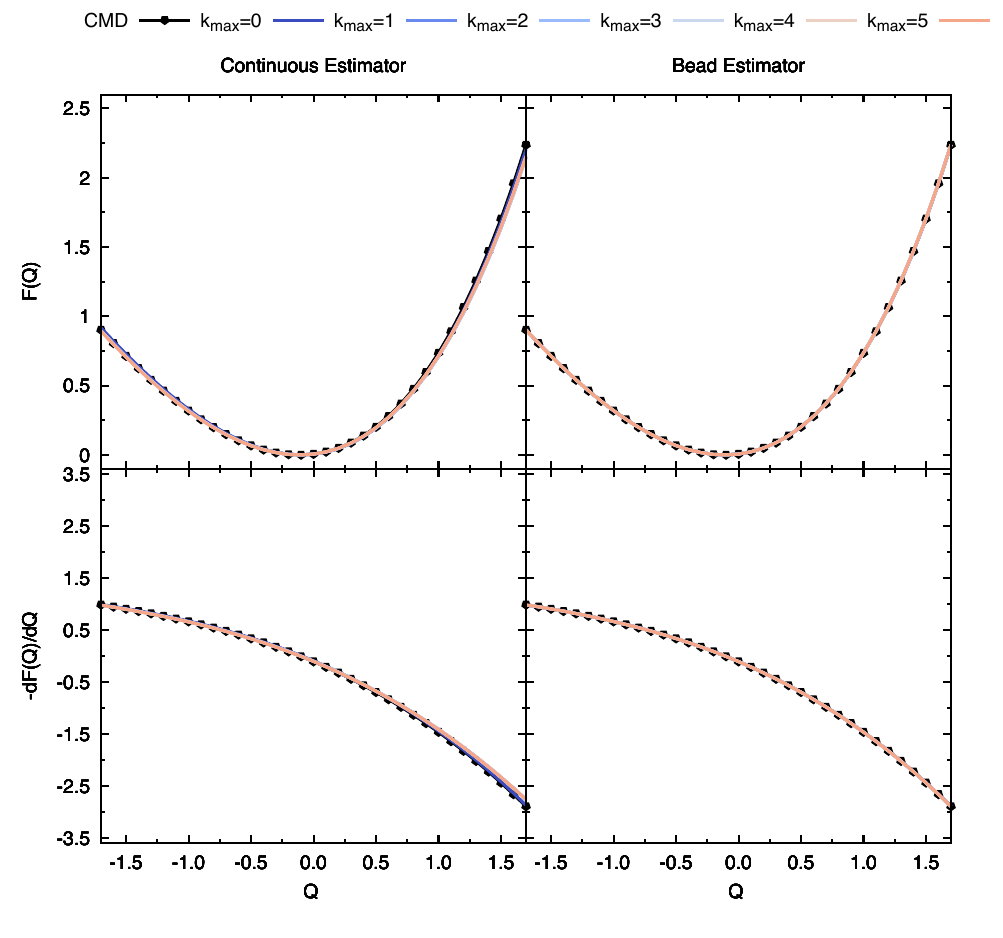}
\end{center}
\caption{Mildly anharmonic oscillator free energy (top) and force (bottom) for 16 beads at $\beta=8$.}
\label{fig:mild-pot-beta8-16bd}
\end{figure}

\newpage
\begin{figure}[h!]
\begin{center}
  \includegraphics[scale=1]{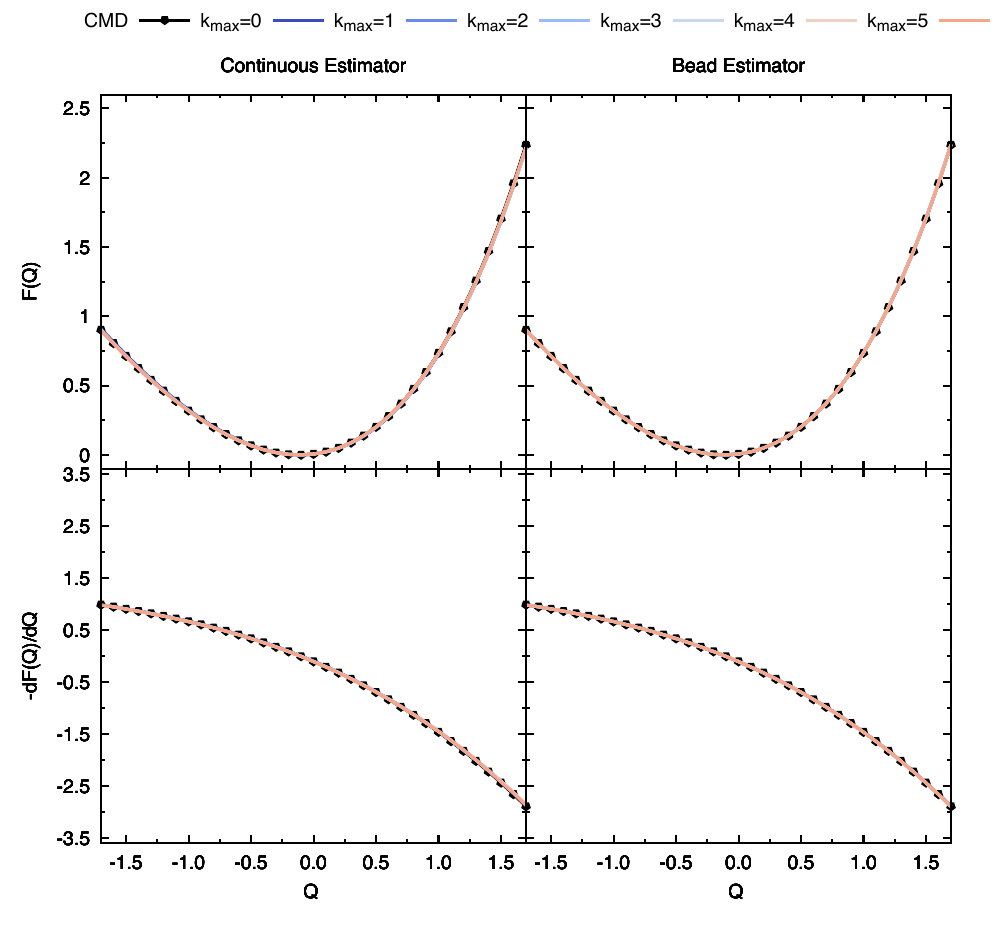}
\end{center}
\caption{Mildly anharmonic oscillator free energy (top) and force (bottom) for 32 beads at $\beta=8$.}
\label{fig:mild-pot-beta8-32bd}
\end{figure}

\begin{figure}[h!]
\begin{center}
  \includegraphics[scale=1]{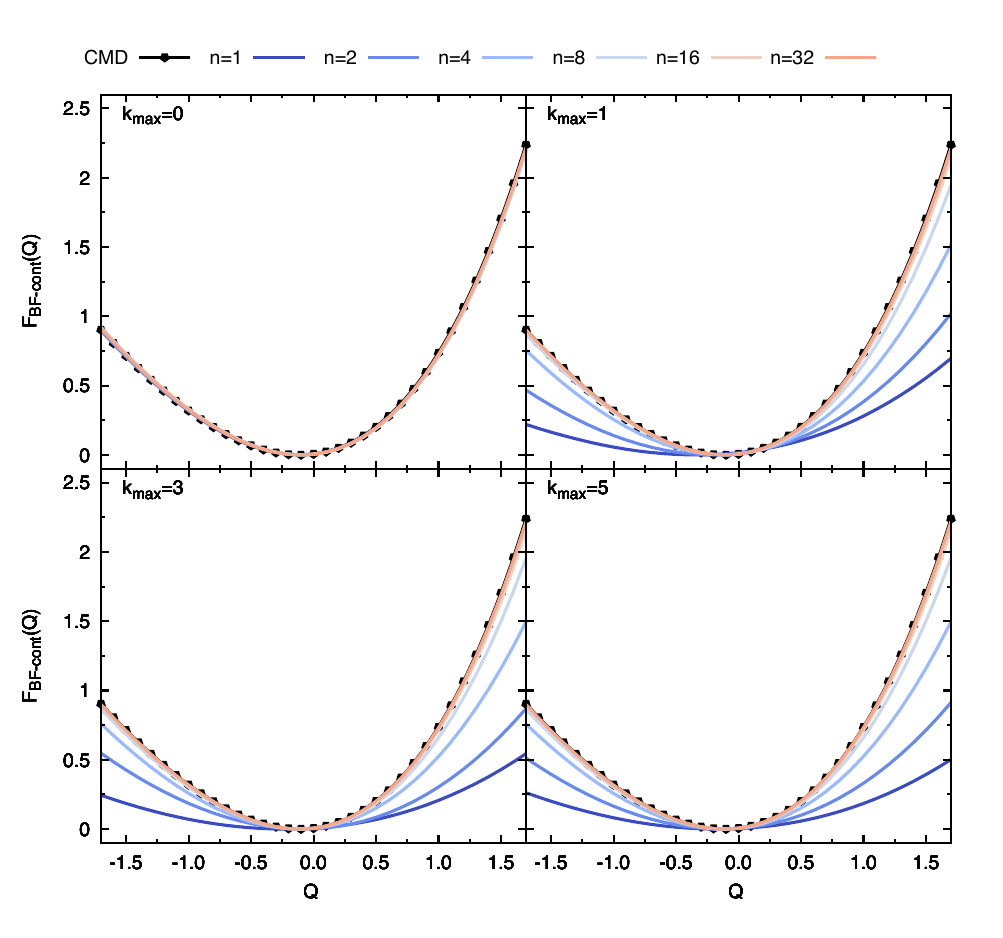}
\end{center}
\caption{Continuous estimated free energy for mildly anharmonic oscillator $ \beta=8 $.}
\label{fig:mild-pot-beta8-cont}
\end{figure}

\newpage
\begin{figure}[h!]
\begin{center}
  \includegraphics[scale=1]{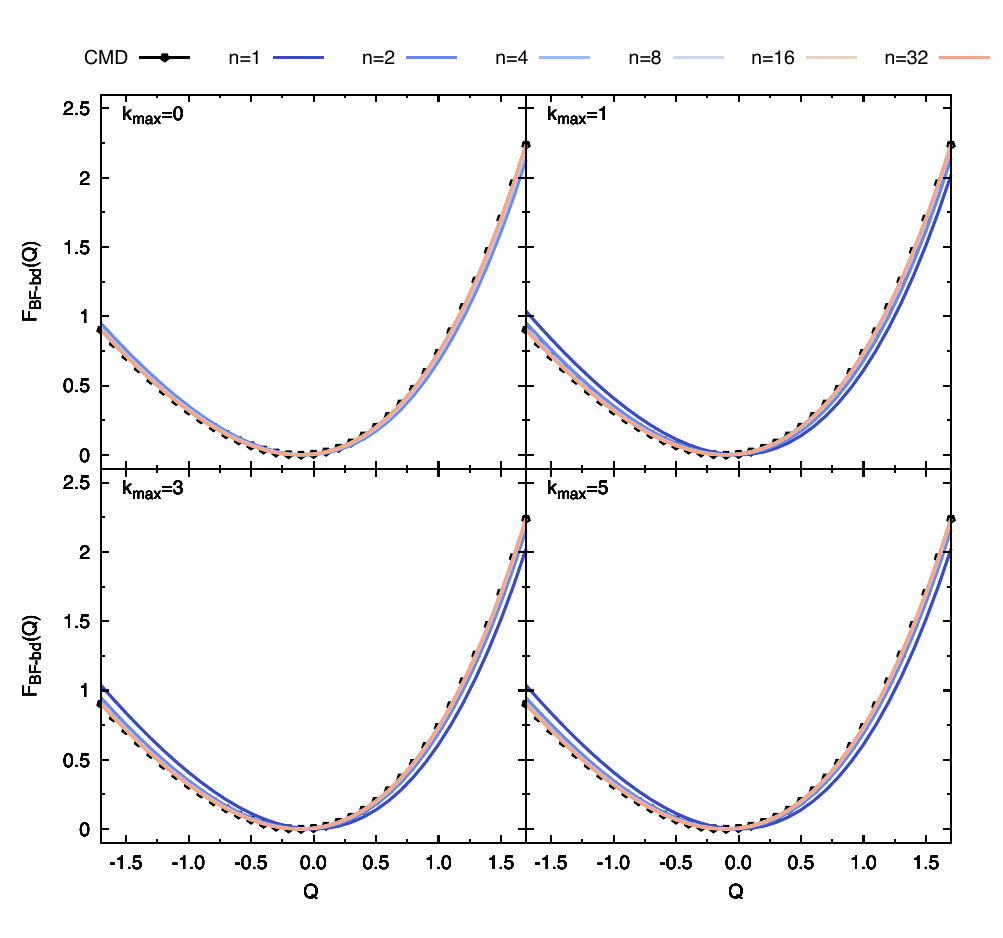}
\end{center}
\caption{Bead estimated free energy for mildly anharmonic oscillator $ \beta=8 $.}
\label{fig:mild-pot-beta8-bead}
\end{figure}

\newpage

\newpage
\begin{figure}[h!]
\begin{center}
  \includegraphics[scale=1]{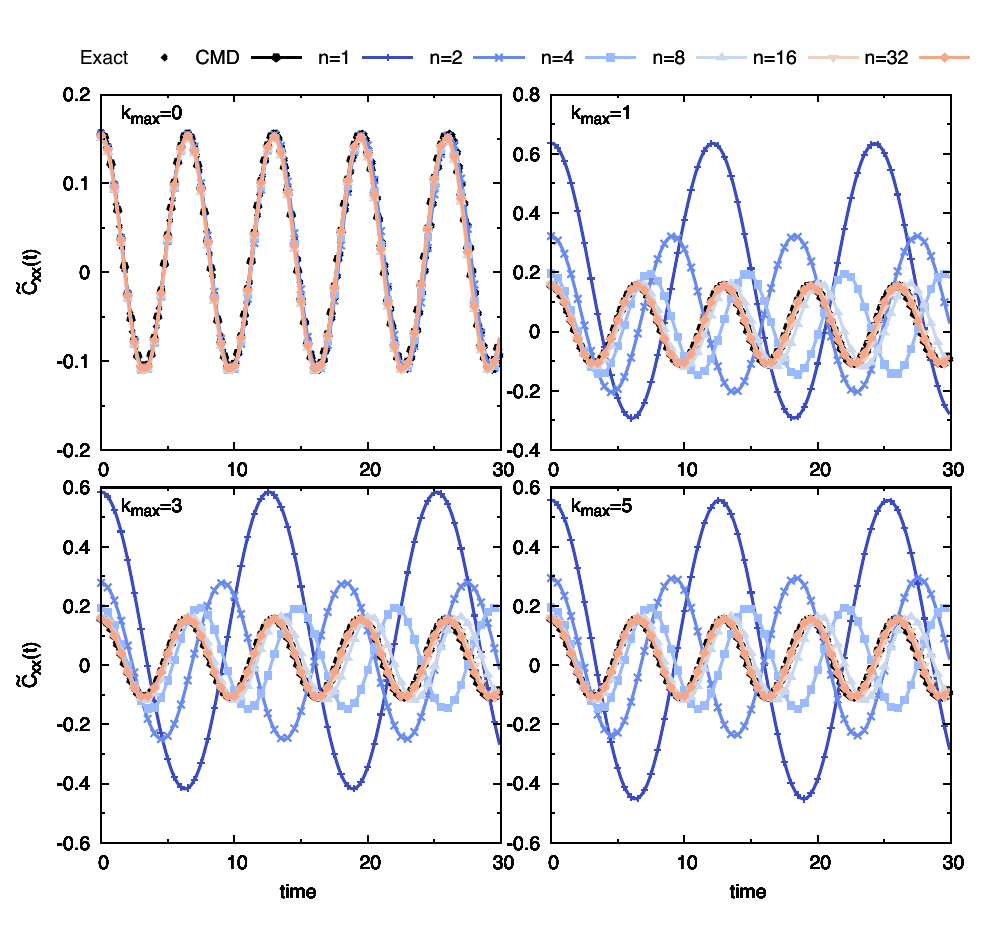}
\end{center}
\caption{Position autocorrelation function from continuous estimated free energy for mildly anharmonic oscillator $ \beta=8 $.}
\label{fig:mild-corr-beta8-cont}
\end{figure}

\newpage
\begin{figure}[h!]
\begin{center}
  \includegraphics[scale=1]{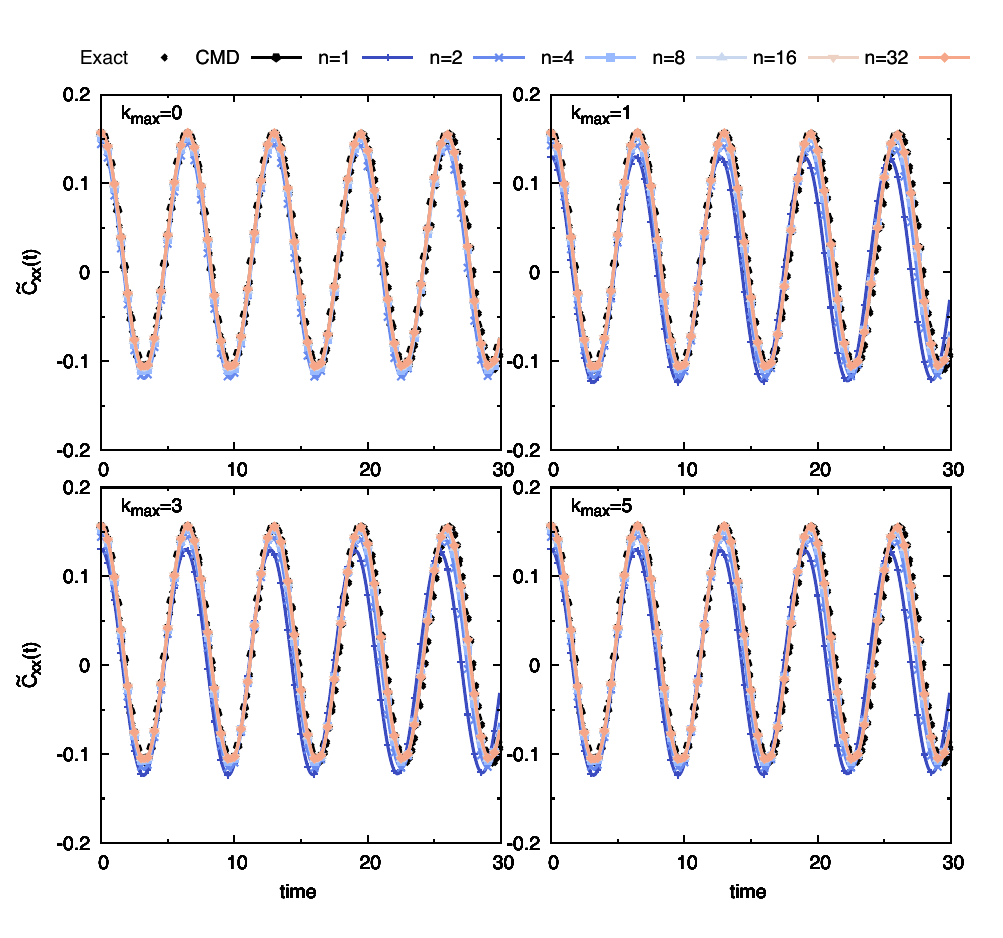}
\end{center}
\caption{Position autocorrelation function from bead estimated free energy for mildly anharmonic oscillator $ \beta=8 $.}
\label{fig:mild-corr-beta8-bead}
\end{figure}

\clearpage
\section{Section S4. Convergence data for the quartic oscillator}
\begin{figure}[h!]
\begin{center}
  \includegraphics[scale=1]{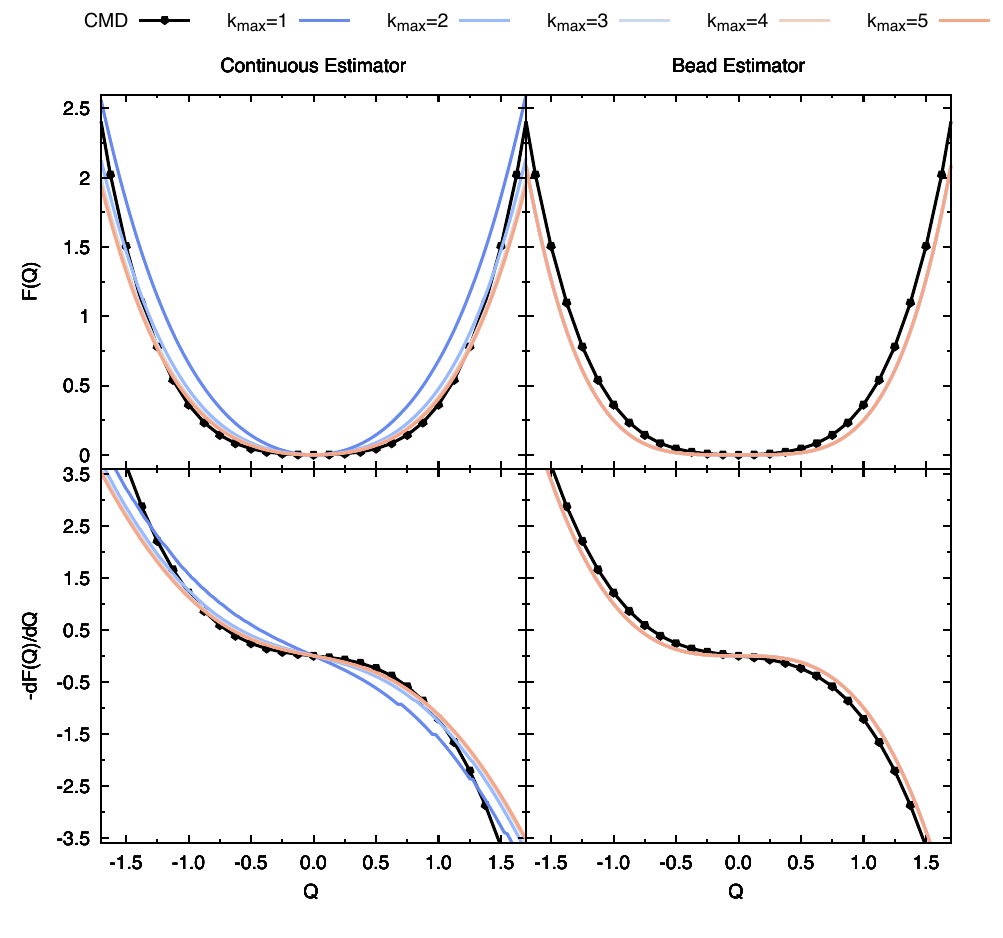}
\end{center}
\caption{Quartic oscillator free energy (top) and force (bottom) for 1 bead at $\beta=1$.}
\label{fig:quart-pot-beta1-1bd}
\end{figure}

\newpage
\begin{figure}[h!]
\begin{center}
  \includegraphics[scale=1]{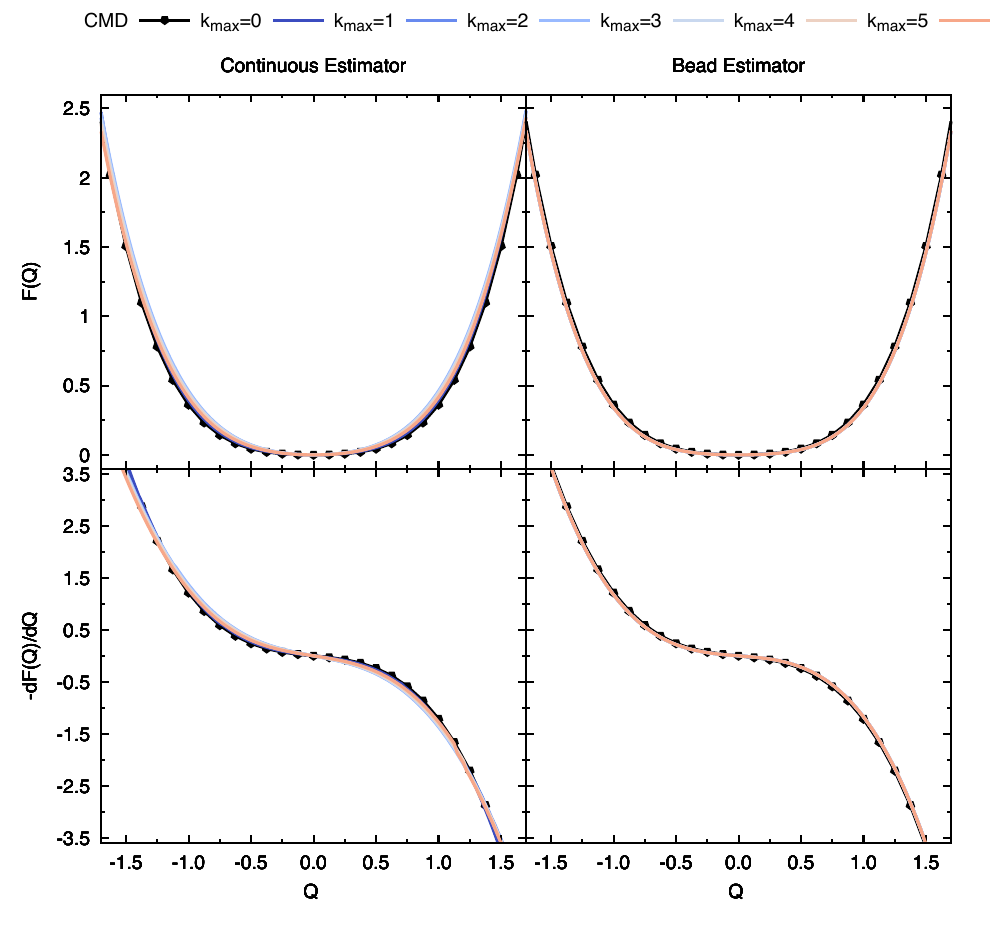}
\end{center}
\caption{Quartic oscillator free energy (top) and force (bottom) for 2 beads at $\beta=1$.}
\label{fig:quart-pot-beta1-2bd}
\end{figure}

\newpage
\begin{figure}[h!]
\begin{center}
  \includegraphics[scale=1]{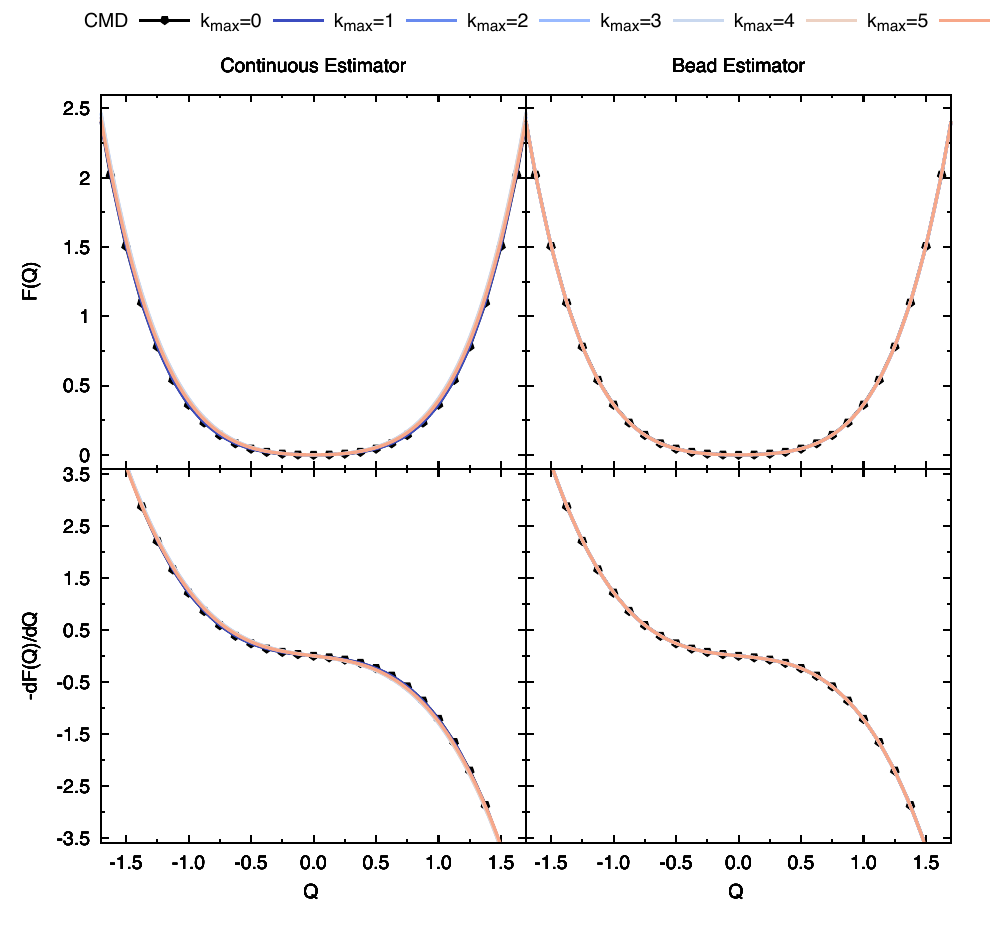}
\end{center}
\caption{Quartic oscillator free energy (top) and force (bottom) for 4 beads at $\beta=1$.}
\label{fig:quart-pot-beta1-4bd}
\end{figure}

\begin{figure}[h!]
\begin{center}
  \includegraphics[scale=1]{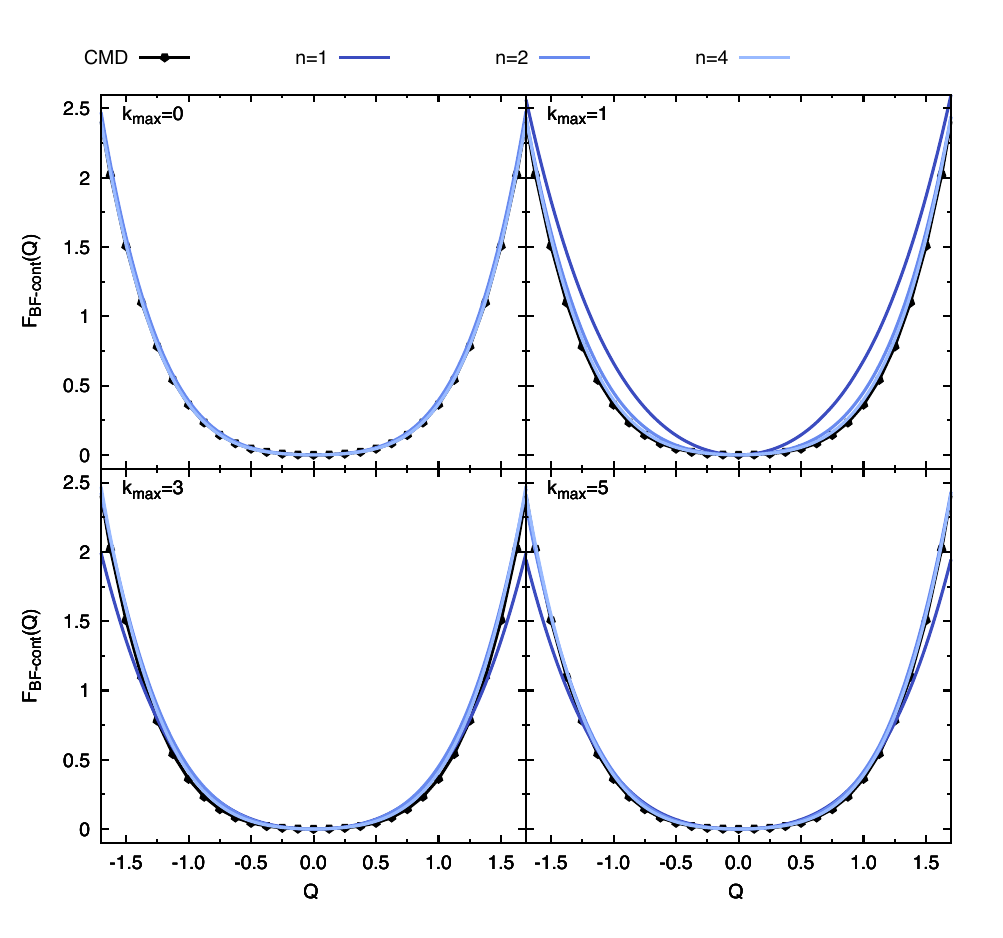}
\end{center}
\caption{Continuous estimated free energy for quartic oscillator $ \beta=1 $.}
\label{fig:quart-pot-beta1-cont}
\end{figure}

\newpage
\begin{figure}[h!]
\begin{center}
  \includegraphics[scale=1]{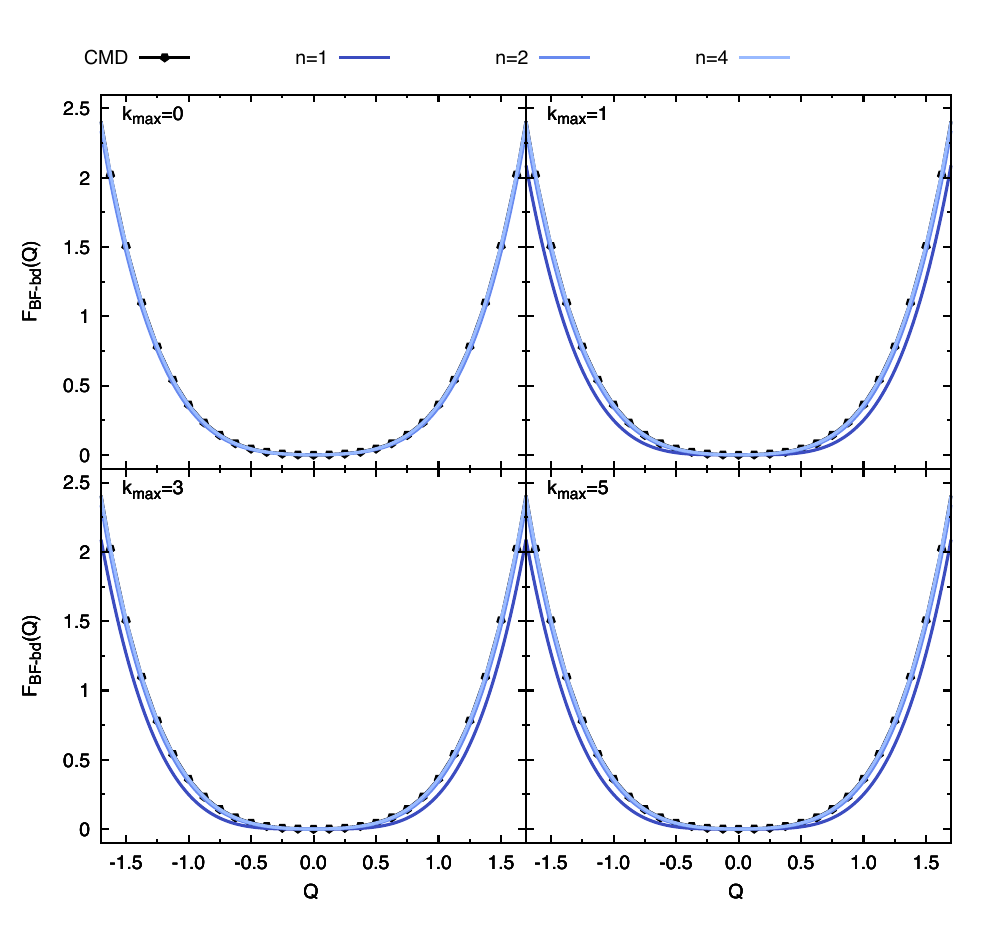}
\end{center}
\caption{Bead estimated free energy for quartic oscillator $ \beta=1 $.}
\label{fig:quart-pot-beta1-bead}
\end{figure}

\newpage

\newpage
\begin{figure}[h!]
\begin{center}
  \includegraphics[scale=1]{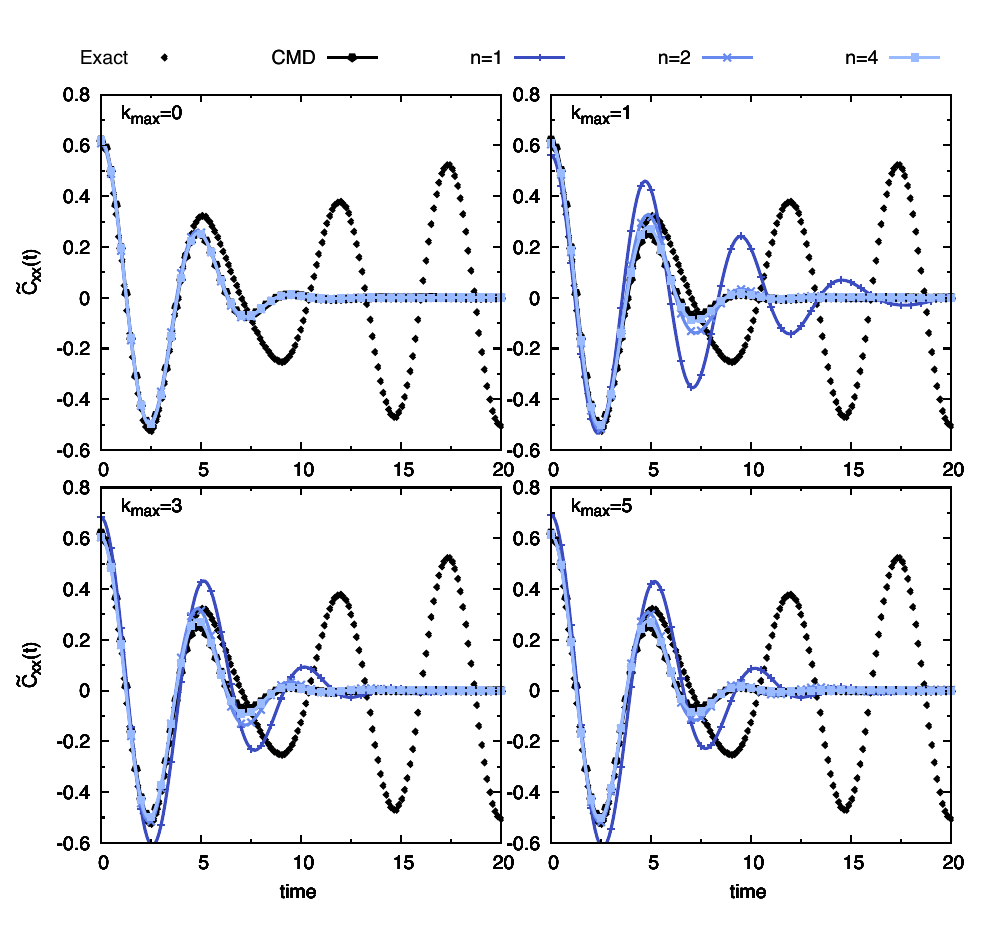}
\end{center}
\caption{Position autocorrelation function from continuous estimated free energy for quartic oscillator $ \beta=1 $.}
\label{fig:quart-corr-beta1-cont}
\end{figure}

\newpage
\begin{figure}[h!]
\begin{center}
  \includegraphics[scale=1]{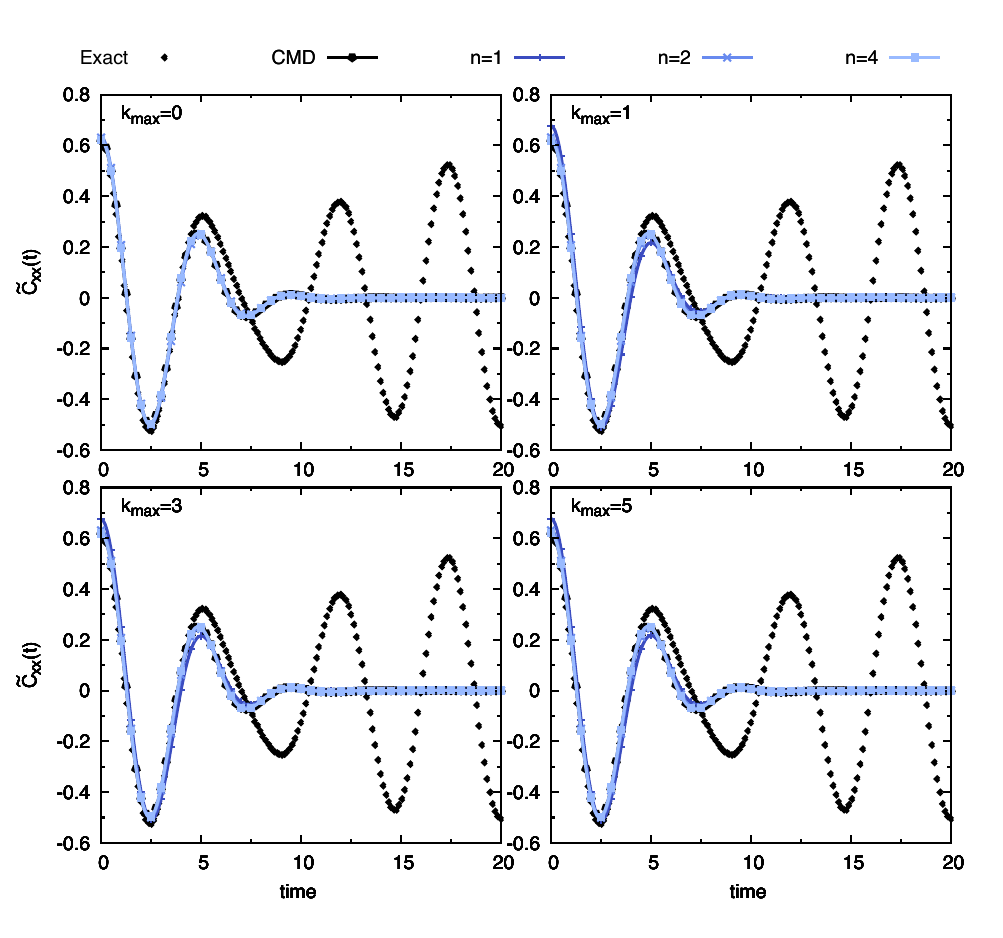}
\end{center}
\caption{Position autocorrelation function from bead estimated free energy for quartic oscillator $ \beta=1 $.}
\label{fig:quart-corr-beta1-bead}
\end{figure}

\newpage
\begin{figure}[h!]
\begin{center}
  \includegraphics[scale=1]{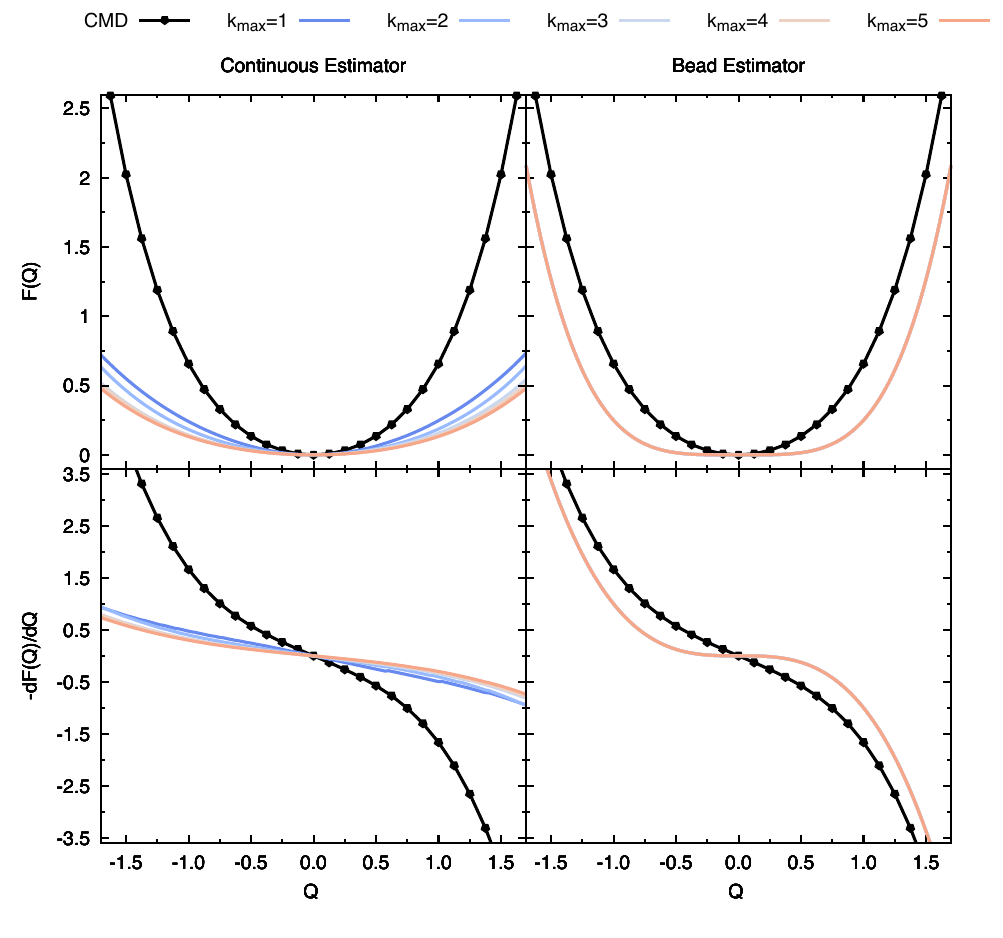}
\end{center}
\caption{Quartic oscillator free energy (top) and force (bottom) for 1 bead at $\beta=8$.}
\label{fig:quart-pot-beta8-1bd}
\end{figure}

\newpage
\begin{figure}[h!]
\begin{center}
  \includegraphics[scale=1]{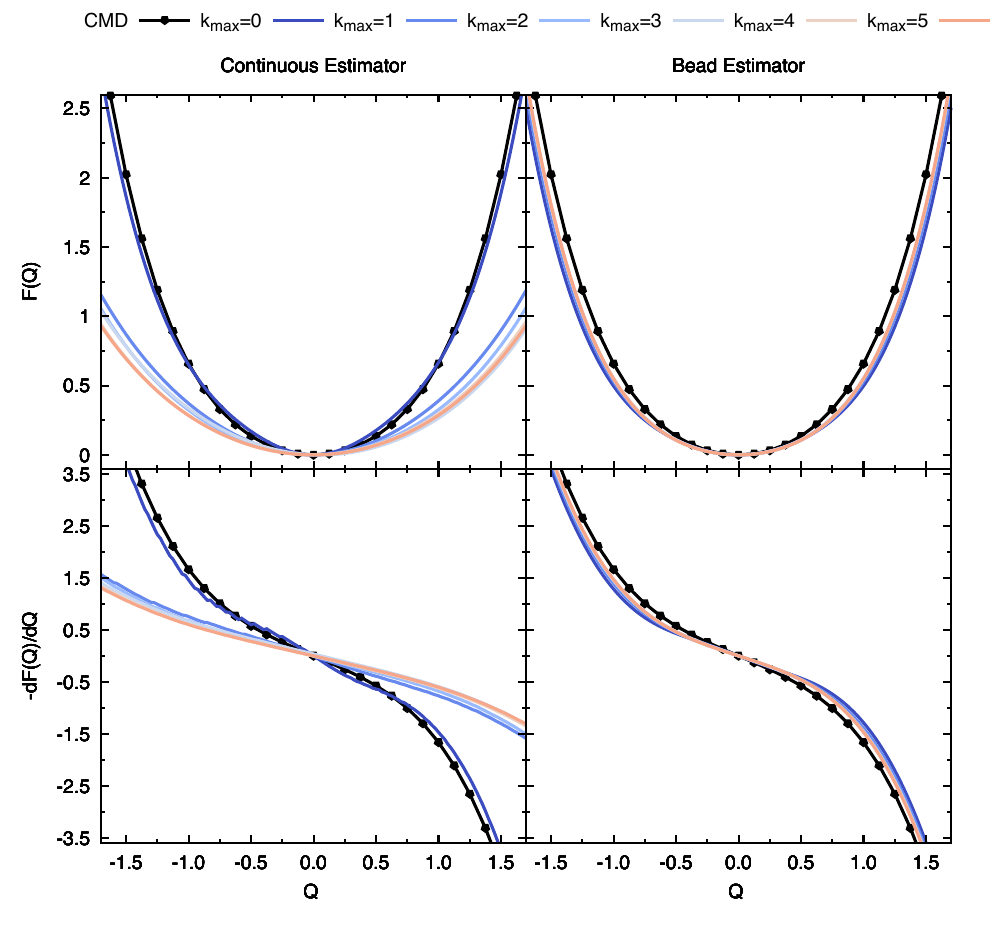}
\end{center}
\caption{Quartic oscillator free energy (top) and force (bottom) for 2 beads at $\beta=8$.}
\label{fig:quart-pot-beta8-2bd}
\end{figure}

\newpage
\begin{figure}[h!]
\begin{center}
  \includegraphics[scale=1]{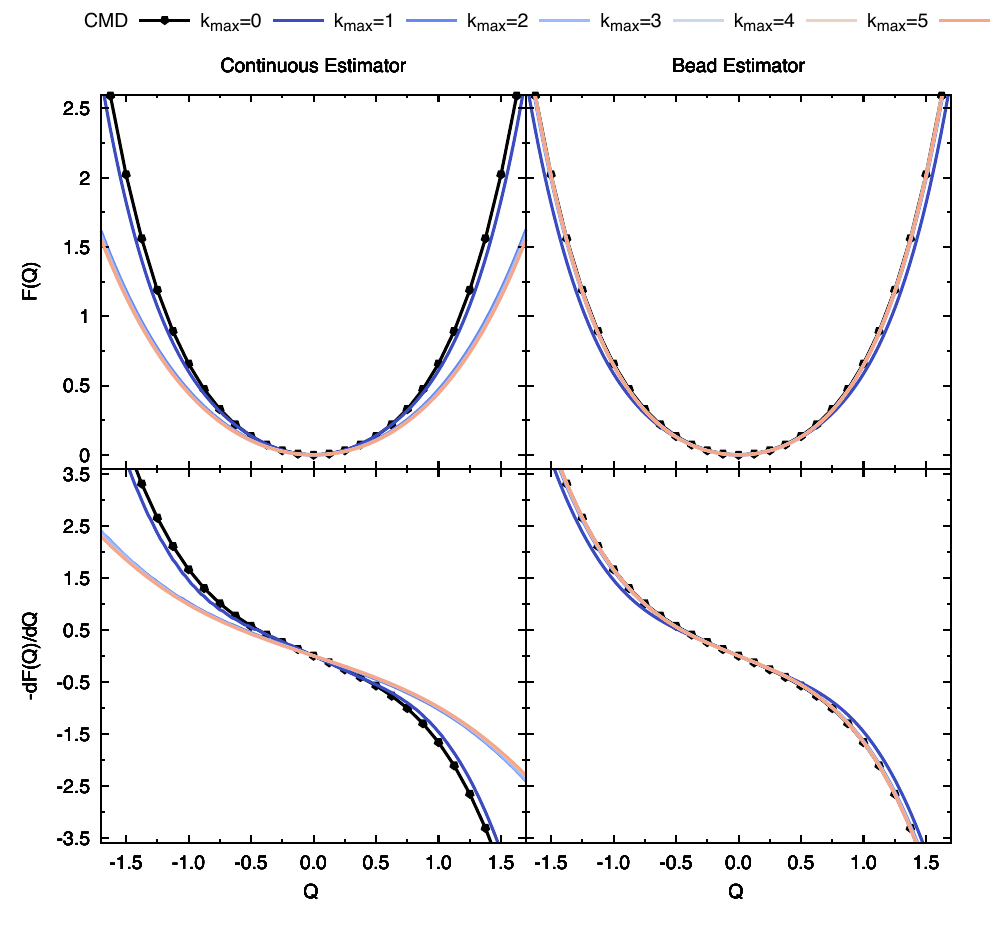}
\end{center}
\caption{Quartic oscillator free energy (top) and force (bottom) for 4 beads at $\beta=8$.}
\label{fig:quart-pot-beta8-4bd}
\end{figure}

\newpage
\begin{figure}[h!]
\begin{center}
  \includegraphics[scale=1]{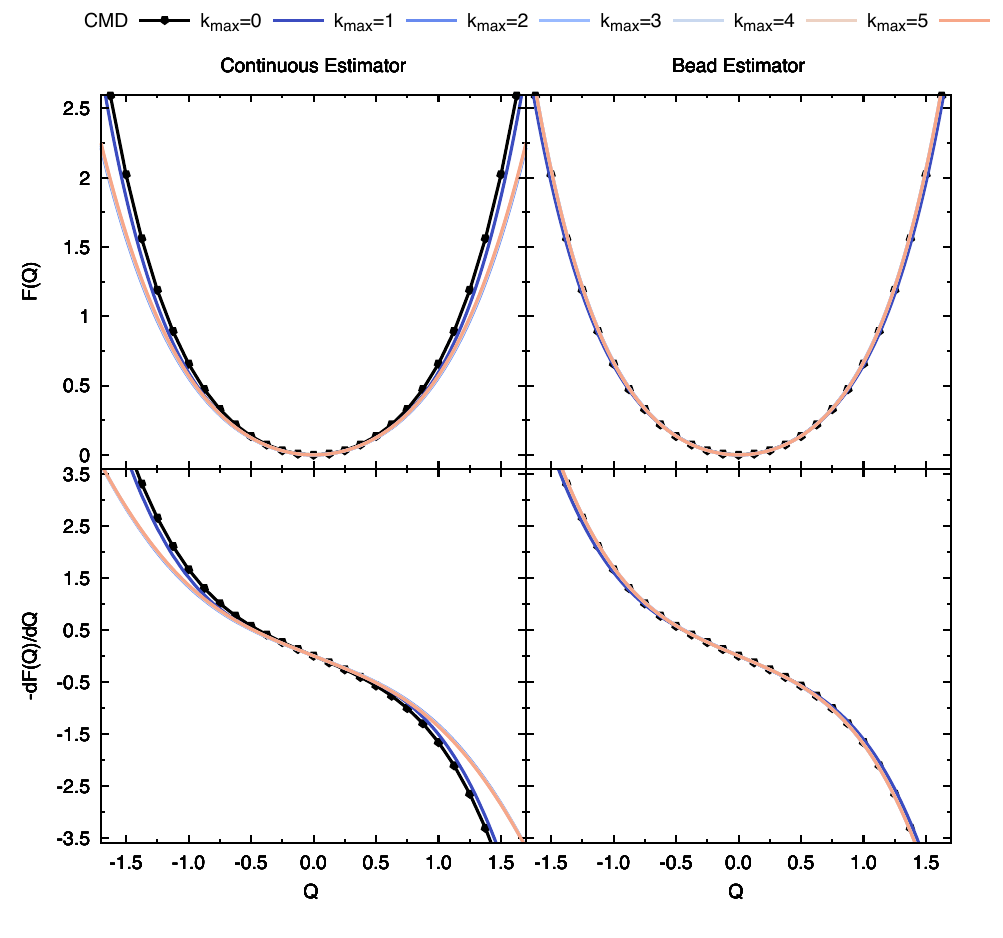}
\end{center}
\caption{Quartic oscillator free energy (top) and force (bottom) for 8 beads at $\beta=8$.}
\label{fig:quart-pot-beta8-8bd}
\end{figure}

\newpage
\begin{figure}[h!]
\begin{center}
  \includegraphics[scale=1]{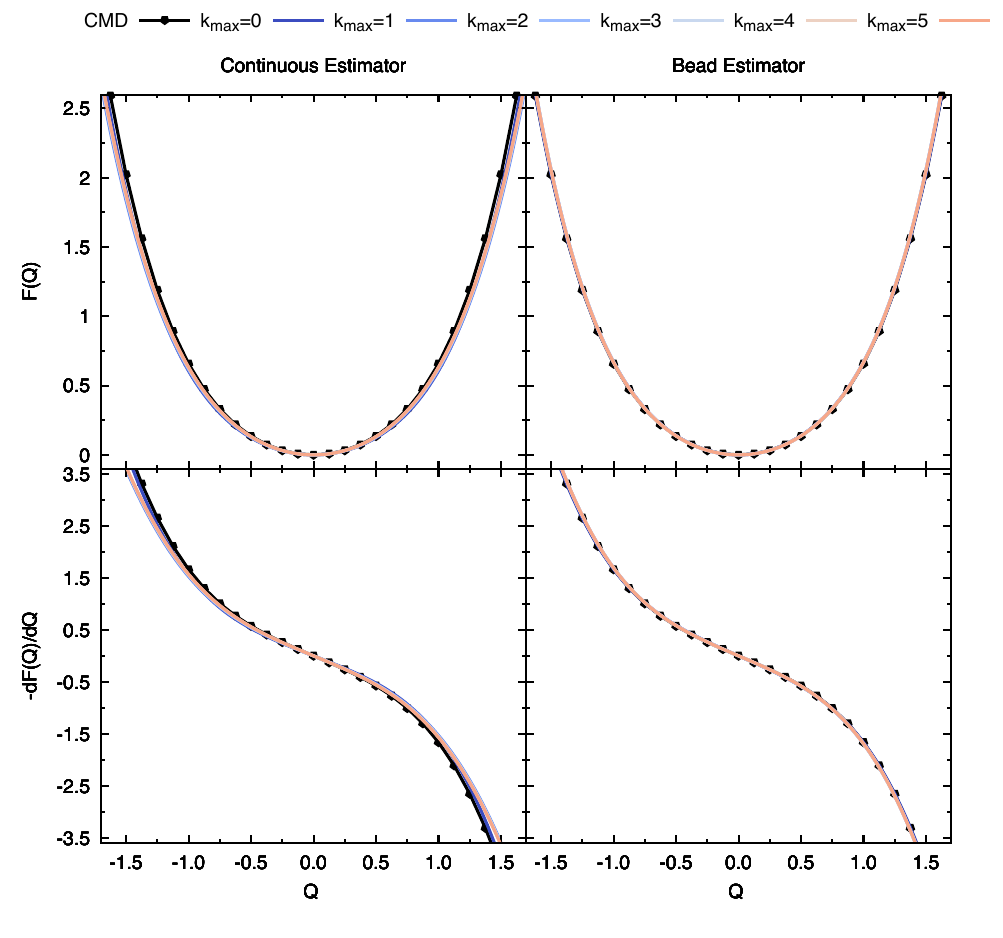}
\end{center}
\caption{Quartic oscillator free energy (top) and force (bottom) for 16 beads at $\beta=8$.}
\label{fig:quart-pot-beta8-16bd}
\end{figure}

\newpage
\begin{figure}[h!]
\begin{center}
  \includegraphics[scale=1]{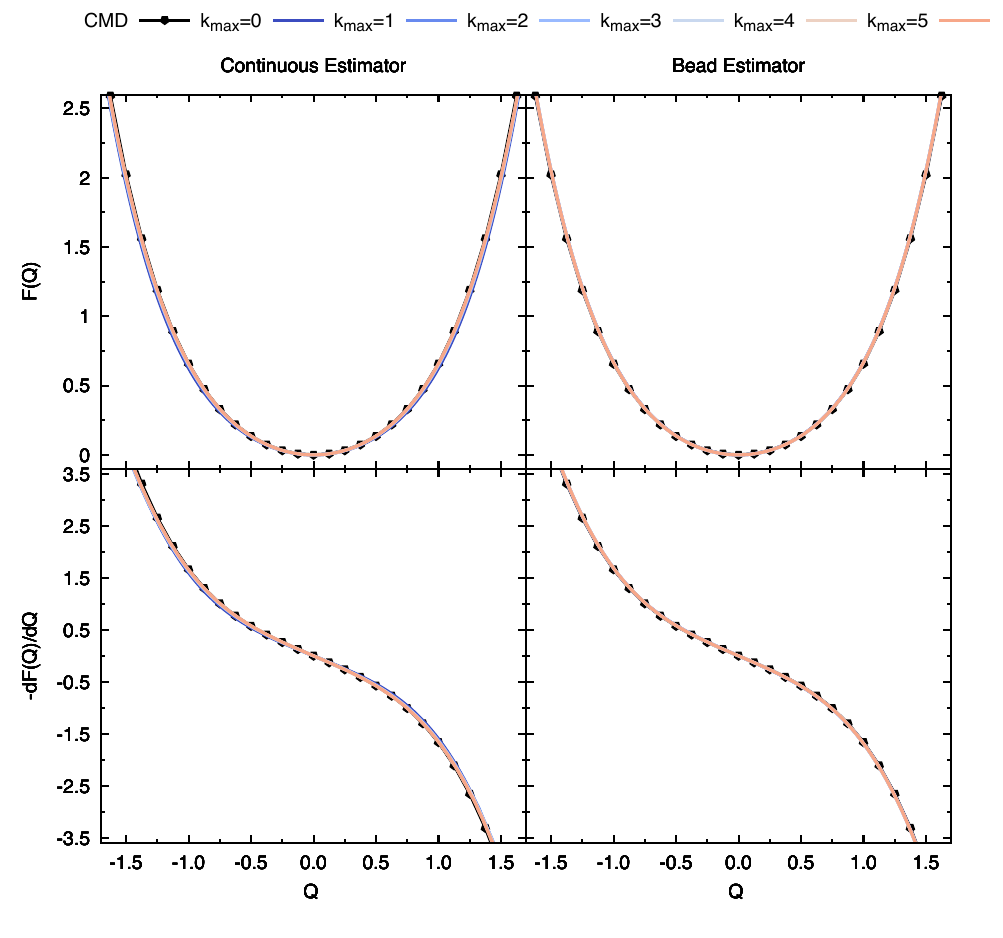}
\end{center}
\caption{Quartic oscillator free energy (top) and force (bottom) for 32 beads at $\beta=8$.}
\label{fig:quart-pot-beta8-32bd}
\end{figure}

\begin{figure}[h!]
\begin{center}
  \includegraphics[scale=1]{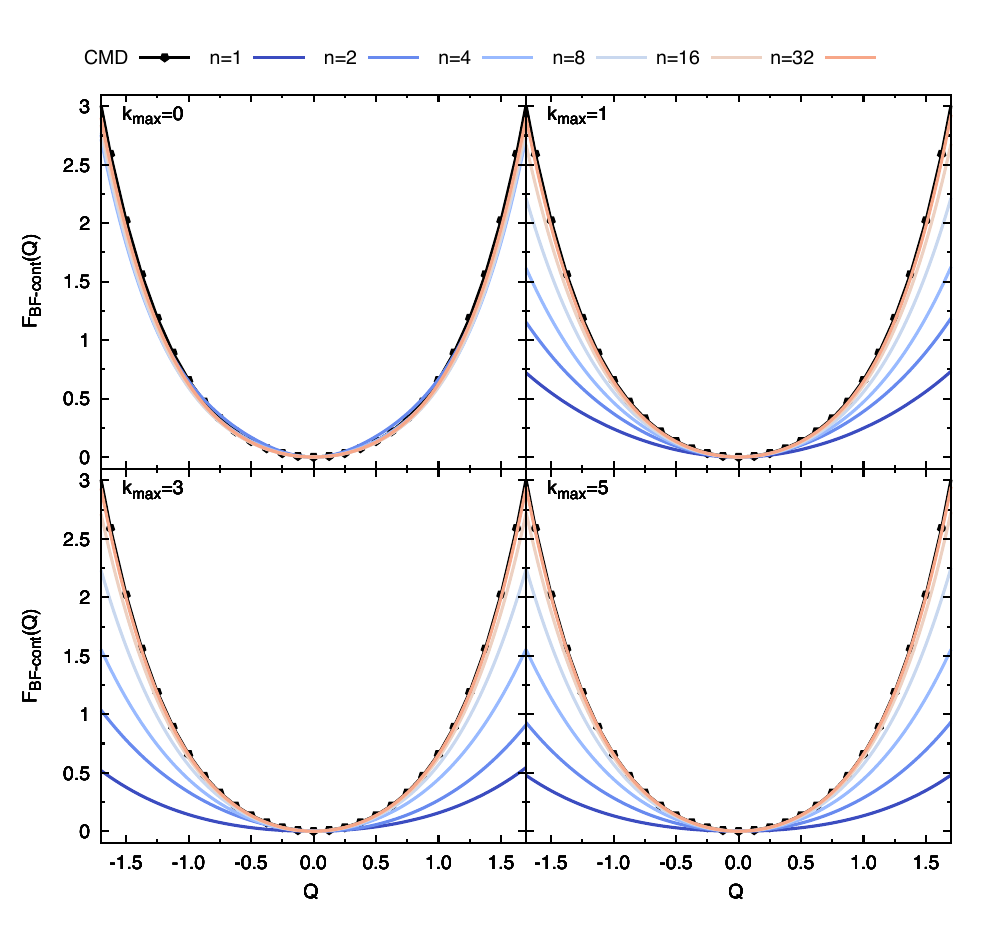}
\end{center}
\caption{Continuous estimated free energy for quartic oscillator $ \beta=8 $.}
\label{fig:quart-pot-beta8-cont}
\end{figure}

\newpage
\begin{figure}[h!]
\begin{center}
  \includegraphics[scale=1]{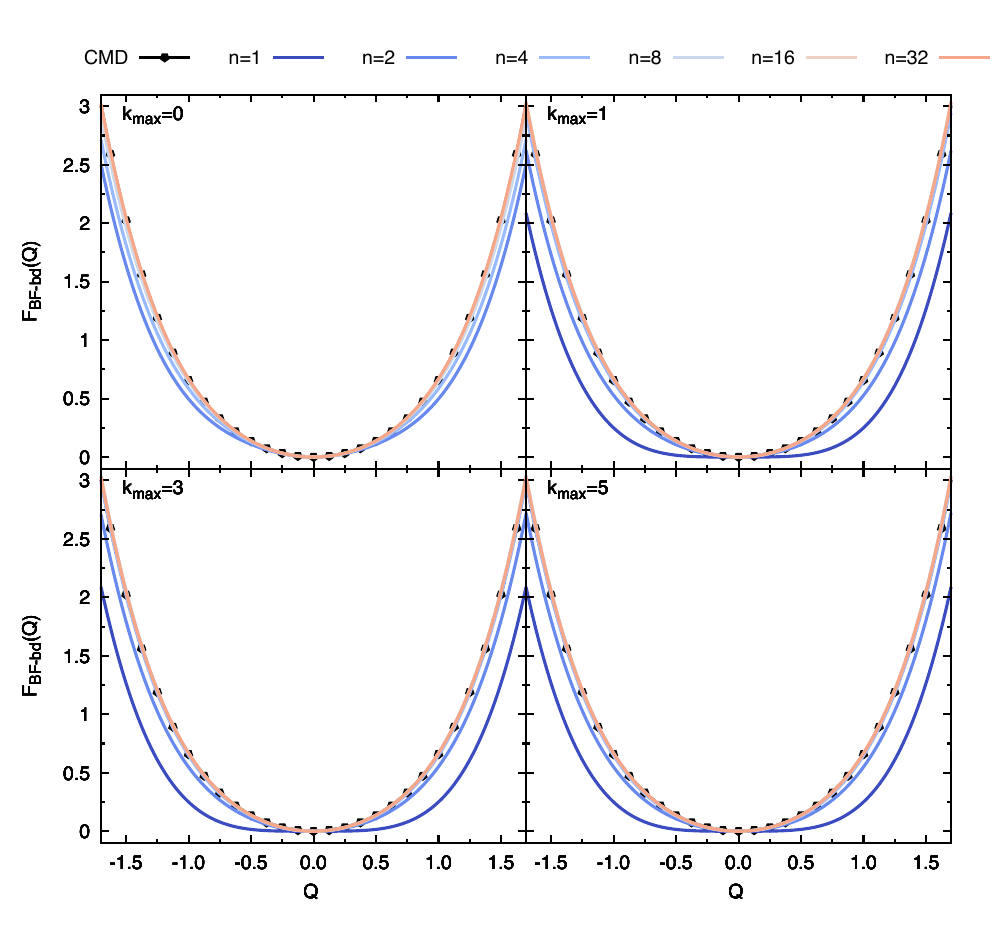}
\end{center}
\caption{Bead estimated free energy for quartic oscillator $ \beta=8 $.}
\label{fig:quart-pot-beta8-bead}
\end{figure}


\newpage
\begin{figure}[h!]
\begin{center}
  \includegraphics[scale=1]{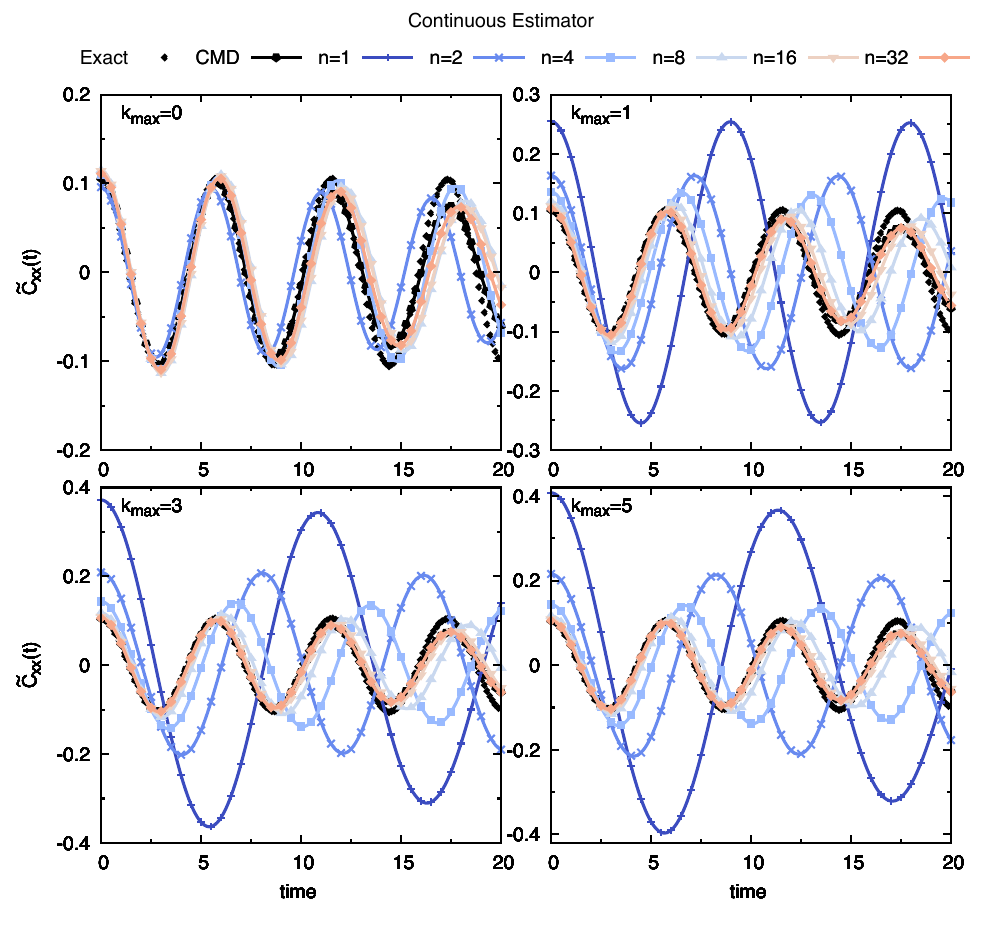}
\end{center}
\caption{Position autocorrelation function from continuous estimated free energy for quartic oscillator $ \beta=8 $.}
\label{fig:quart-corr-beta8-cont}
\end{figure}

\newpage
\begin{figure}[h!]
\begin{center}
  \includegraphics[scale=1]{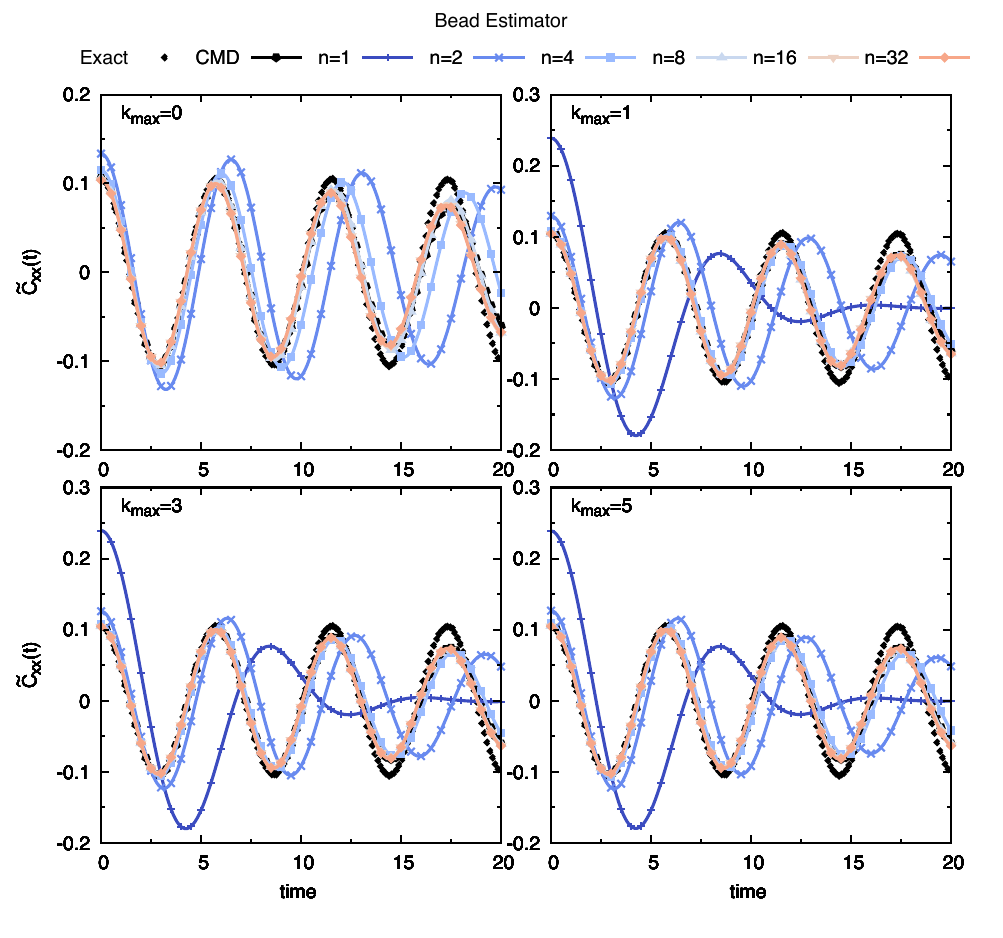}
\end{center}
\caption{Position autocorrelation function from bead estimated free energy for quartic oscillator $ \beta=8 $.}
\label{fig:quart-corr-beta8-bead}
\end{figure}